\documentclass[aip,jcp,reprint,noshowkeys,superscriptaddress]{revtex4-2}
\usepackage{graphicx,dcolumn,bm,xcolor,microtype,multirow,amscd,amsmath,amssymb,amsfonts,physics,wrapfig,bbold,siunitx,xspace,ulem,braket,txfonts}
\usepackage[version=4]{mhchem}

\usepackage[utf8]{inputenc}
\usepackage[T1]{fontenc}
\usepackage{feynmp-auto}
\usepackage{longtable}
\usepackage{booktabs}
\usepackage{hyperref}

\hypersetup{
    colorlinks,
    linkcolor={red!50!black},
    citecolor={red!70!black},
    urlcolor={red!80!black}
}

\usepackage{listings}

\definecolor{codegreen}{rgb}{0.58,0.4,0.2}
\definecolor{codegray}{rgb}{0.5,0.5,0.5}
\definecolor{codepurple}{rgb}{0.25,0.35,0.55}
\definecolor{codeblue}{rgb}{0.30,0.60,0.8}
\definecolor{backcolour}{rgb}{0.98,0.98,0.98}
\definecolor{mygray}{rgb}{0.5,0.5,0.5}

\definecolor{sqred}{rgb}{0.85,0.1,0.1}
\definecolor{sqgreen}{rgb}{0.25,0.65,0.15}
\definecolor{sqorange}{rgb}{0.90,0.50,0.15}
\definecolor{sqblue}{rgb}{0.10,0.3,0.60}

\lstdefinestyle{mystyle}{
    backgroundcolor=\color{backcolour},
    commentstyle=\color{codegreen},
    keywordstyle=\color{codeblue},
    numberstyle=\tiny\color{codegray},
    stringstyle=\color{codepurple},
    basicstyle=\ttfamily\footnotesize,
    breakatwhitespace=false,
    breaklines=true,
    captionpos=b,
    keepspaces=true,
    numbers=left,
    numbersep=5pt,
    numberstyle=\ttfamily\tiny\color{mygray},
    showspaces=false,
    showstringspaces=false,
    showtabs=false,
    tabsize=2
  }

\newcolumntype{d}{D{.}{.}{-1}}

\newcommand{\cre}[1]{a_{#1}^\dagger}
\newcommand{\ani}[1]{a_{#1}}

\newcommand{\UnHam}{Department of Chemistry, University of Hamburg, 22761 Hamburg, Germany; The Hamburg Centre for Ultrafast Imaging (CUI), Hamburg 22761, Germany; Center for Multiscale Theory and Computation, University of M\"unster, 48149 M\"unster, Germany.}

\newcommand{\IPC}{University of M\"unster, Institute of Physical Chemistry, Corrensstraße 28/30, 48149 M\"unster, Germany; Center for Multiscale Theory and Computation, 48149 M\"unster, Germany; International Graduate School BACCARA, 48149 M\"unster.}

\begin{document}	

\title{Renormalization group approach to second-order Green's function theory}

\author{Joshua Krieger}
	\affiliation{\IPC}

\author{Johannes T\"olle}
        \email{jojotoel@gmail.com}
        \affiliation{\UnHam}

\begin{abstract}
In this work, we introduce a new approach for constructing a renormalized and regularized Fock matrix for self-consistent field calculations. The scheme relies on second-order perturbation theory and is conceptually related to quasiparticle self-consistent second-order Green's function theory (GF2).
The regularization is derived within the framework of perturbative similarity renormalization group (SRG) theory.
By optimizing both the regularization and spin-scaling parameters, we introduce three SRG-qsGF2 variants that enable accurate predictions of quasiparticle energies and dipole moments.
Lastly, we demonstrate that formulating second-order perturbation theory for the total electronic energy using the renormalized SRG-qsGF2 Fock matrix as the unperturbed Hamiltonian mitigates divergence problems commonly observed in conventional Møller--Plesset perturbation theory.

\end{abstract}

\maketitle 

\section{Introduction}
The one-particle electronic Green's function is a central quantity for studying (electronic) interacting quantum many-body systems.
It encodes both spectral (dynamical) information (such as ionization potentials and electron affinities) and static ground-state properties (such as electron densities and total energies), c.f.~Refs.~\citenum{fetter2012quantum,schirmer2018many,golzeGWCompendiumPractical2019,marie2024gw}.
However, the exact computation of the Green's function is generally only feasible for systems consisting of a small number of electrons.

Therefore, several approximate procedures for computing the one-particle Green's function have been developed, c.f.~Refs.~\citenum{schirmer1983new,snijdersGF2,nooijen1992coupled,nooijen1993coupled,dahlen2005self}. 
These can be divided into procedures that either approximate the ground-state wave function which is subsequently used in construction of the one-particle Green's function ($\mathbf{G}$) \cite{schirmer1991closed,nooijen1992coupled,nooijen1993coupled,chatterjee2019second}, or approaches that correct the poles of a reference (mean-field) non-interacting Green's function $\mathbf{G}_0$ through the Dyson equation \cite{dyson1949s}
\begin{align}
    \mathbf{G}(\omega) = \mathbf{G}_0(\omega) + \mathbf{G}_0(\omega) \pmb{\Sigma}(\omega) \mathbf{G}(\omega),
\end{align}
where perturbative approximations for the self-energy $\pmb{\Sigma}$ are employed \cite{fetter2012quantum,schirmer2018many}.

A common approximation is the \textit{GW} approximation~\cite{hedinNewMethodCalculating1965}, which has emerged as a standard method for computing quasiparticle energies in both extended \cite{godby1986accurate,hybertsen1986electron,aryasetiawan1998gw} and molecular systems \cite{bruneval2012ionization,van2013gw,van2015gw,golze2018core,forster2020low,bruneval2021gw,marie2024gw}, due to its balance of accuracy and computational cost.
Within the $GW$ approximation, $\pmb{\Sigma}$ is a functional of the interacting Green's function $\mathbf{G}$.
In its most commonly used single-shot variant, $G_0W_0$~\cite{golzeGWCompendiumPractical2019}, $\pmb{\Sigma}$ depends on the initial mean-field non-interacting Green's function $\mathbf{G}_0$.
To circumvent this additional approximation, several self-consistent variants of \textit{GW} have been proposed.
These include fully self-consistent \textit{GW} (sc\textit{GW})~\cite{stanFullySelfconsistentGW2006,stan2009levels}, quasiparticle self-consistent \textit{GW} (qs\textit{GW})~\cite{vanschilfgaardeQuasiparticleSelfConsistentTheory2006,kotaniQuasiparticleSelfconsistentMethod2007,kaplanQuasiParticleSelfConsistentGW2016}, and partially self-consistent schemes such as eigenvalue-only self-consistent \textit{GW} (ev\textit{GW})~\cite{faber2011first,rangel2016evaluating}. 
\\
As an alternative to the $GW$ approximation, second-order Green's function perturbation theory (GF2) \cite{holleboomComparisonMoLler1990,dahlenSelfconsistentSolutionDyson2005,zgidGF2_1,zgidGF2_2,zgidGF2_3}, or its one-shot variant (Dyson) ADC(2) \cite{schirmer1983new,schirmer2018many} is commonly used.
A variety of self-consistent GF2 variants have been presented in the past, including self-consistency in imaginary time and frequency \cite{dahlenSelfconsistentSolutionDyson2005,zgidGF2_1,zgidGF2_2} and in a moment based formulation relying on auxiliary states \cite{boothGF2spectrum,boothAGF2}.
A procedure, to the best of our knowledge not yet extensively discussed in the literature, is the quasiparticle self-consistent GF2 (qsGF2) scheme, c.f.~Ref.~\citenum{LoosGF2}.

A potential reason for this is related to the well-known divergence issues in second-order perturbation theory, when orbital energies approach degeneracy \cite{smigaRegularizedPT2}.
For the second-order ground-state energy within Møller--Plesset (MP2) and second-order Brillouin--Wigner perturbation theory, several regularization schemes have been proposed to mitigate these issues \cite{deltaMP2,RDMP2,kappaOOMP2,evangelistaDSRG,quantum_package,carter2023optimizing,chen2025regularized}.
Regularization strategies for qsGF2 have not been explored so far. 

In this manuscript, we aim to show how the perturbative analysis of the similarity renormalization group (SRG) flow equations \cite{SRG_1,SRG_2,SRG_3} of GF2, as recently applied to the qs\textit{GW} formalism in Ref.~\citenum{marieSRG}, can be used to derive a regularized qsGF2 scheme (SRG-qsGF2).
Subsequently, we will demonstrate that the resulting quasiparticle energies, when combined with a suitable regularization parameter and spin-scaling techniques \cite{scs_1,scs_2,cruz2025performance}, result in mean-absolute and mean-signed errors (MAEs/MSEs) that rival those of self-consistent \textit{GW} variants \cite{scGW_1,scGW_2,gw100_qsgw}.
Furthermore, we demonstrate the good accuracy of the proposed methodology for determining dipole moments for a benchmark set of 46 molecular systems, achieving CCSD level accuracy when compared to experimental values.
Lastly, we make use of the SRG-qsGF2 Fock matrix in a second-order perturbation theory procedure for determining the ground-state electronic energy.
Here, we demonstrate that the divergence issues along the  dissociation coordinate for typical molecular systems are significantly mitigated, extending the applicability of second-order perturbation theory for ground-state energies.

\section{Theory}\label{sec:theo}
\subsection{Second-order Green's Function Perturbation Theory}\label{sec:theo:gf2pt}
Throughout this work we use the common notation for occupied ($i,j,k,\dots$) and virtual ($a,b,c,\dots$) and general ($p,q,r,\dots$) orbital indices.
Making use of this notation, the second-order Green's function perturbation theory (GF2) self-energy is defined as\cite{holleboomComparisonMoLler1990,boothAGF2,boothGF2spectrum}
\begin{align}
    \Sigma_{pq}^{\mathrm{GF2}}(\omega) &= \frac{1}{2}\sum_{ija}\frac{\bra{pa} \ket{ij}\bra{ij}\ket{qa}}{\omega+\epsilon_{a}-\epsilon_{i}-\epsilon_{j}} \nonumber \\
    &+ \frac{1}{2}\sum_{iab}\frac{\bra{pi}\ket{ab}\bra{ab}\ket{qi}}{\omega+\epsilon_{i}-\epsilon_{a}-\epsilon_{b}} \nonumber \\
    &= \left[ \mathbf{W}^{\mathrm{2h1p}}\left(\omega \mathbf{1}-\mathbf{C}^{\mathrm{2h1p}}\right)^{-1}\left(\mathbf{W}^{\mathrm{2h1p}}\right)^{\dagger} \right]_{pq} \nonumber \\
    &+ \left[ \mathbf{W}^{\mathrm{2p1h}}\left(\omega \mathbf{1}-\mathbf{C}^{\mathrm{2p1h}}\right)^{-1}\left(\mathbf{W}^{\mathrm{2p1h}}\right)^{\dagger} \right]_{pq},\label{eq:selfenergy}
\end{align}
with 
\begin{align}
    C_{ija,klb}^{\mathrm{2h1p}} &= (-\epsilon_{a}+\epsilon_{i}+\epsilon_{j})\delta_{ik}\delta_{jl}\delta_{ab},\\
    C_{abi,cdj}^{\mathrm{2p1h}} &= (-\epsilon_{i}+\epsilon_{a}+\epsilon_{b})\delta_{ac}\delta_{bd}\delta_{ij},
\end{align}
and
\begin{align}
    W_{p,ija}^{\mathrm{2h1p}} &= \frac{1}{\sqrt{2}}\bra{pa}\ket{ij},\\
    W_{p,abi}^{\mathrm{2p1h}} &= \frac{1}{\sqrt{2}}\bra{pi}\ket{ab},
\end{align}
where $\epsilon_{p}$ denotes orbital energies.
Furthermore, we use the following convention for the two-electron integrals 
\begin{align}
    \bra{pq} \ket{rs} &= \int d\mathbf{x} \int d\mathbf{x}' \frac{ \phi_p(\mathbf{x}) \phi_r(\mathbf{x}) \phi_q(\mathbf{x}') \phi_s(\mathbf{x}')}{|\mathbf{r} - \mathbf{r}'|}\\
                      &- \int d\mathbf{x} \int d\mathbf{x}' \frac{ \phi_p(\mathbf{x}) \phi_s(\mathbf{x}) \phi_q(\mathbf{x}') \phi_r(\mathbf{x}')}{|\mathbf{r} - \mathbf{r}'|},
\end{align}
assuming real-valued orbitals throughout this work. 

Note that the underlying static effective Hamiltonian, which will be used in Sec.~\ref{sec:theo:srg} is 
\begin{align}
    \mathbf{H}^{\mathrm{GF2}} = \left(
    \begin{matrix}
        \mathbf{F} & \mathbf{W}^{\mathrm{2h1p}} & \mathbf{W}^{\mathrm{2p1h}}\\
        \left(\mathbf{W}^{\mathrm{2h1p}}\right)^{\dagger} & \mathbf{C}^{\mathrm{2h1p}} & \mathbf{0}\\
        \left(\mathbf{W}^{\mathrm{2p1h}}\right)^{\dagger} & \mathbf{0} & \mathbf{C}^{\mathrm{2p1h}}\\
    \end{matrix}\right).
    \label{eq:HGF2}
\end{align}  
The effective Hermitian static self-energy used in the quasiparticle self-consistent (qs) GF2 procedure is\cite{ismail-beigi2017,marieSRG}
\begin{align}
 F^{\mathrm{qsGF2}}_{pq} &= \frac{1}{2} \left[ \Sigma_{pq}^{\mathrm{GF2}}(\epsilon_p)  + \Sigma_{pq}^{\mathrm{GF2}}(\epsilon_q)  \right] \nonumber \\
 &= \frac{1}{4}  \sum_{ija}\frac{\Delta^{pa}_{ij} + \Delta^{qa}_{ij}}{\Delta^{pa}_{ij} \cdot \Delta^{qa}_{ij}} \bra{pa}\ket{ij}\bra{ij}\ket{qa} \nonumber \\
 &+ \frac{1}{4}  \sum_{abi}\frac{\Delta^{pi}_{ab} + \Delta^{qi}_{ab}}{\Delta^{pi}_{ab} \cdot \Delta^{qi}_{ab}} \bra{pi}\ket{ab}\bra{ab}\ket{qi},
 \label{eq:qsGF2F}
\end{align}
with
\begin{align}
    \Delta_{rs\dots}^{pq\dots} = \epsilon_{p} + \epsilon_{q} + \dots - \epsilon_{r} - \epsilon_{s} - \dots~.
    \label{eq:denom}
\end{align}
Augmenting the Hartree--Fock Fock operator with Eq.~(\ref{eq:qsGF2F}) allows to determine the orbitals and orbital energies self-consistently from
\begin{align}
    \underbrace{\left[\hat{f}^\mathrm{HF}+\hat{f}^{\mathrm{\mathrm{qsGF2}}}\right]}_{\hat{f}^{\mathrm{tot}}}\ket{\phi_{p}} = \epsilon_{p}\ket{\phi_{p}},\label{eq:qpequation}
\end{align}
with 
\begin{align}
    \hat{f}^\mathrm{HF} = \sum_{pq} F_{pq}^\mathrm{HF} \cre{p}\ani{q},
\end{align}
where ${F}^\mathrm{HF}_\mathrm{pq}$ denotes the Hartree--Fock Fock matrix,
and 
\begin{align}
    \hat{f}^{\mathrm{qsGF2}} = \sum_{pq} F_{pq}^{\mathrm{qsGF2}} \cre{p}\ani{q}.
\end{align}
\subsection{Similarity renormalization group}\label{sec:theo:srg}
Since orbital energy differences enter the denominator of $F^\mathrm{qsGF2}_{pq}$, it is prone to divergences. 
Several techniques have been proposed to mitigate this issue, cf.~Refs.~\citenum{kappaOOMP2,smigaRegularizedPT2,coveneyQPMP2,shee2021regularizedPT2}.
Here, we make use of the similarity renormalization group (SRG) approach \cite{evangelistaDSRG,SRG_1,SRG_2,SRG_3} to derive a renormalized effective Fock matrix $\tilde{F}(s)$ expression depending on a regularization parameter $s$.
This idea closely follows Ref.~\citenum{marieSRG}, where the SRG framework has been applied to the GW approximation.

The general idea of the SRG approach is the continuous transformation of a Hamiltonian matrix such that certain subspaces become decoupled, i.e.,\cite{kehrein2006}
\begin{align}
    \hat{H}(s) = \hat{U}(s) \hat{H} \hat{U}^\dagger(s),
    \label{eq:transformation}
\end{align}
with $\hat{U}(s)$ denoting a unitary anti-Hermitian transformation matrix [$\hat{U}(s) = -\hat{U}^\dagger (s)$] and $s$ the so-called flow parameter that is defined in the range of $\left[ 0,\infty \right)$. Furthermore, $\hat{U}(0)=1$ so that $\hat{H}$ remains unchanged for $s=0$.

The evolution of Eq.~(\ref{eq:transformation}) follows the ordinary differential equation\cite{kehrein2006}
\begin{align}
    \frac{d \hat{H}(s)}{d s} = \left[\eta,\hat{H}(s) \right],\label{eq:flow}
\end{align}
with 
\begin{align}
    \hat{\eta} = \frac{d \hat{U}(s)}{d s} \hat{U}^\dagger (s). 
\end{align}
In this work, Wegner's canonical generator \cite{SRG_1}
\begin{align}
    \eta(s) = \left[\hat{H}_\mathrm{d}(s),\hat{H}(s)\right]=\left[ \hat{H}_\mathrm{d}(s),\hat{H}_\mathrm{od}(s) \right]
\end{align}
is chosen as the flow generator. 
Wegner's generator partitions the Hamiltonian into a diagonal $(\hat{H}_\mathrm{d})$ and an off-diagonal $(\hat{H}_\mathrm{od})$ part, resulting in a monotonic decrease of the off-diagonal elements.
This holds true as long as $\eta(s)\neq0$.\cite{kehrein2006}
\subsubsection{Perturbative analysis}
Following Refs.~\citenum{evangelistaDSRG,marieSRG}, a perturbative analysis of the flow equations for the effective Hamiltonian [Eq.~(\ref{eq:HGF2})] will be performed in the following.
The diagonal part of the effective Hamiltonian reads 
\begin{align}
    \mathbf{H}_{\mathrm{d}} =    
    \left(
    \begin{matrix}
        \mathbf{F} & \mathbf{0} & \mathbf{0}\\
        \mathbf{0} & \mathbf{C}^{\mathrm{2h1p}} & \mathbf{0}\\
        \mathbf{0} & \mathbf{0} & \mathbf{C}^{\mathrm{2p1h}}\\
    \end{matrix}
    \right),
\end{align}
and the off-diagonal part is 
\begin{align}
    \mathbf{H}_{\mathrm{od}} =     
    \left(
    \begin{matrix}
        \mathbf{0} & \mathbf{W}^{\mathrm{2h1p}} & \mathbf{W}^{\mathrm{2p1h}}\\
        \left(\mathbf{W}^{\mathrm{2h1p}}\right)^{\dagger} & \mathbf{0} & \mathbf{0}\\
        \left(\mathbf{W}^{\mathrm{2p1h}}\right)^{\dagger} & \mathbf{0} & \mathbf{0}\\
    \end{matrix}
    \right).
\end{align}
The perturbative analysis of the SRG equations starts from [compare Eq.~(\ref{eq:HGF2})]
\begin{align}
    \mathbf{H}^{\mathrm{qsGF2}} = \mathbf{H}_{\mathrm{d}} + \lambda \mathbf{H}_{\mathrm{od}},
\end{align}
where $\mathbf{H}_{\mathrm{od}}$ is the perturbation and $\lambda$ denotes the perturbation parameter. 
The SRG Hamiltonian [$\mathbf{H}(s,\lambda)$] and the flow generator [$\pmb{\eta}(s,\lambda)$] can now be expanded in a power series in $\lambda$, and at $N$-th order one finds
\begin{align}
\frac{1}{N!} \left. \frac{d^N \mathbf{H}(s,\lambda)}{d \lambda^N} \right|_{\lambda=0} = \mathbf{H}^{(N)}(s),        
\end{align}
and 
\begin{align}
\frac{1}{N!} \left. \frac{d^N \pmb{\eta}(s,\lambda)}{d \lambda^N} \right|_{\lambda=0} = \pmb{\eta}^{(N)}(s).
\end{align}
Since at zeroth order the perturbation is zero, the zeroth order Hamiltonian is independent of $s$, i.e., $\mathbf{H}^{(0)}(s) = \mathbf{H}^{(0)}(0) = \mathbf{H}_{\mathrm{d}}$. 
Similarly, the zeroth order off-diagonal Hamiltonian ($\mathbf{H}^{(0)}_\mathrm{od} (s) = 0$) and flow generator are zero ($\pmb{\eta}^{(0)}(s) = 0$). 

At second order, one finds 
\begin{align}
    \frac{d \mathbf{H}^{(2)}(s)}{d s} = \left[ \pmb{\eta}^{(1)}(s), \mathbf{H}^{(1)}(s) \right] +  \left[ \pmb{\eta}^{(2)}(s), \mathbf{H}^{(0)}_\mathrm{d} \right].
\end{align}
\\
Following the derivation presented in Ref.~\citenum{marieSRG}, the one-particle/one-hole block $\mathbf{F}^{(2)}(s)$ reads 
\begin{align}
    F_{pq}^{(2)}(s)  &= \frac{1}{2}\sum_{ija}\frac{\Delta^{pa}_{ij} + \Delta^{qa}_{ij}}{\left(\Delta^{pa}_{ij}\right)^2+\left(\Delta^{qa}_{ij}\right)^2} \bra{pa}\ket{ij}\bra{ij}\ket{qa}
    \nonumber \\
    &\times\left[1-e^{-\left[\left(\Delta^{pa}_{ij}\right)^2+\left(\Delta^{qa}_{ij}\right)^2\right]s}\right] \nonumber \\
    &+ \frac{1}{2}\sum_{abi}\frac{\Delta^{pi}_{ab} + \Delta^{qi}_{ab}}{\left(\Delta^{pi}_{ab}\right)^2+\left(\Delta^{qi}_{ab}\right)^2} \bra{pi}\ket{ab}\bra{ab}\ket{qi}
    \nonumber \\
    &\times\left[1-e^{-\left[\left(\Delta^{pi}_{ab}\right)^2+\left(\Delta^{qi}_{ab}\right)^2\right]s}\right]\label{eq:qsGF2SE}.
\end{align}
A detailed derivation is also presented in the SI, Sec.~S1.
From Eq.~(\ref{eq:qsGF2SE}) it is evident that the second-order renormalized Fock matrix ${\mathbf{F}}^{(2)}(s)$ is zero for $s=0$ and has the form of static renormalization of quasiparticle energies similar to Eq.~(\ref{eq:qpequation}) for $s\rightarrow \infty$.
In particular both expressions share the same diagonal elements in this limit.
We therefore denote ${\mathbf{F}}^{(2)}(s)$ as $\mathbf{F}^{\mathrm{SRG-qsGF2}}(s)$ in the following.

Based on $\mathbf{F}^{\mathrm{SRG-qsGF2}}(s)$ a self-consistency procedure similar to Eq.~(\ref{eq:qpequation}) can be defined, resulting in the SRG-qsGF2 method.
The computational cost of the construction of the static self-energy $\mathbf{F}^{\mathrm{SRG-qsGF2}}(s)$ exhibits a formal scaling of $\mathcal{O}(n_{\mathrm{o}}n_{\mathrm{v}}^{2}n_{\mathrm{bf}}^{2})$, where $n_{\mathrm{o}}$ and $n_{\mathrm{v}}$ are the number of occupied and virtual orbitals, and $n_{\mathrm{bf}}$ denotes the total number of basis functions. The transformation of the four center integrals required for $\mathbf{F}^{\mathrm{SRG-qsGF2}}(s)$ into molecular orbital (MO) basis however scales on the order $\mathcal{O}(n_{\mathrm{bf}}^{5})$, resulting in an overall $\mathcal{O}(n_{\mathrm{bf}}^{5})$ scaling of the SRG-qsGF2 method. Improvements to the computational efficiency will be briefly discussed in Sec.~\ref{sec:comp}.
\subsection{Spin-Component Scaling}
Within second-order Møller--Plesset perturbation theory (MP2), the overall accuracy can be drastically improved through scaling of same- and opposite-spin correlation energy contributions ($c_{\mathrm{SS}}$ and $c_{\mathrm{OS}}$)\cite{scs_1,scs_2}.

Similarly\cite{hirata_scs_gf2_2025}, we apply the spin-component scaling (SCS) scheme to the SRG-qsGF2 self-energy.
Dividing the two-electron integrals into same-spin and opposite-spin contributions, the SCS-SRG-qsGF2 spin-integrated self-energy reads 
\begin{align}
    F_{pq}^{\mathrm{SRG-qsGF2}}(s) =& \sum_{ija}\frac{\Delta^{pa}_{ij} + \Delta^{qa}_{ij}}{\left(\Delta^{pa}_{ij}\right)^2+\left(\Delta^{qa}_{ij}\right)^2}\left( pi | aj \right)\nonumber \\
    &\times\left[(c_{\mathrm{SS}}+c_{\mathrm{OS}})\left( qi | aj \right)-c_{\mathrm{SS}} \left( pi | aj \right) \right] \nonumber \\
    &\times\left[1-e^{-\left[\left(\Delta^{pa}_{ij}\right)^2+\left(\Delta^{qa}_{ij}\right)^2\right]s}\right]  \nonumber \\
    +&\sum_{abi}\frac{\Delta^{pi}_{ab} + \Delta^{qi}_{ab}}{\left(\Delta^{pi}_{ab}\right)^2+\left(\Delta^{qi}_{ab}\right)^2}\left( pa | ib \right) \nonumber \\
    &\times\left[(c_{\mathrm{SS}}+c_{\mathrm{OS}})\left( qa | ib \right)-c_{\mathrm{SS}}\left( qb | ia \right) \right]\nonumber \\
     &\times\left[1-e^{-\left[\left(\Delta^{pi}_{ab}\right)^2+\left(\Delta^{qi}_{ab}\right)^2\right]s}\right],
\end{align}
with $(pq|rs)$ denoting the two-electron integrals over spatial orbitals in chemists' notation.
We investigate the choice of $c_{\mathrm{SS}}$ and $c_{\mathrm{OS}}$ in Sec.~\ref{sec:app:qp}.

\subsection{Renormalized Quasiparticle Møller--Plesset Perturbation Theory}
\label{sec:theo:D}
Making use of the renormalized SRG-qsGF2 Fock matrix, an alternative partitioning of one- and two-electron terms in the electronic Hamiltonian can be achieved
\begin{align}
    \hat{H} &= \tilde{f} + \hat{V} \label{eq:HamPartitioning}\\
    \tilde{f} &= \sum_{pq} \left[ F^\mathrm{HF}_{pq} + F^{\mathrm{SRG-qsGF2}}_{pq}(s) \right]  a_p^\dagger a_q \\
    \hat{V} &= \hat{H} - \tilde{f},
\end{align}
allowing for a perturbative treatment of the interaction, similar to Møller--Plesset perturbation theory
\begin{align}
    \hat{H} &= \tilde{f} + \lambda \hat{V}.
\end{align}
In the following, we will not take the $s$ dependence of the orbitals and $\tilde{f}$ (\textit{vide infra}) into account and refer to the perturbation expansion as QP-PT. 
The motivation for this repartitioning is the inclusion of static self-energy-like contributions, which are expected to yield a better zeroth-order description of the system compared to the HF Fock matrix and might accelerate the overall convergence of the perturbation series and mitigate divergence \cite{knowlesConvergence1988,knowles2022perturbation,gombas2024analysis,szabados2024knowles,dittmer2025repartitioning}.

QP-PT0 is given as the sum over the occupied quasiparticle orbital energies, and the first order energy correction (QP-PT1) reads
\begin{align}
    E^{(1)} &= \bra{\Phi}\hat{V}\ket{\Phi} \nonumber \\
    &= \sum_{ij}-\frac{1}{2}\bra{ij}\ket{ij}-F_{ii}^{\mathrm{SRG-qsGF2}}.
\end{align}
At second order (QP-PT2), we obtain
\begin{align}
    E^{(2)} &= \sum_{ia}\frac{|F_{ia}^{\mathrm{SRG-qsGF2}}|^2}{^{\mathrm{QP}} \Delta_a^i} + \frac{1}{4}\sum_{{ij,ab}}\frac{\abs{\bra{ij}\ket{ab}}^{2}}{^{\mathrm{QP}}\Delta_{ab}^{ij}},\label{eq:qsgf2energy}
\end{align}
with $^{\mathrm{QP}}\Delta_{ab\dots}^{ij\dots}$ denoting the quasiparticle orbital energy differences following the definition of Eq.~\ref{eq:denom}.
Thus, the full QP-PT2 energy reads
\begin{align}
    E^{\mathrm{QP-PT2}} &= E^{(0)} + E^{(1)} + E^{(2)} \nonumber \\
    &= E^{\mathrm{HF}} + E^{(2)}.
\end{align}
We would like to point out again that the orbitals used for evaluating the HF energy are not the HF orbitals, but the quasiparticle orbitals obtained within the SRG-qsGF2 procedure.

Although the appearance of quasiparticle energies in the denominators of Eq.~(\ref{eq:qsgf2energy}) is likely to mitigate divergencies in the QP-PT2 energy correction, they could still occur (compare Sec.~\ref{sec:app:pes}).
For MP2 several regularization strategies have been developed in the literature~\cite{kappaOOMP2,smigaRegularizedPT2,coveneyQPMP2,shee2021regularizedPT2}, like the SRG-MP2 formalism of Evangelista \cite{evangelistaDSRG}.
In this case the MP2 energy reads
\begin{align}
    E^{\mathrm{SRG-MP2}}(s) &= \nonumber \\
    &\frac{1}{4}\sum_{ijab}\frac{\abs{\bra{ij}\ket{ab}}^{2}}{\Delta_{ab}^{ij}}\left[1-e^{-2s\left(\Delta_{ab}^{ij}\right)^2}\right].
\end{align}
Since the SRG formalism was already adapted to the qsGF2 self-energy, it will be also used for regularizing the QP-PT2 energy correction. 
Following the derivation of Ref.~\citenum{evangelistaDSRG}, the SRG-QP-PT2 correlation energy is 
\begin{align}
    &E^{\mathrm{SRG-QP-PT2}}(s)=  \nonumber \\
    &\sum_{ia}\frac{\abs{F_{ia}^{\mathrm{SRG-qsGF2}}}^{2}}{^{\mathrm{QP}}\Delta_{a}^{i}}\left[1-e^{-2s\left(^{\mathrm{QP}}\Delta_a^i\right)^2}\right] \nonumber\\
    &+ \frac{1}{4}\sum_{ijab}\frac{\abs{\bra{ij}\ket{ab}}^{2}}{^{\mathrm{QP}}\Delta_{ab}^{ij}}\left[1-e^{-2s\left(^{\mathrm{QP}}\Delta_{ab}^{ij}\right)^2}\right].\label{eq:e2srg}
\end{align} 
A derivation is also presented in the SI, Sec.~S2.
\section{Computational Details}
\label{sec:comp}
The methods outlined in Sec.~\ref{sec:theo} was implemented within the RI approximation~\cite{densfit_1,densfit_2} in a development version of \textsc{Serenity} \cite{serenity_1,serenity_2,serenity_1.6.1}, and will be made publicly available in the upcoming release 1.7.0.

The SRG-qsGF2 with the RI approximation has a formal scaling of $\mathcal{O}(n_{\mathrm{o}}n_{\mathrm{v}}n_{\mathrm{bf}}^{2}n_{\mathrm{aux}})$, where $n_{\mathrm{o}}$ and $n_{\mathrm{v}}$ are the number of occupied and virtual orbitals, $n_{\mathrm{bf}}$ the number of basis functions and $n_{\mathrm{aux}}$ the number of basis functions in the auxiliary basis.

To reduce the number of auxiliary basis functions, we make use of naturally auxiliary functions (NAF) \cite{nafs_1,nafs_2,tolle2024accelerating}.
We have found that an eigenvalue threshold of $10^{-2}$ yields an error in the quasiparticle energies on the same order as the error introduced by the RI approximation.
We will therefore choose the eigenvalue threshold of $10^{-2}$ in all calculations using NAFs, unless stated otherwise.
A detailed discussion of using the NAF approximation in the present context, related to thresholds and starting point dependence can be found in the SI, Sec.~S3.
For all qsGF2 calculations, performed in this work, we employ the DIIS~\cite{diis1,diis2,diis3} method for convergence acceleration and make use of the following convergence thresholds for self-consistent field procedures: i) $5 \times 10^{-8}~\mathrm{a.u.}$
for the change in energy; ii) $1 \times 10^{-8}~\mathrm{a.u.}$ for the root mean square deviation (RMSD) of the
density matrix; iii) and a threshold of $5 \times 10^{-7}~\mathrm{a.u.}$ for the largest coefficient in the DIIS $\left[\mathbf{F},\mathbf{P}\right]$ error vector~\cite{diis2}.
Throughout Hartree atomic units (a.u.) are used.

\section{Application}\label{sec:app}

\subsection{Quasiparticle energies}\label{sec:app:qp}

First, we will investigate the performance of the SRG-qsGF2 method for calculating principal quasiparticle energies (IPs and EAs) and investigate the role of the flow-parameter $s$.
For this, we will make use of the GW50 benchmark set~\cite{marieSRG}, which consists of 50 atomic and small molecular systems.
Reference $\Delta$CCSD(T) ionization potentials (IPs) and electron affinities (EAs) have been taken from Ref.~\citenum{marieSRG}.
All calculations on the GW50 benchmark set were performed in aug-cc-pVTZ basis set\cite{augccpvtz_1,augccpvtz_2,augccpvtz_3,augccpvtz_4} and corresponding auxiliary basis set \cite{pVTZ_RI_1,pvtz_RI_2}, except for systems containing lithium, sodium, or potassium.
In these cases, the def2-TZVPPD\cite{def2_RI,def2svpd_tzvppd_qzvppd_RI} auxiliary basis set is used.

Table~\ref{tab:srg-maes} displays the MAEs of the IPs and EAs for different values of the SRG parameter $s$. 
From Eq.~(\ref{eq:qsGF2SE}) it can be seen that the regularization vanishes for $s\to\infty$, while for $s = 0$, the second-order contribution to the Hamiltonian completely vanishes, yielding the Hartree--Fock quasiparticle energies.
Ideally, $s$ is chosen as large as possible while still ensuring the convergence of the calculations.
However, already for $s=10$ the calculation for the ozone molecule does not converge, and the number of non-converging calculations increases further for larger values of $s$.
At the same time, the MAEs of the IPs and EAs significantly increase when going from $s=1$ to $s=10$, and decrease again for higher values of $s$. 
Examples of convergent and divergent molecular systems with respect to the choice of $s$ can be found in the SI, Sec .~S4A.
Note that the $s$-dependence for SRG-qs$GW$ is significantly smaller~\cite{marieSRG}.
\begin{table}[htbp]
    \centering
    \caption{Number of converged (conv.) SRG-qsGF2 calculations, mean absolute errors (MAE) and mean signed error (MSE) for ionization potentials (IPs) and electron affinities (EAs) in eV relative to $\Delta$CCSD(T) reference values for the GW50 benchmark set in aug-cc-pVTZ basis set.}
    \label{tab:srg-maes}
    \begin{tabular}{lcccccccc}
        \toprule
        \midrule
        $s$ & \multicolumn{2}{c}{1} & \multicolumn{2}{c}{10} & \multicolumn{2}{c}{100} & \multicolumn{2}{c}{1000}\\
        \midrule
        conv. & \multicolumn{2}{c}{50} & \multicolumn{2}{c}{49} & \multicolumn{2}{c}{45} & \multicolumn{2}{c}{31} \\
        \midrule
        & IP & EA & IP & EA & IP & EA & IP & EA\\
        \midrule
        MAE & 0.446 & 0.189 & 1.388 & 0.256 & 1.325 & 0.243 & 1.255 & 0.156 \\
        MSE & $-$0.347 & $-$0.005 & $-$1.388 & 0.083 & $-$1.325 & 0.114 & $-$1.255 & -0.005 \\
        \midrule
        \bottomrule
    \end{tabular}
\end{table}

Following the ideas of regularization strategies used in the context of orbital-optimized MP2 \cite{kappaOOMP2}, we will use $s$ as an empirical parameter to minimize the MAE of the IPs and EAs for the GW50 benchmark set. 
Details of the optimization process are presented in the SI, Sec.~S4B.
Optimization of the SRG-parameter (without employing spin scaling) in order to balance the overall error in the IPs and EAs for the GW50 benchmark set shows that the IP error exhibits a significantly larger dependence on $s$ than the EAs. The optimal value for $s$ with respect to the IP MAE is $s=0.525$, resulting in a MAE of $0.25~\mathrm{eV}$ for IPs and an MAE of $0.20~\mathrm{eV}$ for EAs. 
The resulting method will be denoted as SRG-qsGF2 in the following.

Next, we investigate the possibility of further reducing the error in the quasiparticle energies by applying the spin component scaling formalism~\cite{scs_1,scs_2} within the SRG-qsGF2 procedure.
Similarly to the SRG parameter optimization, the variation of the EA MAE across different parameter choices is significantly smaller than the IP MAE (SI, Sec.~S4B).
We therefore focus on minimizing the IP error first.
The optimization of the SCS parameters and the SRG parameter with respect to the IP error yields $c_\mathrm{SS} = 0.0$, $c_{\mathrm{OS}} = 1.0$, and $s = 1.4$ as the optimal set of parameters, and will thus be denoted SRG-SOS-qsGF2. This choice of parameters shows a minimal IP MAE of $0.13~\mathrm{eV}$, however, the EA MAE increases to $0.26~\mathrm{eV}$ from $0.20~\mathrm{eV}$.
We note that a minimal EA MAE of $0.189~\mathrm{eV}$ is recorded at $c_{\mathrm{SS}} = c_{\mathrm{OS}} = s = 1.0$, however, the IP MAE increases to $0.447~\mathrm{eV}$, hence, we do not consider optimizing for the EA error to be expedient.
While optimizing for the EA MAE alone might not be feasible, searching for a set of parameters that considers both the IP and EA errors was also explored.
A set of parameters that balances IP and EA MAEs can be obtained by normalizing the errors according to the maximum observed IP and EA MAE, respectively
\begin{equation}
    \begin{aligned}
        F^{\mathrm{MAE}}(\mathbf{c}) = \frac{\mathrm{IP}^{\mathrm{MAE}}(\mathbf{c})}{max(\mathrm{IP}^{\mathrm{MAE}}(\mathbf{c}))} + \frac{\mathrm{EA}^{\mathrm{MAE}}(\mathbf{c})}{max(\mathrm{EA}^{\mathrm{MAE}}(\mathbf{c}))}.
    \end{aligned}
\end{equation}
Here, $\mathbf{c}$ is a vector of the three parameters $\mathbf{c} = (\begin{matrix} c_\mathrm{SS} & c_\mathrm{OS} & s \end{matrix})$
The optimized parameter become $\mathbf{c} = (\begin{matrix}
    c_\mathrm{SS} & c_\mathrm{OS} & s
\end{matrix}) = (\begin{matrix}
    0.6 & 1.0 & 0.7
\end{matrix})$, resulting in an IP MAE of $0.21~\mathrm{eV}$ and an EA MAE of $0.20~\mathrm{eV}$. 
This parametrization will be referred to as SRG-SCS-qsGF2 in the following.

The optimal spin-scaling factors found for both SRG-SOS-qsGF2 and SRG-SCS-qsGF2 differ significantly from the commonly used scaling factors in the case of SCS/SOS-MP2\cite{scs_1,scs_2,sos}.
These are $c_{\mathrm{SS}}=0.33$ and $c_{\mathrm{OS}}=1.20$ for SCS-MP2 and $c_{\mathrm{SS}} = 0.00$, $c_{\mathrm{OS}} = 1.30$ for SOS-MP2. 
However, the SRG-SOS-qsGF2 factors match the IP optimized $\Delta$MP2-SOS(IP)\cite{deltamp2_1,deltamp2_2,deltamp2_3,deltamp2_4} parameters ($c_{\mathrm{CS}} = 0.0$, $c_{\mathrm{OS}} = 1.0$). 
In the case of SRG-SCS-qsGF2, both $c_{\mathrm{SS}} = 0.6$ and $c_{\mathrm{OS}} = 1.0$ are each approximately 1.5 times larger than the $\Delta$MP2(IP) parameters ($c_{\mathrm{SS}} = 0.38$ and $c_{\mathrm{OS}} = 0.67$) of Refs.~\citenum{deltamp2_1,deltamp2_2,deltamp2_3,deltamp2_4}.
This difference could be related to the stronger regularization in the case of the SRG-SCS-qsGF2 variant, which results in an increased suppression of contributions with small denominators of $\mathbf{F}^{\mathrm{SRG-qsGF2}}$, requiring larger SCS parameters to recover the lost contribution.

\begin{figure}
    \includegraphics[width=.99\columnwidth]{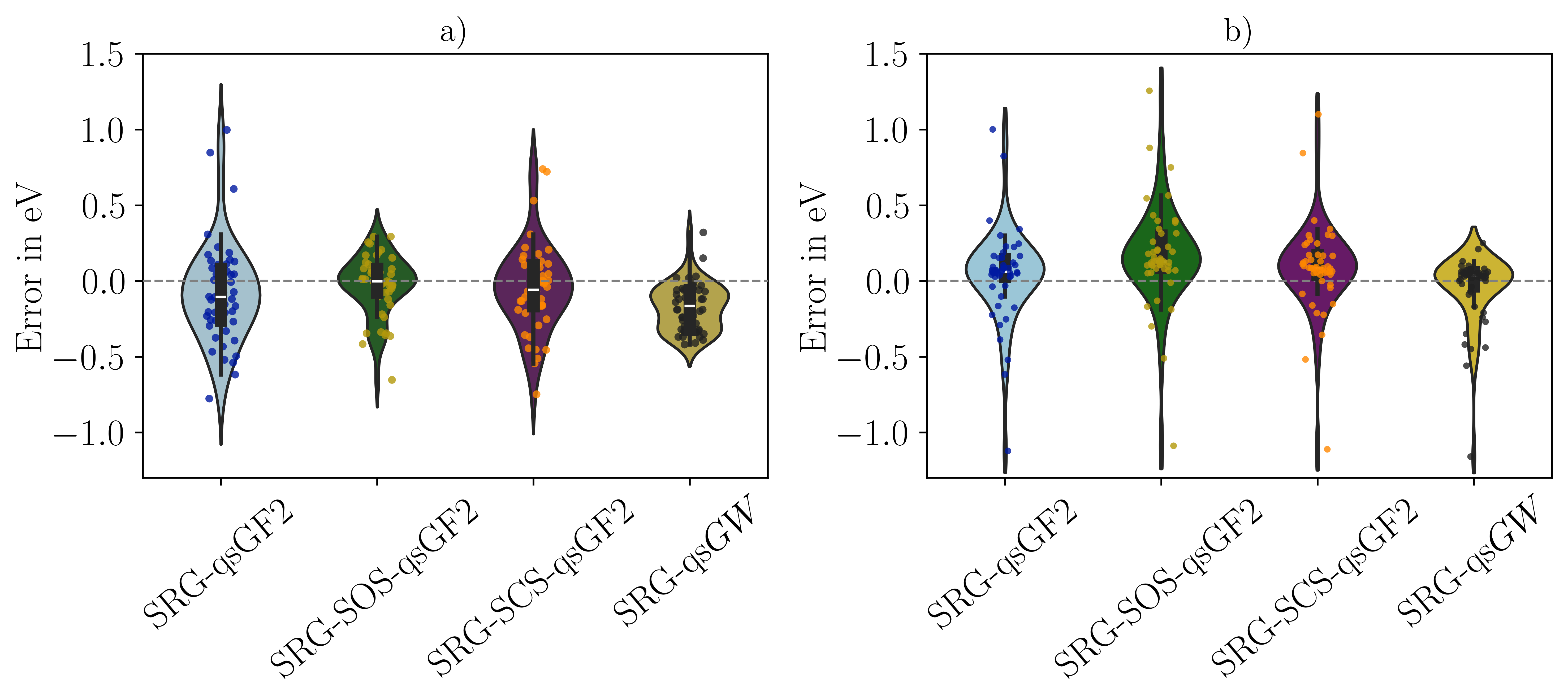}
    \caption{Violinplots of the HOMO a) and LUMO energy b) error distributions in the GW50 benchmark set for SRG-qsGF2, SRG-SOS-qsGF2 and SRG-SCS-qsGF2, as well as SRG-qs\textit{GW}\cite{marieSRG}. The data for SRG-qs\textit{GW} are taken from \citenum{marieSRG}.
    All qsGF2 calculations were performed using the aug-cc-pVTZ basis\cite{augccpvtz_1,augccpvtz_2,augccpvtz_3,augccpvtz_4}.}\label{fig:gw50violin}
\end{figure}

Fig.~\ref{fig:gw50violin} depicts the IP and EA error distributions with respect to $\Delta$CCSD(T) of the three qsGF2 parametrizations (SRG-qsGF2, SRG-SOS-qsGF2, SRG-SCS-qsGF2) for the GW50 benchmark set.
Additionally, the error distribution for the SRG-qs\textit{GW} procedure of Ref.~\citenum{marieSRG} ($s= 1000$) is presented, exhibiting an IP and EA MAE of $0.19$/$0.12$ eV. 
It can be seen that the SRG-qsGF2 procedure results in a wider spread in the deviations for IPs and EAs when compared to SRG-qs\textit{GW}, manifesting in increased MAEs. 
In contrast, SRG-SOS-qsGF2 exhibits a similar error distribution for the IPs, but a wider spread for EAs.
Overall, the spread in the EA errors remains fairly consistent between the qsGF2 variants.
For all three SRG-qsGF2 variants, the largest deviations in IPs and EAs are caused by systems that were particularly prone to divergencies in qsGF2.
For example, in the case of SRG-SCS-qsGF2, the largest outliers in the IPs and EAs arise for MgO ($-0.75$~eV) and C$_{2}$H$_{4}$ ($1.11$~eV), respectively.
Both systems start diverging already for smaller $s$. 
In the case of MgO, a small $s$ parameter actually increases the IP error of MgO compared to HF and only improves the IP for larger $s$ [see Fig.~S2~c) in the SI].
While not quite matching the accuracy of the SRG-qs\textit{GW} procedure, SRG-qsGF2, and particularly its spin-scaled variants, yield reasonably accurate IPs and EAs at reduced computational cost [$\mathcal{O}(n^{5})$ compared to $\mathcal{O}(n^{6})$ for SRG-qs\textit{GW} within their sum-over-states expressions].

\begin{figure*}
    \centering
    \includegraphics[width=\textwidth]{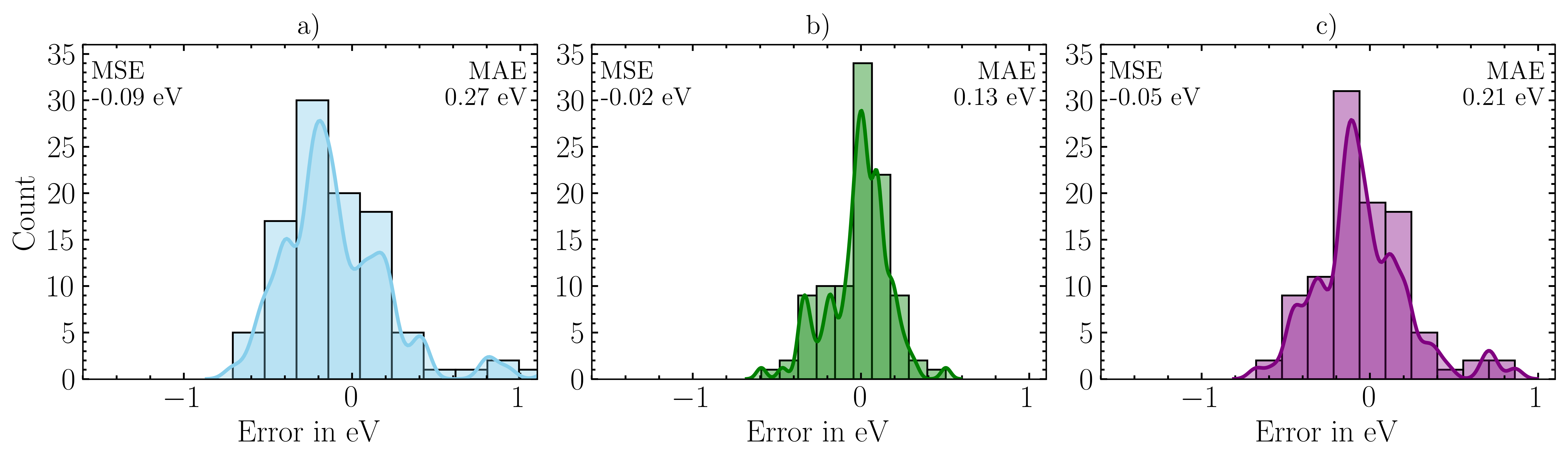}
    \caption{Histogram of the errors with respect to $\Delta$CCSD(T) for the IPs of the  GW100 benchmark set \cite{bruneval2021gw,GW100}: a) SRG-qsGF2, b) SRG-SOS-qsGF2 and c) SRG-SCS-qsGF2. All calculations were performed using the def2-TZVPP basis set.~\cite{def2svp_tzvp_tzvppd,def2_RI}}
    \label{fig:GW100_hist}
\end{figure*}

Next, we investigate the transferability of the parametrizations, deduced from the GW50 benchmark set, for the three SRG-qsGF2 variants against the full GW100 benchmark set.
In this case, the def2-TZVPP basis set\cite{def2svp_tzvp_tzvppd} with the corresponding auxiliary basis set has been employed.\cite{def2_RI,def2svpd_tzvppd_qzvppd_RI}. 
The results for the GW100 benchmark set were evaluated using the updated CCSD(T) reference energies presented in Ref.~\citenum{bruneval2021gw}. 
All SRG-qsGF2 results were obtained using the revised geometries for vinyl bromide and phenole from Ref~\citenum{van2015gw}, while the original geometries were used for {$G_0$$W_0$}@HF, qs\textit{GW}, and sc$GW$. Furthermore, in the case of sc$GW$, IP energies for seven systems were unavailable.
The resulting metrics for the deviations in the IPs with respect to $\Delta$CCSD(T) reference values are presented in Tab.~\ref{tab:GW100}, and the error distributions for the IPs are shown in Fig.~\ref{fig:GW100_hist}. Tab.~\ref{tab:GW100} also contains reference $G_{0}W_{0}$@HF, qs\textit{GW} (not regularized) and sc\textit{GW} results of Refs.~\citenum{gw100_qsgw}.
From Fig.~\ref{fig:GW100_hist} and Tab.~\ref{tab:GW100} it is evident that the performance of all three SRG-qsGF2 variants is retained when moving from the GW50 to the full GW100 benchmark set.
The resulting MAEs for the IPs are $0.27~\mathrm{eV}$ in the case of SRG-qsGF2, $0.13~\mathrm{eV}$ for SRG-SOS-qsGF2, and $0.21~\mathrm{eV}$ for SRG-SCS-qsGF2.
When comparing SRG-qsGF2 and SRG-SCS-qsGF2 to \textit{G$_0$W$_0$}@HF and qs\textit{GW} (without regularization), comparable MAEs are observed. 
The SRG-SOS-qsGF2 variant results in the lowest overall MAE, RMSD, and standard deviation of error (SDE).
Furthermore, the SRG-qsGF2 MAE is comparable to the sc$GW$ MAE, while the two other SRG-qsGF2 parametrizations perform better. 
Notably, a smaller MSE for all three SRG-qsGF2 variants compared to \textit{G$_0$W$_0$}@HF, qs\textit{GW}, and sc$GW$ is found, indicating that the error distribution largely centered around zero, compare also Fig.~\ref{fig:GW100_hist}. 

\begin{table}[htbp]
    \centering
    \caption{Metrics of the principal IP errors (in eV) of GW100 benchmark with respect to $\Delta$CCSD(T) for SRG-sqGF2, SRG-SOS-qsGF2, SRG-SCS-qs-GF2, $G_{0}W_{0}$@HF\cite{hybertsen1986electron}, qs\textit{GW}~\cite{kaplanQuasiParticleSelfConsistentGW2016} and sc$GW$~\cite{scGW_1,scGW_2}. Calculations were performed using the def2-TZVPP basis set~\cite{def2svp_tzvp_tzvppd,def2_RI}.}
    \label{tab:GW100}
    \begin{tabular}{lcccc}
        \toprule
        \midrule
        Method & MAE & MSE & RMSD & SDE \\
        \midrule
        SRG-qsGF2 & 0.27 & --0.09 & 0.34 & 0.33 \\
        SRG-SOS-qsGF2 & 0.13 & --0.02 & 0.18 & 0.18 \\
        SRG-SCS-qsGF2 & 0.21 & --0.05 & 0.28 & 0.28 \\
        \textit{G$_{0}$W$_{0}$}@HF & 0.30 & --0.26 & 0.36 & 0.24 \\
        qs\textit{GW} & 0.21 & --0.18 & 0.27 & 0.20 \\
        sc\textit{GW} & 0.29 & --0.27 & 0.36 & 0.25 \\
        \midrule
        \bottomrule 
    \end{tabular}
\end{table}

Having established three parameterized SRG-qsGF2 variants for the GW50 benchmark set and explored the transferability to the larger GW100 benchmark set, the performance beyond quasiparticle energies will be assessed next.

\subsection{Dipole moments}\label{sec:app:dip}
Having parametrized three different SRG-qsGF2 variants with respect to quasiparticle energies, we will now assess their performance for calculating molecular dipole moments.
For this, the dipole moments of several prototypical small molecular systems are investigated in a first step.
Calculations were performed using the def2-QZVPPD basis set~\cite{def2svpd_tzvppd_qzvppd_RI} with geometries taken from Ref.~\citenum{sekinoMolecularHyperpolarizabilities1993}.

Table~\ref{tab:res:dip:1} compares the errors in the dipole moments calculated using the different SRG-qsGF2 parametrizations, as well as dipole moments from HF and CCSD(T)~\cite{sekinoMolecularHyperpolarizabilities1993}.
All SRG-qsGF2 parametrizations significantly improve upon the HF results, resulting in MAEs comparable to those of CCSD(T) when compared to experiment.
Notably, all three SRG-qsGF2 variants correctly predicted the sign of the CO dipole moment, a notoriously difficult problem for HF theory~\cite{dipoleCO}.
\begin{table}[htbp]
    \centering
    \caption{Comparison of the errors (in Debye) of dipole moments relative to experimental values~\cite{adamoAccurateStaticPolarizabilities1999} for CCSD(T)~\cite{sekinoMolecularHyperpolarizabilities1993}, HF, SRG-qsGF2, SRG-SCS-qsGF2, and SRG-SOS-qsGF2.
    Calculations were performed using the def2-QZVPPD basis set.~\cite{def2svpd_tzvppd_qzvppd_RI}}
    \label{tab:res:dip:1}
    \begin{tabular}{lcccccc}
        \toprule
        \midrule
        molecule & exp. & CCSD(T) & HF & \shortstack{SRG-\\qsGF2} & \shortstack{SRG-SCS-\\qsGF2} & \shortstack{SRG-SOS-\\qsGF2} \\
        \midrule
        HF & 1.82 & --0.04 & 0.10 & --0.04 & --0.03 & --0.01 \\
        CO & 0.11 & 0.04 & --0.38 & 0.13 & 0.10 & 0.04 \\
        H$_2$O & 1.85 & -0.02 & 0.13 & --0.03 & 0.02 & 0.03 \\
        NH$_3$ & 1.47 & 0.03 & 0.12 & 0.05 & 0.05 & 0.06 \\
        H$_2$S & 0.97 & 0.01 & 0.19 & 0.04 & 0.03 & 0.02 \\
        \midrule
        MAE & & 0.03 & 0.17 & 0.06 & 0.05 & 0.03 \\
        MSE & & 0.00 & 0.02 & 0.03 & 0.03 & 0.03 \\
        max & & 0.04 & 0.38 & 0.13 & 0.10 & 0.06 \\
        \midrule
        \bottomrule
        \end{tabular}
\end{table}

To further assess the performance of the different SRG-qSGF2 variants in predicting dipole moments, we benchmarked them against a larger set of 37 molecular systems.
This benchmark set has been previously employed to evaluate the accuracy of various electronic structure methods to determine dipole moments \cite{dipole_benchmark}. 
The set comprises primarily small organic molecules along with several diatomic species, including those listed in Table~\ref{tab:res:dip:1}.
Since the geometries were not provided in Ref.~\citenum{dipole_benchmark}, we generated them as described below:
The set of 37 geometries was obtained via MP2 geometry optimizations with \textsc{ORCA 6.0.1}~\cite{orca_1,orca_2,orca_3,orca_4,orca_5} of the minimum energy conformers generated by CREST~\cite{crest_1,crest_2,crest_3}. The OO-MP2 results in this benchmark were also calculated using \textsc{ORCA 6.0.1}.
The complete results and the included molecules can be found in Tab.~S2 in the SI.

First, we compare dipole moments on the newly generated using MP2 to the MP2 results from Ref.~\citenum{dipole_benchmark}.
We find an overall mean absolute difference between the two of $0.009$~Debye.
The largest deviation of $0.14~\mathrm{Debye}$ is found for cytosine.
Notably, the MP2 dipole moments from the geometries used in this study are in better agreement with the experimental values of Ref.~\citenum{dipole_benchmark}, yielding a MAE of $0.093~\mathrm{Debye}$ compared to $0.098~\mathrm{Debye}$ as reported in Ref.~\citenum{dipole_benchmark} (Table~\ref{tab:res:dip:2}). 
Given this overall small deviation, we include the comparison to CCSD of Ref.~\citenum{dipole_benchmark} in the present study.

The MAEs and RMSDs of the three SRG-qsGF2 methods relative to experimental reference values are summarized in Table~\ref{tab:res:dip:2}.
All three SRG-qsGF2 variants provide a more accurate description of dipole moments than OO-MP2.
While SRG-qsGF2 delivers dipole moments of comparable quality to MP2, the spin-component-scaled variants exhibit substantially smaller deviations from experiment.
In particular, SRG-SOS-qsGF2 not only performs best among the three parametrizations, but also outperforms all other methods listed in Table~\ref{tab:res:dip:2}.
\begin{table}[h!]
    \centering
    \caption{MAE and RMSD in Debye w.r.t to experimental dipole moments taken from Ref.~\citenum{dipole_benchmark} (cc-pVTZ\cite{augccpvtz_1,augccpvtz_2,augccpvtz_3,augccpvtz_4}).}
    \label{tab:res:dip:2}
    \begin{tabular}{lcc}
        \toprule
        \midrule
        method & MAE & RMSD\\
        \midrule
        CCSD$^\mathrm{a)}$ & 0.079 & 0.133\\
        SRG-qsGF2 & 0.096 & 0.149\\
        SRG-SCS-qsGF2 & 0.088 & 0.142\\
        SRG-SOS-qsGF2 & 0.079 & 0.129\\
        OO-MP2 & 0.113 & 0.129\\
        MP2 & 0.093 & 0.150 \\
        MP2$^\mathrm{a)}$ & 0.098 & 0.168\\
        \midrule
        \bottomrule
    \end{tabular}
    \\
    \footnotesize{a) taken from Ref.~\citenum{dipole_benchmark}}
\end{table}

\subsection{Potential Energy Surfaces}\label{sec:app:pes}

Next, we investigate the ability of the SRG-qsGF2 in combination with second-order perturbation theory (PT2, Sec.~\ref{sec:theo:D}) to describe the dissociation curves of H$_2$, N$_2$, and H$_{10}$.
For this, we make use of the Hamiltonian partitioning of Eq.~(\ref{eq:HamPartitioning}).
The resulting methods will be denoted as QP($M$)-PT2, with $M$ referring to the three parametrization for the underlying SRG-qsGF2 approach also used of the PT2 correction, i.e., QP-PT2 in the case of SRG-qsGF2, QP(SCS)-PT2 in the case of SRG-SCS-qsGF2, and QP(SOS)-PT2 in the case of SRG-SOS-qsGF2.
Additionally, we consider SRG-qsGF2 in combination with second-order perturbation theory for the limit $s\rightarrow \infty$, denoted as QP($\infty$)-PT2.
Furthermore, we compare the resulting potential energy surfaces (PES) to CCSD, MP2, OO-MP2~\cite{OOMP2_1,OO-MP2_1} and its regularized variant $\kappa$-OO-MP2~\cite{kappaOOMP2,kappaSigmaOOMP2}.
H$_{2}$ calculations were performed using the cc-pVDZ basis set~\cite{augccpvtz_1,augccpvtz_2,augccpvtz_3,augccpvtz_4,pVTZ_RI_1,pvtz_RI_2}, the sto-3g~\cite{sto-Ng,sto-ng_1,sto-ng_2} basis set for N$_{2}$, and cc-pVTZ~\cite{augccpvtz_1,augccpvtz_2,augccpvtz_3,augccpvtz_4,pVTZ_RI_1,pvtz_RI_2} for H$_{10}$.
The CCSD calculations for H$_{2}$ and H$_{10}$ were performed using \textsc{Serenity} 1.6.3.~\cite{serenity_1,serenity_2,serenity_1.6.1}
CCSD calculations for N$_{2}$ and all OO-MP2 calculations were performed with \textsc{ORCA}~\cite{orca_1,orca_2,orca_3,orca_4,orca_5} 6.0.1, and the $\kappa$-OO-MP2 calculations were performed with \textsc{QChem 6.1}~\cite{qchem}.
Reference dissociation curves were obtained from full configuration interaction (FCI) calculations for H$_2$ and N$_2$ through \textsc{PySCF}~\cite{pyscf_1,pyscf_2,libcint_pyscf_3}, and MRCI+Q calculations for H$_{10}$ taken from Ref.~\citenum{mrci}.
All calculations are based on restricted Hartree--Fock (RHF) reference wave functions.

\begin{figure*}
    \centering
    \includegraphics[width=\linewidth]{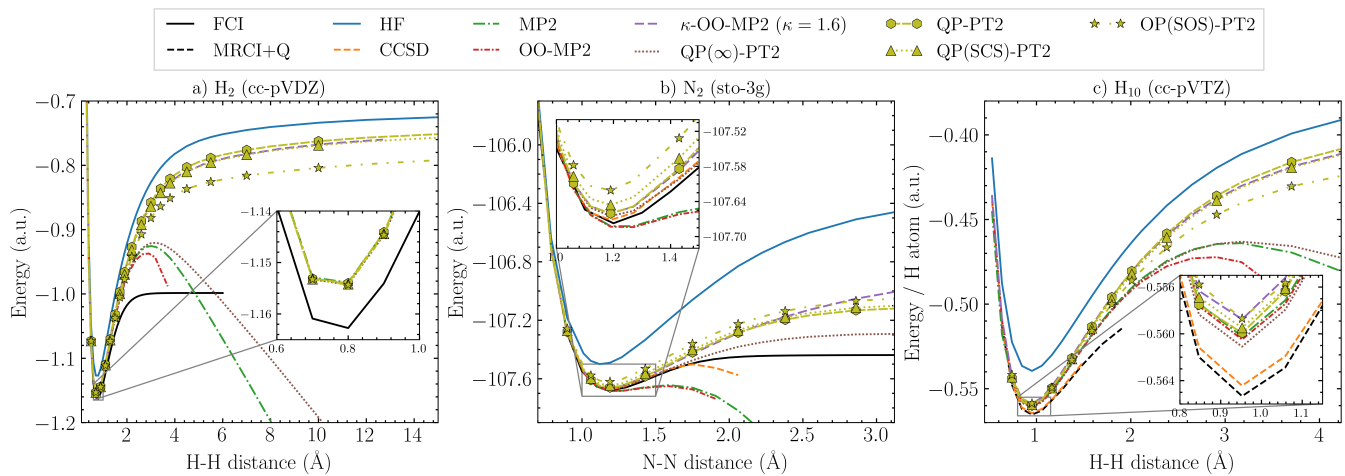}
    \caption{
        Dissociation curves for a)~H$_{2}$ (cc-pVDZ), b)~N$_{2}$ (sto-3g) and c) a linear H$_{10}$ (cc-pVTZ). 
        The MRCI+Q results were taken from Ref.~\citenum{mrci}
    }
    \label{fig:pes_curves} 
\end{figure*} 
The dissociation curves are shown in Figure~\ref{fig:pes_curves}.
Figure~\ref{fig:pes_curves}~a) depicts the ground-state dissociation curve of H$_{2}$.
Around the equilibrium distance, all second-order methods give similar results, which are about $10^{-2}~\mathrm{Ha}$ higher in energy than the FCI energy.
Note that CCSD is identical to FCI in this case and is not explicitly shown.
At larger distances (around 3~\AA) MP2 starts to diverge and fails to qualitatively describe the PES. 
OO-MP2 diverges at shorter distances, and the optimization procedure did not converge beyond 4~\AA.
This behavior is cured by $\kappa$-OO-MP2, which does not diverge, but exhibits discontinuities beyond 12~\AA.

The unregularized expression for QP($\infty$)-PT2 also diverges, but at a slightly larger distance when compared to MP2.
The QP(\textit{M})-PT2 energies for the H$_{2}$ system are independent of $c_{\mathrm{SS}}$, as the single occupied orbital results in the cancellation of the same-spin contribution, and since the parameter sets all use $c_{\mathrm{OS}} = 1$, the only quantitative differences in the potential energy surfaces are due to the different choice of regularization parameter $s$, together with the reference orbitals and quasiparticle energies from the underlying SRG-qsGF2 calculations. 
In the case of QP(SOS)-PT2, the regularization parameter is largest with $s=1.4$, hence, the energy difference to the FCI results is the smallest.

In case of the N$_{2}$ dissociation [Figure~\ref{fig:pes_curves} b)], OO-MP2 and MP2 diverge at relatively short internuclear distances.
Also, CCSD starts diverging at a distance of around 1.8~\AA\, and the amplitude optimization becomes unstable beyond 2.1~\AA.
Interestingly, the QP($\infty$)-PT2 energies do not diverge along the potential energy curve and result in the overall best agreement with the FCI reference. 
In contrast, the regularized QP(\textit{M})-PT2 energies start to differ around 1.2~\AA.
The regularized energies (QP(\textit{M})-PT2 and $\kappa$-OO-MP2) appear to approach a constant value in the dissociation limit.
However, the QP(\textit{M})-PT2 variants provide a better qualitative description compared to the $\kappa$-OO-MP2 energies, plateauing at shorter distances.
Within the QP(\textit{M})-PT2 procedures, an offset for QP(SOS)-PT2 is observed, whereas QP-PT2 and QP(SCS)-PT2 result in overall similar energies.

For the linear H$_{10}$ molecule, the dissociation curves of the different methods show a qualitatively similar behavior as for the dissociation of H$_{2}$.
Around the equilibrium distance, the CCSD energy (per hydrogen atom) agrees within  $~$ 1~mHA of the MRCI+Q reference, while the various second-order approaches deviate by 5 to 8~mHa per hydrogen atom. 
However, at distances larger than 1.5~\AA, the CCSD calculations fails to converge, while the MP2, OO-MP2, and QP($\infty$)-PT2 energies start diverging at distances beyond 3~\AA.
OO-MP2 becomes numerically unstable beyond a bond length of 3.5~\AA.
In contrast, QP(\textit{M})-PT2 and $\kappa$-OO-MP2 yield non-divergent PESs, with the dissociation behavior mimicking that of H$_2$. 

\section{Conclusion and Outlook}\label{sec:conclusion}
In this work, we investigate the performance of the quasiparticle self-consistent GF2 (qsGF2) method in combination with a regularization procedure deduced from the perturbative analysis at second-order within the similarity renormalization group (SRG) framework, denoted as SRG-qsGF2.

By investigating the convergence behavior of SRG-qsGF2 with respect to the regularization parameter $s$ for the GW50 benchmark set, we decided to choose $s$ such that the MAE for IPs and EAs is minimized within this benchmark set. 
Furthermore, we have explored the possibility of applying spin-component scaling (SCS) procedures within the SRG-qsGF2 framework to further reduce the error, resulting in three different SRG-qsGF2 parametrizations, which we denoted as SRG-qsGF2, SRG-SCS-qsGF2, and SRG-SOS-qsGF2.
We assessed the transferability of these parametrizations by benchmarking error in IPs of the full GW100 benchmark set, finding similar performance when compared to the GW50 benchmark set.
Notably, the SRG-SOS-qsGF2 parametrization yields lower MAE, RMSD, and SDE than \textit{G$_0$W$_0$}@HF, qs\textit{GW} and sc$GW$ \cite{kaplanQuasiParticleSelfConsistentGW2016}.

In the next step, we investigated the performance of the different SRG-qsGF2 parametrizations for calculating molecular dipole moments for prototypical small molecules as well as a larger set of 37 small molecular systems, finding that all three parametrizations significantly improve upon HF results, yielding dipole moments that are comparable to CCSD. Furthermore, we found an improved description when compared to MP2 and OO-MP2.
This highlights the promise of the SRG-qsGF2 procedure for the description of molecular properties.

Finally, we investigated the usage of the effective SRG-qsGF2 Hamiltonian [Eq.~(\ref{eq:HamPartitioning})] in second-order perturbation theory. 
By redefining the fluctuation potential, this approach is free of double counting and allows for the usage of quasiparticle energies and orbitals instead of Hartree--Fock ones in regular MP2. 
Applying the same regularization procedure to the energy correction, the QP-PT2 energies yield descriptions of comparable accuracy to other second-order methods, and cure the divergent behavior, as well as show improved stability at larger bond lengths.

Having established the SRG-qsGF2 framework, several avenues for future research open up.
For example, the application of the SRG-qsGF2 procedure for calculating excited states via either a $\Delta$SCF approach or linear-response theory is an interesting direction to pursue \cite{gilbert2008self,paetow2025excited}.
Particularly, leveraging the Hamiltonian partitioning of Eq.~(\ref{eq:HamPartitioning}) in combination with CC2\cite{christiansen1995second}/ADC(2)\cite{trofimov1995efficient} might improve the description of excitation energies.
Furthermore,  reducing the overall computational cost of the SRG-qsGF2 procedure via the usage of efficient integral decomposition techniques such as tensor hypercontraction (THC) \cite{hohenstein2012tensor,parrish2012tensor,matthews2020improved} or alternative compression approaches is of interest \cite{boothAGF2,boothGF2spectrum,scott2023moment,backhouse2025self,tolle2024ab}.

\section*{Acknowledgement}
% \section*{Funding}
J.T. acknowledges funding from the Fonds der Chemischen Industrie (FCI) via a Liebig fellowship and support by the Cluster of Excellence “CUI: Advanced Imaging of Matter” of the Deutsche Forschungsgemeinschaft (DFG) (EXC 2056, funding ID 390715994).
J.K. acknowledges funding from by the Ministry of Culture and Science of the State North Rhine--Westphalia within the International Graduate School for Battery Chemistry, Characterization, Analysis, Recycling and Application (BACCARA).
%%%%%%%%%%%%%%%%%%%%%%%%%%%%%%%%
\section*{References}
%%%%%%%%%%%%%%%%%%%%%%%%%%%%%


\begin{thebibliography}{124}%
\makeatletter
\providecommand \@ifxundefined [1]{%
 \@ifx{#1\undefined}
}%
\providecommand \@ifnum [1]{%
 \ifnum #1\expandafter \@firstoftwo
 \else \expandafter \@secondoftwo
 \fi
}%
\providecommand \@ifx [1]{%
 \ifx #1\expandafter \@firstoftwo
 \else \expandafter \@secondoftwo
 \fi
}%
\providecommand \natexlab [1]{#1}%
\providecommand \enquote  [1]{``#1''}%
\providecommand \bibnamefont  [1]{#1}%
\providecommand \bibfnamefont [1]{#1}%
\providecommand \citenamefont [1]{#1}%
\providecommand \href@noop [0]{\@secondoftwo}%
\providecommand \href [0]{\begingroup \@sanitize@url \@href}%
\providecommand \@href[1]{\@@startlink{#1}\@@href}%
\providecommand \@@href[1]{\endgroup#1\@@endlink}%
\providecommand \@sanitize@url [0]{\catcode `\\12\catcode `\$12\catcode `\&12\catcode `\#12\catcode `\^12\catcode `\_12\catcode `\%12\relax}%
\providecommand \@@startlink[1]{}%
\providecommand \@@endlink[0]{}%
\providecommand \url  [0]{\begingroup\@sanitize@url \@url }%
\providecommand \@url [1]{\endgroup\@href {#1}{\urlprefix }}%
\providecommand \urlprefix  [0]{URL }%
\providecommand \Eprint [0]{\href }%
\providecommand \doibase [0]{https://doi.org/}%
\providecommand \selectlanguage [0]{\@gobble}%
\providecommand \bibinfo  [0]{\@secondoftwo}%
\providecommand \bibfield  [0]{\@secondoftwo}%
\providecommand \translation [1]{[#1]}%
\providecommand \BibitemOpen [0]{}%
\providecommand \bibitemStop [0]{}%
\providecommand \bibitemNoStop [0]{.\EOS\space}%
\providecommand \EOS [0]{\spacefactor3000\relax}%
\providecommand \BibitemShut  [1]{\csname bibitem#1\endcsname}%
\let\auto@bib@innerbib\@empty
%</preamble>
\bibitem [{\citenamefont {Fetter}\ and\ \citenamefont {Walecka}(2012)}]{fetter2012quantum}%
  \BibitemOpen
  \bibfield  {author} {\bibinfo {author} {\bibfnamefont {A.~L.}\ \bibnamefont {Fetter}}\ and\ \bibinfo {author} {\bibfnamefont {J.~D.}\ \bibnamefont {Walecka}},\ }\href@noop {} {\emph {\bibinfo {title} {Quantum theory of many-particle systems}}}\ (\bibinfo  {publisher} {Courier Corporation},\ \bibinfo {year} {2012})\BibitemShut {NoStop}%
\bibitem [{\citenamefont {Schirmer}(2018)}]{schirmer2018many}%
  \BibitemOpen
  \bibfield  {author} {\bibinfo {author} {\bibfnamefont {J.}~\bibnamefont {Schirmer}},\ }\href {https://doi.org/10.1007/978-3-319-93602-4} {\emph {\bibinfo {title} {Many-body methods for atoms, molecules and clusters}}},\ Vol.~\bibinfo {volume} {94}\ (\bibinfo  {publisher} {Springer},\ \bibinfo {year} {2018})\BibitemShut {NoStop}%
\bibitem [{\citenamefont {Golze}, \citenamefont {Dvorak},\ and\ \citenamefont {Rinke}(2019)}]{golzeGWCompendiumPractical2019}%
  \BibitemOpen
  \bibfield  {author} {\bibinfo {author} {\bibfnamefont {D.}~\bibnamefont {Golze}}, \bibinfo {author} {\bibfnamefont {M.}~\bibnamefont {Dvorak}},\ and\ \bibinfo {author} {\bibfnamefont {P.}~\bibnamefont {Rinke}},\ }\bibfield  {title} {\enquote {\bibinfo {title} {The {{GW Compendium}}: {{A Practical Guide}} to {{Theoretical Photoemission Spectroscopy}}},}\ }\href {https://doi.org/10.3389/fchem.2019.00377} {\bibfield  {journal} {\bibinfo  {journal} {Front. Chem.}\ }\textbf {\bibinfo {volume} {7}},\ \bibinfo {pages} {377} (\bibinfo {year} {2019})}\BibitemShut {NoStop}%
\bibitem [{\citenamefont {Marie}, \citenamefont {Ammar},\ and\ \citenamefont {Loos}(2024)}]{marie2024gw}%
  \BibitemOpen
  \bibfield  {author} {\bibinfo {author} {\bibfnamefont {A.}~\bibnamefont {Marie}}, \bibinfo {author} {\bibfnamefont {A.}~\bibnamefont {Ammar}},\ and\ \bibinfo {author} {\bibfnamefont {P.-F.}\ \bibnamefont {Loos}},\ }\bibfield  {title} {\enquote {\bibinfo {title} {The gw approximation: A quantum chemistry perspective},}\ }in\ \href {https://doi.org/10.1016/bs.aiq.2024.04.001} {\emph {\bibinfo {booktitle} {Adv. Quantum Chem.}}},\ Vol.~\bibinfo {volume} {90}\ (\bibinfo  {publisher} {Elsevier},\ \bibinfo {year} {2024})\ pp.\ \bibinfo {pages} {157--184}\BibitemShut {NoStop}%
\bibitem [{\citenamefont {Schirmer}, \citenamefont {Cederbaum},\ and\ \citenamefont {Walter}(1983)}]{schirmer1983new}%
  \BibitemOpen
  \bibfield  {author} {\bibinfo {author} {\bibfnamefont {J.}~\bibnamefont {Schirmer}}, \bibinfo {author} {\bibfnamefont {L.~S.}\ \bibnamefont {Cederbaum}},\ and\ \bibinfo {author} {\bibfnamefont {O.}~\bibnamefont {Walter}},\ }\bibfield  {title} {\enquote {\bibinfo {title} {New approach to the one-particle green's function for finite fermi systems},}\ }\href {https://doi.org/10.1103/PhysRevA.28.1237} {\bibfield  {journal} {\bibinfo  {journal} {Phys. Rev. A}\ }\textbf {\bibinfo {volume} {28}},\ \bibinfo {pages} {1237} (\bibinfo {year} {1983})}\BibitemShut {NoStop}%
\bibitem [{\citenamefont {Holleboom}\ and\ \citenamefont {Snijders}(1990{\natexlab{a}})}]{snijdersGF2}%
  \BibitemOpen
  \bibfield  {author} {\bibinfo {author} {\bibfnamefont {L.~J.}\ \bibnamefont {Holleboom}}\ and\ \bibinfo {author} {\bibfnamefont {J.~G.}\ \bibnamefont {Snijders}},\ }\bibfield  {title} {\enquote {\bibinfo {title} {A comparison between the {{Mo}}/ller--{{Plesset}} and {{Green}}'s function perturbative approaches to the calculation of the correlation energy in the many-electron problem},}\ }\href {https://doi.org/10.1063/1.459578} {\bibfield  {journal} {\bibinfo  {journal} {J. Chem. Phys.}\ }\textbf {\bibinfo {volume} {93}},\ \bibinfo {pages} {5826--5837} (\bibinfo {year} {1990}{\natexlab{a}})}\BibitemShut {NoStop}%
\bibitem [{\citenamefont {Nooijen}\ and\ \citenamefont {Snijders}(1992)}]{nooijen1992coupled}%
  \BibitemOpen
  \bibfield  {author} {\bibinfo {author} {\bibfnamefont {M.}~\bibnamefont {Nooijen}}\ and\ \bibinfo {author} {\bibfnamefont {J.~G.}\ \bibnamefont {Snijders}},\ }\bibfield  {title} {\enquote {\bibinfo {title} {Coupled cluster approach to the single-particle green's function},}\ }\href {https://doi.org/10.1002/qua.560440808} {\bibfield  {journal} {\bibinfo  {journal} {Int. J. Quantum Chem.}\ }\textbf {\bibinfo {volume} {44}},\ \bibinfo {pages} {55--83} (\bibinfo {year} {1992})}\BibitemShut {NoStop}%
\bibitem [{\citenamefont {Nooijen}\ and\ \citenamefont {Snijders}(1993)}]{nooijen1993coupled}%
  \BibitemOpen
  \bibfield  {author} {\bibinfo {author} {\bibfnamefont {M.}~\bibnamefont {Nooijen}}\ and\ \bibinfo {author} {\bibfnamefont {J.~G.}\ \bibnamefont {Snijders}},\ }\bibfield  {title} {\enquote {\bibinfo {title} {Coupled cluster green's function method: Working equations and applications},}\ }\href {https://doi.org/10.1002/qua.560480103} {\bibfield  {journal} {\bibinfo  {journal} {Int. J. Quantum Chem.}\ }\textbf {\bibinfo {volume} {48}},\ \bibinfo {pages} {15--48} (\bibinfo {year} {1993})}\BibitemShut {NoStop}%
\bibitem [{\citenamefont {Dahlen}\ and\ \citenamefont {van Leeuwen}(2005)}]{dahlen2005self}%
  \BibitemOpen
  \bibfield  {author} {\bibinfo {author} {\bibfnamefont {N.~E.}\ \bibnamefont {Dahlen}}\ and\ \bibinfo {author} {\bibfnamefont {R.}~\bibnamefont {van Leeuwen}},\ }\bibfield  {title} {\enquote {\bibinfo {title} {Self-consistent solution of the dyson equation for atoms and molecules within a conserving approximation},}\ }\href {https://doi.org/10.1063/1.1884965} {\bibfield  {journal} {\bibinfo  {journal} {J. Chem. Phys.}\ }\textbf {\bibinfo {volume} {122}} (\bibinfo {year} {2005}),\ 10.1063/1.1884965}\BibitemShut {NoStop}%
\bibitem [{\citenamefont {Schirmer}(1991)}]{schirmer1991closed}%
  \BibitemOpen
  \bibfield  {author} {\bibinfo {author} {\bibfnamefont {J.}~\bibnamefont {Schirmer}},\ }\bibfield  {title} {\enquote {\bibinfo {title} {Closed-form intermediate representations of many-body propagators and resolvent matrices},}\ }\href {https://doi.org/10.1103/PhysRevA.43.4647} {\bibfield  {journal} {\bibinfo  {journal} {Phys. Rev. A}\ }\textbf {\bibinfo {volume} {43}},\ \bibinfo {pages} {4647} (\bibinfo {year} {1991})}\BibitemShut {NoStop}%
\bibitem [{\citenamefont {Chatterjee}\ and\ \citenamefont {Sokolov}(2019)}]{chatterjee2019second}%
  \BibitemOpen
  \bibfield  {author} {\bibinfo {author} {\bibfnamefont {K.}~\bibnamefont {Chatterjee}}\ and\ \bibinfo {author} {\bibfnamefont {A.~Y.}\ \bibnamefont {Sokolov}},\ }\bibfield  {title} {\enquote {\bibinfo {title} {Second-order multireference algebraic diagrammatic construction theory for photoelectron spectra of strongly correlated systems},}\ }\href {https://doi.org/10.1021/acs.jctc.9b00528} {\bibfield  {journal} {\bibinfo  {journal} {J. Chem. Theory Comput.}\ }\textbf {\bibinfo {volume} {15}},\ \bibinfo {pages} {5908--5924} (\bibinfo {year} {2019})}\BibitemShut {NoStop}%
\bibitem [{\citenamefont {Dyson}(1949)}]{dyson1949s}%
  \BibitemOpen
  \bibfield  {author} {\bibinfo {author} {\bibfnamefont {F.~J.}\ \bibnamefont {Dyson}},\ }\bibfield  {title} {\enquote {\bibinfo {title} {The s matrix in quantum electrodynamics},}\ }\href {https://doi.org/10.1103/PhysRev.75.1736} {\bibfield  {journal} {\bibinfo  {journal} {Phys. Rev.}\ }\textbf {\bibinfo {volume} {75}},\ \bibinfo {pages} {1736} (\bibinfo {year} {1949})}\BibitemShut {NoStop}%
\bibitem [{\citenamefont {Hedin}(1965)}]{hedinNewMethodCalculating1965}%
  \BibitemOpen
  \bibfield  {author} {\bibinfo {author} {\bibfnamefont {L.}~\bibnamefont {Hedin}},\ }\bibfield  {title} {\enquote {\bibinfo {title} {New {{Method}} for {{Calculating}} the {{One-Particle Green}}'s {{Function}} with {{Application}} to the {{Electron-Gas Problem}}},}\ }\href {https://doi.org/10.1103/PhysRev.139.A796} {\bibfield  {journal} {\bibinfo  {journal} {Phys. Rev.}\ }\textbf {\bibinfo {volume} {139}},\ \bibinfo {pages} {A796--A823} (\bibinfo {year} {1965})}\BibitemShut {NoStop}%
\bibitem [{\citenamefont {Godby}, \citenamefont {Schl{\"u}ter},\ and\ \citenamefont {Sham}(1986)}]{godby1986accurate}%
  \BibitemOpen
  \bibfield  {author} {\bibinfo {author} {\bibfnamefont {R.}~\bibnamefont {Godby}}, \bibinfo {author} {\bibfnamefont {M.}~\bibnamefont {Schl{\"u}ter}},\ and\ \bibinfo {author} {\bibfnamefont {L.}~\bibnamefont {Sham}},\ }\bibfield  {title} {\enquote {\bibinfo {title} {Accurate exchange-correlation potential for silicon and its discontinuity on addition of an electron},}\ }\href {https://doi.org/10.1103/PhysRevLett.56.2415} {\bibfield  {journal} {\bibinfo  {journal} {Phys. Rev. Lett.}\ }\textbf {\bibinfo {volume} {56}},\ \bibinfo {pages} {2415} (\bibinfo {year} {1986})}\BibitemShut {NoStop}%
\bibitem [{\citenamefont {Hybertsen}\ and\ \citenamefont {Louie}(1986)}]{hybertsen1986electron}%
  \BibitemOpen
  \bibfield  {author} {\bibinfo {author} {\bibfnamefont {M.~S.}\ \bibnamefont {Hybertsen}}\ and\ \bibinfo {author} {\bibfnamefont {S.~G.}\ \bibnamefont {Louie}},\ }\bibfield  {title} {\enquote {\bibinfo {title} {Electron correlation in semiconductors and insulators: Band gaps and quasiparticle energies},}\ }\href {https://doi.org/10.1103/PhysRevB.34.5390} {\bibfield  {journal} {\bibinfo  {journal} {Phys. Rev. B}\ }\textbf {\bibinfo {volume} {34}},\ \bibinfo {pages} {5390} (\bibinfo {year} {1986})}\BibitemShut {NoStop}%
\bibitem [{\citenamefont {Aryasetiawan}\ and\ \citenamefont {Gunnarsson}(1998)}]{aryasetiawan1998gw}%
  \BibitemOpen
  \bibfield  {author} {\bibinfo {author} {\bibfnamefont {F.}~\bibnamefont {Aryasetiawan}}\ and\ \bibinfo {author} {\bibfnamefont {O.}~\bibnamefont {Gunnarsson}},\ }\bibfield  {title} {\enquote {\bibinfo {title} {The gw method},}\ }\href {https://doi.org/10.1088/0034-4885/61/3/002} {\bibfield  {journal} {\bibinfo  {journal} {Rep. Prog. Phys.}\ }\textbf {\bibinfo {volume} {61}},\ \bibinfo {pages} {237} (\bibinfo {year} {1998})}\BibitemShut {NoStop}%
\bibitem [{\citenamefont {Bruneval}(2012)}]{bruneval2012ionization}%
  \BibitemOpen
  \bibfield  {author} {\bibinfo {author} {\bibfnamefont {F.}~\bibnamefont {Bruneval}},\ }\bibfield  {title} {\enquote {\bibinfo {title} {Ionization energy of atoms obtained from gw self-energy or from random phase approximation total energies},}\ }\href {https://doi.org/10.1063/1.4718428} {\bibfield  {journal} {\bibinfo  {journal} {J. Chem. Phys}\ }\textbf {\bibinfo {volume} {136}} (\bibinfo {year} {2012}),\ 10.1063/1.4718428}\BibitemShut {NoStop}%
\bibitem [{\citenamefont {van Setten}, \citenamefont {Weigend},\ and\ \citenamefont {Evers}(2013)}]{van2013gw}%
  \BibitemOpen
  \bibfield  {author} {\bibinfo {author} {\bibfnamefont {M.~J.}\ \bibnamefont {van Setten}}, \bibinfo {author} {\bibfnamefont {F.}~\bibnamefont {Weigend}},\ and\ \bibinfo {author} {\bibfnamefont {F.}~\bibnamefont {Evers}},\ }\bibfield  {title} {\enquote {\bibinfo {title} {The gw-method for quantum chemistry applications: Theory and implementation},}\ }\href {https://doi.org/10.1021/ct300648t} {\bibfield  {journal} {\bibinfo  {journal} {J. Chem. Theory Comput.}\ }\textbf {\bibinfo {volume} {9}},\ \bibinfo {pages} {232--246} (\bibinfo {year} {2013})}\BibitemShut {NoStop}%
\bibitem [{\citenamefont {Van~Setten}\ \emph {et~al.}(2015{\natexlab{a}})\citenamefont {Van~Setten}, \citenamefont {Caruso}, \citenamefont {Sharifzadeh}, \citenamefont {Ren}, \citenamefont {Scheffler}, \citenamefont {Liu}, \citenamefont {Lischner}, \citenamefont {Lin}, \citenamefont {Deslippe}, \citenamefont {Louie} \emph {et~al.}}]{van2015gw}%
  \BibitemOpen
  \bibfield  {author} {\bibinfo {author} {\bibfnamefont {M.~J.}\ \bibnamefont {Van~Setten}}, \bibinfo {author} {\bibfnamefont {F.}~\bibnamefont {Caruso}}, \bibinfo {author} {\bibfnamefont {S.}~\bibnamefont {Sharifzadeh}}, \bibinfo {author} {\bibfnamefont {X.}~\bibnamefont {Ren}}, \bibinfo {author} {\bibfnamefont {M.}~\bibnamefont {Scheffler}}, \bibinfo {author} {\bibfnamefont {F.}~\bibnamefont {Liu}}, \bibinfo {author} {\bibfnamefont {J.}~\bibnamefont {Lischner}}, \bibinfo {author} {\bibfnamefont {L.}~\bibnamefont {Lin}}, \bibinfo {author} {\bibfnamefont {J.~R.}\ \bibnamefont {Deslippe}}, \bibinfo {author} {\bibfnamefont {S.~G.}\ \bibnamefont {Louie}}, \emph {et~al.},\ }\bibfield  {title} {\enquote {\bibinfo {title} {Gw 100: Benchmarking g 0 w 0 for molecular systems},}\ }\href {https://doi.org/10.1021/acs.jctc.5b00453} {\bibfield  {journal} {\bibinfo  {journal} {J. Chem. Theory Comput.}\ }\textbf {\bibinfo {volume} {11}},\ \bibinfo {pages} {5665--5687} (\bibinfo {year} {2015}{\natexlab{a}})}\BibitemShut {NoStop}%
\bibitem [{\citenamefont {Golze}\ \emph {et~al.}(2018)\citenamefont {Golze}, \citenamefont {Wilhelm}, \citenamefont {Van~Setten},\ and\ \citenamefont {Rinke}}]{golze2018core}%
  \BibitemOpen
  \bibfield  {author} {\bibinfo {author} {\bibfnamefont {D.}~\bibnamefont {Golze}}, \bibinfo {author} {\bibfnamefont {J.}~\bibnamefont {Wilhelm}}, \bibinfo {author} {\bibfnamefont {M.~J.}\ \bibnamefont {Van~Setten}},\ and\ \bibinfo {author} {\bibfnamefont {P.}~\bibnamefont {Rinke}},\ }\bibfield  {title} {\enquote {\bibinfo {title} {Core-level binding energies from gw: An efficient full-frequency approach within a localized basis},}\ }\href {https://doi.org/10.1021/acs.jctc.8b00458} {\bibfield  {journal} {\bibinfo  {journal} {J. Chem. Theory Comput.}\ }\textbf {\bibinfo {volume} {14}},\ \bibinfo {pages} {4856--4869} (\bibinfo {year} {2018})}\BibitemShut {NoStop}%
\bibitem [{\citenamefont {F\"orster}\ and\ \citenamefont {Visscher}(2020)}]{forster2020low}%
  \BibitemOpen
  \bibfield  {author} {\bibinfo {author} {\bibfnamefont {A.}~\bibnamefont {F\"orster}}\ and\ \bibinfo {author} {\bibfnamefont {L.}~\bibnamefont {Visscher}},\ }\bibfield  {title} {\enquote {\bibinfo {title} {Low-order scaling g 0 w 0 by pair atomic density fitting},}\ }\href {https://doi.org/10.1021/acs.jctc.0c00693} {\bibfield  {journal} {\bibinfo  {journal} {J. Chem. Theory Comput.}\ }\textbf {\bibinfo {volume} {16}},\ \bibinfo {pages} {7381--7399} (\bibinfo {year} {2020})}\BibitemShut {NoStop}%
\bibitem [{\citenamefont {Bruneval}, \citenamefont {Dattani},\ and\ \citenamefont {Van~Setten}(2021)}]{bruneval2021gw}%
  \BibitemOpen
  \bibfield  {author} {\bibinfo {author} {\bibfnamefont {F.}~\bibnamefont {Bruneval}}, \bibinfo {author} {\bibfnamefont {N.}~\bibnamefont {Dattani}},\ and\ \bibinfo {author} {\bibfnamefont {M.~J.}\ \bibnamefont {Van~Setten}},\ }\bibfield  {title} {\enquote {\bibinfo {title} {The {{GW Miracle}} in {{Many-Body Perturbation Theory}} for the {{Ionization Potential}} of {{Molecules}}},}\ }\href {https://doi.org/10.3389/fchem.2021.749779} {\bibfield  {journal} {\bibinfo  {journal} {Front. Chem.}\ }\textbf {\bibinfo {volume} {9}},\ \bibinfo {pages} {749779} (\bibinfo {year} {2021})}\BibitemShut {NoStop}%
\bibitem [{\citenamefont {Stan}, \citenamefont {Dahlen},\ and\ \citenamefont {Leeuwen}(2006)}]{stanFullySelfconsistentGW2006}%
  \BibitemOpen
  \bibfield  {author} {\bibinfo {author} {\bibfnamefont {A.}~\bibnamefont {Stan}}, \bibinfo {author} {\bibfnamefont {N.~E.}\ \bibnamefont {Dahlen}},\ and\ \bibinfo {author} {\bibfnamefont {R.~V.}\ \bibnamefont {Leeuwen}},\ }\bibfield  {title} {\enquote {\bibinfo {title} {Fully self-consistent {{{\emph{GW}}}} calculations for atoms and molecules},}\ }\href {https://doi.org/10.1209/epl/i2006-10266-6} {\bibfield  {journal} {\bibinfo  {journal} {Europhys. Lett.}\ }\textbf {\bibinfo {volume} {76}},\ \bibinfo {pages} {298--304} (\bibinfo {year} {2006})}\BibitemShut {NoStop}%
\bibitem [{\citenamefont {Stan}, \citenamefont {Dahlen},\ and\ \citenamefont {Van~Leeuwen}(2009)}]{stan2009levels}%
  \BibitemOpen
  \bibfield  {author} {\bibinfo {author} {\bibfnamefont {A.}~\bibnamefont {Stan}}, \bibinfo {author} {\bibfnamefont {N.~E.}\ \bibnamefont {Dahlen}},\ and\ \bibinfo {author} {\bibfnamefont {R.}~\bibnamefont {Van~Leeuwen}},\ }\bibfield  {title} {\enquote {\bibinfo {title} {Levels of self-consistency in the gw approximation},}\ }\href {https://doi.org/10.1063/1.3089567} {\bibfield  {journal} {\bibinfo  {journal} {J. Chem. Phys.}\ }\textbf {\bibinfo {volume} {130}} (\bibinfo {year} {2009}),\ 10.1063/1.3089567}\BibitemShut {NoStop}%
\bibitem [{\citenamefont {{van~Schilfgaarde}}, \citenamefont {Kotani},\ and\ \citenamefont {Faleev}(2006)}]{vanschilfgaardeQuasiparticleSelfConsistentTheory2006}%
  \BibitemOpen
  \bibfield  {author} {\bibinfo {author} {\bibfnamefont {M.}~\bibnamefont {{van~Schilfgaarde}}}, \bibinfo {author} {\bibfnamefont {T.}~\bibnamefont {Kotani}},\ and\ \bibinfo {author} {\bibfnamefont {S.}~\bibnamefont {Faleev}},\ }\bibfield  {title} {\enquote {\bibinfo {title} {Quasiparticle {{Self-Consistent G W Theory}}},}\ }\href {https://doi.org/10.1103/PhysRevLett.96.226402} {\bibfield  {journal} {\bibinfo  {journal} {Phys. Rev. Lett.}\ }\textbf {\bibinfo {volume} {96}},\ \bibinfo {pages} {226402} (\bibinfo {year} {2006})}\BibitemShut {NoStop}%
\bibitem [{\citenamefont {Kotani}, \citenamefont {Van~Schilfgaarde},\ and\ \citenamefont {Faleev}(2007)}]{kotaniQuasiparticleSelfconsistentMethod2007}%
  \BibitemOpen
  \bibfield  {author} {\bibinfo {author} {\bibfnamefont {T.}~\bibnamefont {Kotani}}, \bibinfo {author} {\bibfnamefont {M.}~\bibnamefont {Van~Schilfgaarde}},\ and\ \bibinfo {author} {\bibfnamefont {S.~V.}\ \bibnamefont {Faleev}},\ }\bibfield  {title} {\enquote {\bibinfo {title} {Quasiparticle self-consistent {{G W}} method: {{A}} basis for the independent-particle approximation},}\ }\href {https://doi.org/10.1103/PhysRevB.76.165106} {\bibfield  {journal} {\bibinfo  {journal} {Phys. Rev. B}\ }\textbf {\bibinfo {volume} {76}},\ \bibinfo {pages} {165106} (\bibinfo {year} {2007})}\BibitemShut {NoStop}%
\bibitem [{\citenamefont {Kaplan}\ \emph {et~al.}(2016)\citenamefont {Kaplan}, \citenamefont {Harding}, \citenamefont {Seiler}, \citenamefont {Weigend}, \citenamefont {Evers},\ and\ \citenamefont {Van~Setten}}]{kaplanQuasiParticleSelfConsistentGW2016}%
  \BibitemOpen
  \bibfield  {author} {\bibinfo {author} {\bibfnamefont {F.}~\bibnamefont {Kaplan}}, \bibinfo {author} {\bibfnamefont {M.~E.}\ \bibnamefont {Harding}}, \bibinfo {author} {\bibfnamefont {C.}~\bibnamefont {Seiler}}, \bibinfo {author} {\bibfnamefont {F.}~\bibnamefont {Weigend}}, \bibinfo {author} {\bibfnamefont {F.}~\bibnamefont {Evers}},\ and\ \bibinfo {author} {\bibfnamefont {M.~J.}\ \bibnamefont {Van~Setten}},\ }\bibfield  {title} {\enquote {\bibinfo {title} {Quasi-{{Particle Self-Consistent}} {{{\emph{GW}}}} for {{Molecules}}},}\ }\href {https://doi.org/10.1021/acs.jctc.5b01238} {\bibfield  {journal} {\bibinfo  {journal} {J. Chem. Theory Comput.}\ }\textbf {\bibinfo {volume} {12}},\ \bibinfo {pages} {2528--2541} (\bibinfo {year} {2016})}\BibitemShut {NoStop}%
\bibitem [{\citenamefont {Faber}\ \emph {et~al.}(2011)\citenamefont {Faber}, \citenamefont {Attaccalite}, \citenamefont {Olevano}, \citenamefont {Runge},\ and\ \citenamefont {Blase}}]{faber2011first}%
  \BibitemOpen
  \bibfield  {author} {\bibinfo {author} {\bibfnamefont {C.}~\bibnamefont {Faber}}, \bibinfo {author} {\bibfnamefont {C.}~\bibnamefont {Attaccalite}}, \bibinfo {author} {\bibfnamefont {V.}~\bibnamefont {Olevano}}, \bibinfo {author} {\bibfnamefont {E.}~\bibnamefont {Runge}},\ and\ \bibinfo {author} {\bibfnamefont {X.}~\bibnamefont {Blase}},\ }\bibfield  {title} {\enquote {\bibinfo {title} {First-principles gw calculations for dna and rna nucleobases},}\ }\href {https://doi.org/10.1103/PhysRevB.83.115123} {\bibfield  {journal} {\bibinfo  {journal} {Phys. Rev. B}\ }\textbf {\bibinfo {volume} {83}},\ \bibinfo {pages} {115123} (\bibinfo {year} {2011})}\BibitemShut {NoStop}%
\bibitem [{\citenamefont {Rangel}\ \emph {et~al.}(2016)\citenamefont {Rangel}, \citenamefont {Hamed}, \citenamefont {Bruneval},\ and\ \citenamefont {Neaton}}]{rangel2016evaluating}%
  \BibitemOpen
  \bibfield  {author} {\bibinfo {author} {\bibfnamefont {T.}~\bibnamefont {Rangel}}, \bibinfo {author} {\bibfnamefont {S.~M.}\ \bibnamefont {Hamed}}, \bibinfo {author} {\bibfnamefont {F.}~\bibnamefont {Bruneval}},\ and\ \bibinfo {author} {\bibfnamefont {J.~B.}\ \bibnamefont {Neaton}},\ }\bibfield  {title} {\enquote {\bibinfo {title} {Evaluating the gw approximation with ccsd (t) for charged excitations across the oligoacenes},}\ }\href {https://doi.org/10.1021/acs.jctc.6b00163} {\bibfield  {journal} {\bibinfo  {journal} {J. Chem. Theory Comput.}\ }\textbf {\bibinfo {volume} {12}},\ \bibinfo {pages} {2834--2842} (\bibinfo {year} {2016})}\BibitemShut {NoStop}%
\bibitem [{\citenamefont {Holleboom}\ and\ \citenamefont {Snijders}(1990{\natexlab{b}})}]{holleboomComparisonMoLler1990}%
  \BibitemOpen
  \bibfield  {author} {\bibinfo {author} {\bibfnamefont {L.~J.}\ \bibnamefont {Holleboom}}\ and\ \bibinfo {author} {\bibfnamefont {J.~G.}\ \bibnamefont {Snijders}},\ }\bibfield  {title} {\enquote {\bibinfo {title} {A comparison between the {{Mo}}/ller--{{Plesset}} and {{Green}}'s function perturbative approaches to the calculation of the correlation energy in the many-electron problem},}\ }\href {https://doi.org/10.1063/1.459578} {\bibfield  {journal} {\bibinfo  {journal} {J. Chem. Phys.}\ }\textbf {\bibinfo {volume} {93}},\ \bibinfo {pages} {5826--5837} (\bibinfo {year} {1990}{\natexlab{b}})}\BibitemShut {NoStop}%
\bibitem [{\citenamefont {Dahlen}\ and\ \citenamefont {Van~Leeuwen}(2005)}]{dahlenSelfconsistentSolutionDyson2005}%
  \BibitemOpen
  \bibfield  {author} {\bibinfo {author} {\bibfnamefont {N.~E.}\ \bibnamefont {Dahlen}}\ and\ \bibinfo {author} {\bibfnamefont {R.}~\bibnamefont {Van~Leeuwen}},\ }\bibfield  {title} {\enquote {\bibinfo {title} {Self-consistent solution of the {{Dyson}} equation for atoms and molecules within a conserving approximation},}\ }\href {https://doi.org/10.1063/1.1884965} {\bibfield  {journal} {\bibinfo  {journal} {J. Chem. Phys.}\ }\textbf {\bibinfo {volume} {122}},\ \bibinfo {pages} {164102} (\bibinfo {year} {2005})}\BibitemShut {NoStop}%
\bibitem [{\citenamefont {Phillips}\ and\ \citenamefont {Zgid}(2014)}]{zgidGF2_1}%
  \BibitemOpen
  \bibfield  {author} {\bibinfo {author} {\bibfnamefont {J.~J.}\ \bibnamefont {Phillips}}\ and\ \bibinfo {author} {\bibfnamefont {D.}~\bibnamefont {Zgid}},\ }\bibfield  {title} {\enquote {\bibinfo {title} {Communication: {{The}} description of strong correlation within self-consistent {{Green}}'s function second-order perturbation theory},}\ }\href {https://doi.org/10.1063/1.4884951} {\bibfield  {journal} {\bibinfo  {journal} {J. Chem. Phys.}\ }\textbf {\bibinfo {volume} {140}},\ \bibinfo {pages} {241101} (\bibinfo {year} {2014})}\BibitemShut {NoStop}%
\bibitem [{\citenamefont {Rusakov}\ and\ \citenamefont {Zgid}(2016)}]{zgidGF2_2}%
  \BibitemOpen
  \bibfield  {author} {\bibinfo {author} {\bibfnamefont {A.~A.}\ \bibnamefont {Rusakov}}\ and\ \bibinfo {author} {\bibfnamefont {D.}~\bibnamefont {Zgid}},\ }\bibfield  {title} {\enquote {\bibinfo {title} {Self-consistent second-order {{Green}}'s function perturbation theory for periodic systems},}\ }\href {https://doi.org/10.1063/1.4940900} {\bibfield  {journal} {\bibinfo  {journal} {J. Chem. Phys.}\ }\textbf {\bibinfo {volume} {144}},\ \bibinfo {pages} {054106} (\bibinfo {year} {2016})}\BibitemShut {NoStop}%
\bibitem [{\citenamefont {Welden}, \citenamefont {Rusakov},\ and\ \citenamefont {Zgid}(2016)}]{zgidGF2_3}%
  \BibitemOpen
  \bibfield  {author} {\bibinfo {author} {\bibfnamefont {A.~R.}\ \bibnamefont {Welden}}, \bibinfo {author} {\bibfnamefont {A.~A.}\ \bibnamefont {Rusakov}},\ and\ \bibinfo {author} {\bibfnamefont {D.}~\bibnamefont {Zgid}},\ }\bibfield  {title} {\enquote {\bibinfo {title} {Exploring connections between statistical mechanics and {{Green}}'s functions for realistic systems: {{Temperature}} dependent electronic entropy and internal energy from a self-consistent second-order {{Green}}'s function},}\ }\href {https://doi.org/10.1063/1.4967449} {\bibfield  {journal} {\bibinfo  {journal} {J. Chem. Phys.}\ }\textbf {\bibinfo {volume} {145}},\ \bibinfo {pages} {204106} (\bibinfo {year} {2016})}\BibitemShut {NoStop}%
\bibitem [{\citenamefont {Backhouse}\ and\ \citenamefont {Booth}(2020)}]{boothGF2spectrum}%
  \BibitemOpen
  \bibfield  {author} {\bibinfo {author} {\bibfnamefont {O.~J.}\ \bibnamefont {Backhouse}}\ and\ \bibinfo {author} {\bibfnamefont {G.~H.}\ \bibnamefont {Booth}},\ }\bibfield  {title} {\enquote {\bibinfo {title} {Efficient {{Excitations}} and {{Spectra}} within a {{Perturbative Renormalization Approach}}},}\ }\href {https://doi.org/10.1021/acs.jctc.0c00701} {\bibfield  {journal} {\bibinfo  {journal} {J. Chem. Theory Comput.}\ }\textbf {\bibinfo {volume} {16}},\ \bibinfo {pages} {6294--6304} (\bibinfo {year} {2020})}\BibitemShut {NoStop}%
\bibitem [{\citenamefont {Backhouse}, \citenamefont {Nusspickel},\ and\ \citenamefont {Booth}(2020)}]{boothAGF2}%
  \BibitemOpen
  \bibfield  {author} {\bibinfo {author} {\bibfnamefont {O.~J.}\ \bibnamefont {Backhouse}}, \bibinfo {author} {\bibfnamefont {M.}~\bibnamefont {Nusspickel}},\ and\ \bibinfo {author} {\bibfnamefont {G.~H.}\ \bibnamefont {Booth}},\ }\bibfield  {title} {\enquote {\bibinfo {title} {Wave {{Function Perspective}} and {{Efficient Truncation}} of {{Renormalized Second-Order Perturbation Theory}}},}\ }\href {https://doi.org/10.1021/acs.jctc.9b01182} {\bibfield  {journal} {\bibinfo  {journal} {J. Chem. Theory Comput.}\ }\textbf {\bibinfo {volume} {16}},\ \bibinfo {pages} {1090--1104} (\bibinfo {year} {2020})}\BibitemShut {NoStop}%
\bibitem [{\citenamefont {Monino}\ and\ \citenamefont {Loos}(2023)}]{LoosGF2}%
  \BibitemOpen
  \bibfield  {author} {\bibinfo {author} {\bibfnamefont {E.}~\bibnamefont {Monino}}\ and\ \bibinfo {author} {\bibfnamefont {P.-F.}\ \bibnamefont {Loos}},\ }\bibfield  {title} {\enquote {\bibinfo {title} {Connections and performances of {{Green}}'s function methods for charged and neutral excitations},}\ }\href {https://doi.org/10.1063/5.0159853} {\bibfield  {journal} {\bibinfo  {journal} {J. Chem. Phys.}\ }\textbf {\bibinfo {volume} {159}},\ \bibinfo {pages} {034105} (\bibinfo {year} {2023})}\BibitemShut {NoStop}%
\bibitem [{\citenamefont {Sawicki}\ \emph {et~al.}(2025)\citenamefont {Sawicki}, \citenamefont {Triglione}, \citenamefont {Jana},\ and\ \citenamefont {{\'S}miga}}]{smigaRegularizedPT2}%
  \BibitemOpen
  \bibfield  {author} {\bibinfo {author} {\bibfnamefont {I.}~\bibnamefont {Sawicki}}, \bibinfo {author} {\bibfnamefont {V.}~\bibnamefont {Triglione}}, \bibinfo {author} {\bibfnamefont {S.}~\bibnamefont {Jana}},\ and\ \bibinfo {author} {\bibfnamefont {S.}~\bibnamefont {{\'S}miga}},\ }\bibfield  {title} {\enquote {\bibinfo {title} {An {{Analysis}} of {{Regularized Second-Order Energy Expressions}} in the {{Context}} of {{Post-HF}} and {{KS-DFT Calculations}}: {{What Do We Gain}} and {{What Do We Lose}}?}}\ }\href {https://doi.org/10.1021/acs.jctc.4c01547} {\bibfield  {journal} {\bibinfo  {journal} {J. Chem. Theory Comput.}\ }\textbf {\bibinfo {volume} {21}},\ \bibinfo {pages} {2928--2941} (\bibinfo {year} {2025})}\BibitemShut {NoStop}%
\bibitem [{\citenamefont {St{\"u}ck}\ and\ \citenamefont {{Head-Gordon}}(2013)}]{deltaMP2}%
  \BibitemOpen
  \bibfield  {author} {\bibinfo {author} {\bibfnamefont {D.}~\bibnamefont {St{\"u}ck}}\ and\ \bibinfo {author} {\bibfnamefont {M.}~\bibnamefont {{Head-Gordon}}},\ }\bibfield  {title} {\enquote {\bibinfo {title} {Regularized orbital-optimized second-order perturbation theory},}\ }\href {https://doi.org/10.1063/1.4851816} {\bibfield  {journal} {\bibinfo  {journal} {J. Chem. Phys.}\ }\textbf {\bibinfo {volume} {139}},\ \bibinfo {pages} {244109} (\bibinfo {year} {2013})}\BibitemShut {NoStop}%
\bibitem [{\citenamefont {Ohnishi}, \citenamefont {Ishimura},\ and\ \citenamefont {{Ten-no}}(2014)}]{RDMP2}%
  \BibitemOpen
  \bibfield  {author} {\bibinfo {author} {\bibfnamefont {Y.-y.}\ \bibnamefont {Ohnishi}}, \bibinfo {author} {\bibfnamefont {K.}~\bibnamefont {Ishimura}},\ and\ \bibinfo {author} {\bibfnamefont {S.}~\bibnamefont {{Ten-no}}},\ }\bibfield  {title} {\enquote {\bibinfo {title} {Interaction {{Energy}} of {{Large Molecules}} from {{Restrained Denominator MP2-F12}}},}\ }\href {https://doi.org/10.1021/ct500738g} {\bibfield  {journal} {\bibinfo  {journal} {J. Chem. Theory Comput.}\ }\textbf {\bibinfo {volume} {10}},\ \bibinfo {pages} {4857--4861} (\bibinfo {year} {2014})}\BibitemShut {NoStop}%
\bibitem [{\citenamefont {Lee}\ and\ \citenamefont {{Head-Gordon}}(2018)}]{kappaOOMP2}%
  \BibitemOpen
  \bibfield  {author} {\bibinfo {author} {\bibfnamefont {J.}~\bibnamefont {Lee}}\ and\ \bibinfo {author} {\bibfnamefont {M.}~\bibnamefont {{Head-Gordon}}},\ }\bibfield  {title} {\enquote {\bibinfo {title} {Regularized {{Orbital-Optimized Second-Order M{\o}ller}}--{{Plesset Perturbation Theory}}: {{A Reliable Fifth-Order-Scaling Electron Correlation Model}} with {{Orbital Energy Dependent Regularizers}}},}\ }\href {https://doi.org/10.1021/acs.jctc.8b00731} {\bibfield  {journal} {\bibinfo  {journal} {J. Chem. Theory Comput.}\ }\textbf {\bibinfo {volume} {14}},\ \bibinfo {pages} {5203--5219} (\bibinfo {year} {2018})}\BibitemShut {NoStop}%
\bibitem [{\citenamefont {Evangelista}(2014)}]{evangelistaDSRG}%
  \BibitemOpen
  \bibfield  {author} {\bibinfo {author} {\bibfnamefont {F.~A.}\ \bibnamefont {Evangelista}},\ }\bibfield  {title} {\enquote {\bibinfo {title} {A driven similarity renormalization group approach to quantum many-body problems},}\ }\href {https://doi.org/10.1063/1.4890660} {\bibfield  {journal} {\bibinfo  {journal} {J. Chem. Phys.}\ }\textbf {\bibinfo {volume} {141}},\ \bibinfo {pages} {054109} (\bibinfo {year} {2014})}\BibitemShut {NoStop}%
\bibitem [{\citenamefont {Garniron}\ \emph {et~al.}(2019)\citenamefont {Garniron}, \citenamefont {Applencourt}, \citenamefont {Gasperich}, \citenamefont {Benali}, \citenamefont {Fert{\'e}}, \citenamefont {Paquier}, \citenamefont {Pradines}, \citenamefont {Assaraf}, \citenamefont {Reinhardt}, \citenamefont {Toulouse} \emph {et~al.}}]{quantum_package}%
  \BibitemOpen
  \bibfield  {author} {\bibinfo {author} {\bibfnamefont {Y.}~\bibnamefont {Garniron}}, \bibinfo {author} {\bibfnamefont {T.}~\bibnamefont {Applencourt}}, \bibinfo {author} {\bibfnamefont {K.}~\bibnamefont {Gasperich}}, \bibinfo {author} {\bibfnamefont {A.}~\bibnamefont {Benali}}, \bibinfo {author} {\bibfnamefont {A.}~\bibnamefont {Fert{\'e}}}, \bibinfo {author} {\bibfnamefont {J.}~\bibnamefont {Paquier}}, \bibinfo {author} {\bibfnamefont {B.}~\bibnamefont {Pradines}}, \bibinfo {author} {\bibfnamefont {R.}~\bibnamefont {Assaraf}}, \bibinfo {author} {\bibfnamefont {P.}~\bibnamefont {Reinhardt}}, \bibinfo {author} {\bibfnamefont {J.}~\bibnamefont {Toulouse}}, \emph {et~al.},\ }\bibfield  {title} {\enquote {\bibinfo {title} {Quantum package 2.0: An open-source determinant-driven suite of programs},}\ }\href@noop {} {\bibfield  {journal} {\bibinfo  {journal} {J. Chem. Theory Comput.}\ }\textbf {\bibinfo {volume} {15}},\ \bibinfo {pages} {3591--3609} (\bibinfo {year} {2019})}\BibitemShut {NoStop}%
\bibitem [{\citenamefont {Carter-Fenk}, \citenamefont {Shee},\ and\ \citenamefont {Head-Gordon}(2023)}]{carter2023optimizing}%
  \BibitemOpen
  \bibfield  {author} {\bibinfo {author} {\bibfnamefont {K.}~\bibnamefont {Carter-Fenk}}, \bibinfo {author} {\bibfnamefont {J.}~\bibnamefont {Shee}},\ and\ \bibinfo {author} {\bibfnamefont {M.}~\bibnamefont {Head-Gordon}},\ }\bibfield  {title} {\enquote {\bibinfo {title} {Optimizing the regularization in size-consistent second-order brillouin-wigner perturbation theory},}\ }\href {https://doi.org/10.1063/5.0174923} {\bibfield  {journal} {\bibinfo  {journal} {J. Chem. Phys.}\ }\textbf {\bibinfo {volume} {159}} (\bibinfo {year} {2023}),\ 10.1063/5.0174923}\BibitemShut {NoStop}%
\bibitem [{\citenamefont {Chen}\ \emph {et~al.}(2025)\citenamefont {Chen}, \citenamefont {Zhang}, \citenamefont {Dinh}, \citenamefont {Rettig},\ and\ \citenamefont {Lee}}]{chen2025regularized}%
  \BibitemOpen
  \bibfield  {author} {\bibinfo {author} {\bibfnamefont {M.-F.}\ \bibnamefont {Chen}}, \bibinfo {author} {\bibfnamefont {J.}~\bibnamefont {Zhang}}, \bibinfo {author} {\bibfnamefont {H.~Q.}\ \bibnamefont {Dinh}}, \bibinfo {author} {\bibfnamefont {A.}~\bibnamefont {Rettig}},\ and\ \bibinfo {author} {\bibfnamefont {J.}~\bibnamefont {Lee}},\ }\bibfield  {title} {\enquote {\bibinfo {title} {Regularized perturbation theory for ab initio solids},}\ }\href {https://doi.org/10.1021/acs.jpclett.5c02731} {\bibfield  {journal} {\bibinfo  {journal} {J. Phys. Chem. Lett.}\ }\textbf {\bibinfo {volume} {16}},\ \bibinfo {pages} {11373--11381} (\bibinfo {year} {2025})}\BibitemShut {NoStop}%
\bibitem [{\citenamefont {Wegner}(1994)}]{SRG_1}%
  \BibitemOpen
  \bibfield  {author} {\bibinfo {author} {\bibfnamefont {F.}~\bibnamefont {Wegner}},\ }\bibfield  {title} {\enquote {\bibinfo {title} {Flow-equations for {{Hamiltonians}}},}\ }\href {https://doi.org/10.1002/andp.19945060203} {\bibfield  {journal} {\bibinfo  {journal} {Ann. Phys.}\ }\textbf {\bibinfo {volume} {506}},\ \bibinfo {pages} {77--91} (\bibinfo {year} {1994})}\BibitemShut {NoStop}%
\bibitem [{\citenamefont {G{\l}azek}\ and\ \citenamefont {Wilson}(1993)}]{SRG_2}%
  \BibitemOpen
  \bibfield  {author} {\bibinfo {author} {\bibfnamefont {S.~D.}\ \bibnamefont {G{\l}azek}}\ and\ \bibinfo {author} {\bibfnamefont {K.~G.}\ \bibnamefont {Wilson}},\ }\bibfield  {title} {\enquote {\bibinfo {title} {Renormalization of {{Hamiltonians}}},}\ }\href {https://doi.org/10.1103/PhysRevD.48.5863} {\bibfield  {journal} {\bibinfo  {journal} {Phys. Rev. D}\ }\textbf {\bibinfo {volume} {48}},\ \bibinfo {pages} {5863--5872} (\bibinfo {year} {1993})}\BibitemShut {NoStop}%
\bibitem [{\citenamefont {Glazek}\ and\ \citenamefont {Wilson}(1994)}]{SRG_3}%
  \BibitemOpen
  \bibfield  {author} {\bibinfo {author} {\bibfnamefont {S.~D.}\ \bibnamefont {Glazek}}\ and\ \bibinfo {author} {\bibfnamefont {K.~G.}\ \bibnamefont {Wilson}},\ }\bibfield  {title} {\enquote {\bibinfo {title} {Perturbative renormalization group for {{Hamiltonians}}},}\ }\href {https://doi.org/10.1103/PhysRevD.49.4214} {\bibfield  {journal} {\bibinfo  {journal} {Phys. Rev. D}\ }\textbf {\bibinfo {volume} {49}},\ \bibinfo {pages} {4214--4218} (\bibinfo {year} {1994})}\BibitemShut {NoStop}%
\bibitem [{\citenamefont {Marie}\ and\ \citenamefont {Loos}(2023)}]{marieSRG}%
  \BibitemOpen
  \bibfield  {author} {\bibinfo {author} {\bibfnamefont {A.}~\bibnamefont {Marie}}\ and\ \bibinfo {author} {\bibfnamefont {P.-F.}\ \bibnamefont {Loos}},\ }\bibfield  {title} {\enquote {\bibinfo {title} {A {{Similarity Renormalization Group Approach}} to {{Green}}'s {{Function Methods}}},}\ }\href {https://doi.org/10.1021/acs.jctc.3c00281} {\bibfield  {journal} {\bibinfo  {journal} {J. Chem. Theory Comput.}\ }\textbf {\bibinfo {volume} {19}},\ \bibinfo {pages} {3943--3957} (\bibinfo {year} {2023})}\BibitemShut {NoStop}%
\bibitem [{\citenamefont {Grimme}(2003)}]{scs_1}%
  \BibitemOpen
  \bibfield  {author} {\bibinfo {author} {\bibfnamefont {S.}~\bibnamefont {Grimme}},\ }\bibfield  {title} {\enquote {\bibinfo {title} {Improved second-order {{M{\o}ller}}--{{Plesset}} perturbation theory by separate scaling of parallel- and antiparallel-spin pair correlation energies},}\ }\href {https://doi.org/10.1063/1.1569242} {\bibfield  {journal} {\bibinfo  {journal} {J. Chem. Phys.}\ }\textbf {\bibinfo {volume} {118}},\ \bibinfo {pages} {9095--9102} (\bibinfo {year} {2003})}\BibitemShut {NoStop}%
\bibitem [{\citenamefont {Grimme}, \citenamefont {Goerigk},\ and\ \citenamefont {Fink}(2012)}]{scs_2}%
  \BibitemOpen
  \bibfield  {author} {\bibinfo {author} {\bibfnamefont {S.}~\bibnamefont {Grimme}}, \bibinfo {author} {\bibfnamefont {L.}~\bibnamefont {Goerigk}},\ and\ \bibinfo {author} {\bibfnamefont {R.~F.}\ \bibnamefont {Fink}},\ }\bibfield  {title} {\enquote {\bibinfo {title} {Spin-component-scaled electron correlation methods},}\ }\href {https://doi.org/10.1002/wcms.1110} {\bibfield  {journal} {\bibinfo  {journal} {WIREs Comput. Mol. Sci.}\ }\textbf {\bibinfo {volume} {2}},\ \bibinfo {pages} {886--906} (\bibinfo {year} {2012})}\BibitemShut {NoStop}%
\bibitem [{\citenamefont {Cruz}\ \emph {et~al.}(2025{\natexlab{a}})\citenamefont {Cruz}, \citenamefont {Opoku}, \citenamefont {{\'S}miga}, \citenamefont {Grabowski}, \citenamefont {Ortiz},\ and\ \citenamefont {Hirata}}]{cruz2025performance}%
  \BibitemOpen
  \bibfield  {author} {\bibinfo {author} {\bibfnamefont {J.~C.}\ \bibnamefont {Cruz}}, \bibinfo {author} {\bibfnamefont {E.}~\bibnamefont {Opoku}}, \bibinfo {author} {\bibfnamefont {S.}~\bibnamefont {{\'S}miga}}, \bibinfo {author} {\bibfnamefont {I.}~\bibnamefont {Grabowski}}, \bibinfo {author} {\bibfnamefont {J.~V.}\ \bibnamefont {Ortiz}},\ and\ \bibinfo {author} {\bibfnamefont {S.}~\bibnamefont {Hirata}},\ }\bibfield  {title} {\enquote {\bibinfo {title} {Performance of the spin-component-scaled methods for energy bands},}\ }\href {https://doi.org/10.1080/00268976.2025.2504545} {\bibfield  {journal} {\bibinfo  {journal} {Mol. Phys.}\ ,\ \bibinfo {pages} {e2504545}} (\bibinfo {year} {2025}{\natexlab{a}})}\BibitemShut {NoStop}%
\bibitem [{\citenamefont {Caruso}\ \emph {et~al.}(2013)\citenamefont {Caruso}, \citenamefont {Rinke}, \citenamefont {Ren}, \citenamefont {Rubio},\ and\ \citenamefont {Scheffler}}]{scGW_1}%
  \BibitemOpen
  \bibfield  {author} {\bibinfo {author} {\bibfnamefont {F.}~\bibnamefont {Caruso}}, \bibinfo {author} {\bibfnamefont {P.}~\bibnamefont {Rinke}}, \bibinfo {author} {\bibfnamefont {X.}~\bibnamefont {Ren}}, \bibinfo {author} {\bibfnamefont {A.}~\bibnamefont {Rubio}},\ and\ \bibinfo {author} {\bibfnamefont {M.}~\bibnamefont {Scheffler}},\ }\bibfield  {title} {\enquote {\bibinfo {title} {Self-consistent {{G W}} : {{All-electron}} implementation with localized basis functions},}\ }\href {https://doi.org/10.1103/PhysRevB.88.075105} {\bibfield  {journal} {\bibinfo  {journal} {Phys. Rev. B}\ }\textbf {\bibinfo {volume} {88}},\ \bibinfo {pages} {075105} (\bibinfo {year} {2013})}\BibitemShut {NoStop}%
\bibitem [{\citenamefont {Caruso}\ \emph {et~al.}(2012)\citenamefont {Caruso}, \citenamefont {Rinke}, \citenamefont {Ren}, \citenamefont {Scheffler},\ and\ \citenamefont {Rubio}}]{scGW_2}%
  \BibitemOpen
  \bibfield  {author} {\bibinfo {author} {\bibfnamefont {F.}~\bibnamefont {Caruso}}, \bibinfo {author} {\bibfnamefont {P.}~\bibnamefont {Rinke}}, \bibinfo {author} {\bibfnamefont {X.}~\bibnamefont {Ren}}, \bibinfo {author} {\bibfnamefont {M.}~\bibnamefont {Scheffler}},\ and\ \bibinfo {author} {\bibfnamefont {A.}~\bibnamefont {Rubio}},\ }\bibfield  {title} {\enquote {\bibinfo {title} {Unified description of ground and excited states of finite systems: {{The}} self-consistent {{G W}} approach},}\ }\href {https://doi.org/10.1103/PhysRevB.86.081102} {\bibfield  {journal} {\bibinfo  {journal} {Phys. Rev. B}\ }\textbf {\bibinfo {volume} {86}},\ \bibinfo {pages} {081102} (\bibinfo {year} {2012})}\BibitemShut {NoStop}%
\bibitem [{\citenamefont {Caruso}\ \emph {et~al.}(2016)\citenamefont {Caruso}, \citenamefont {Dauth}, \citenamefont {Van~Setten},\ and\ \citenamefont {Rinke}}]{gw100_qsgw}%
  \BibitemOpen
  \bibfield  {author} {\bibinfo {author} {\bibfnamefont {F.}~\bibnamefont {Caruso}}, \bibinfo {author} {\bibfnamefont {M.}~\bibnamefont {Dauth}}, \bibinfo {author} {\bibfnamefont {M.~J.}\ \bibnamefont {Van~Setten}},\ and\ \bibinfo {author} {\bibfnamefont {P.}~\bibnamefont {Rinke}},\ }\bibfield  {title} {\enquote {\bibinfo {title} {Benchmark of {{{\emph{GW}}}} {{Approaches}} for the {{{\emph{GW}}}} 100 {{Test Set}}},}\ }\href {https://doi.org/10.1021/acs.jctc.6b00774} {\bibfield  {journal} {\bibinfo  {journal} {J. Chem. Theory Comput.}\ }\textbf {\bibinfo {volume} {12}},\ \bibinfo {pages} {5076--5087} (\bibinfo {year} {2016})}\BibitemShut {NoStop}%
\bibitem [{\citenamefont {{Ismail-Beigi}}(2017)}]{ismail-beigi2017}%
  \BibitemOpen
  \bibfield  {author} {\bibinfo {author} {\bibfnamefont {S.}~\bibnamefont {{Ismail-Beigi}}},\ }\bibfield  {title} {\enquote {\bibinfo {title} {Justifying quasiparticle self-consistent schemes via gradient optimization in {{Baym}}--{{Kadanoff}} theory},}\ }\href {https://doi.org/10.1088/1361-648X/aa7803} {\bibfield  {journal} {\bibinfo  {journal} {J. Phys.: Condens. Matter}\ }\textbf {\bibinfo {volume} {29}},\ \bibinfo {pages} {385501} (\bibinfo {year} {2017})}\BibitemShut {NoStop}%
\bibitem [{\citenamefont {Coveney}\ and\ \citenamefont {Tew}(2023)}]{coveneyQPMP2}%
  \BibitemOpen
  \bibfield  {author} {\bibinfo {author} {\bibfnamefont {C.~J.~N.}\ \bibnamefont {Coveney}}\ and\ \bibinfo {author} {\bibfnamefont {D.~P.}\ \bibnamefont {Tew}},\ }\bibfield  {title} {\enquote {\bibinfo {title} {A {{Regularized Second-Order Correlation Method}} from {{Green}}'s {{Function Theory}}},}\ }\href {https://doi.org/10.1021/acs.jctc.3c00246} {\bibfield  {journal} {\bibinfo  {journal} {J. Chem. Theory Comput.}\ }\textbf {\bibinfo {volume} {19}},\ \bibinfo {pages} {3915--3928} (\bibinfo {year} {2023})}\BibitemShut {NoStop}%
\bibitem [{\citenamefont {Shee}\ \emph {et~al.}(2021{\natexlab{a}})\citenamefont {Shee}, \citenamefont {Loipersberger}, \citenamefont {Rettig}, \citenamefont {Lee},\ and\ \citenamefont {{Head-Gordon}}}]{shee2021regularizedPT2}%
  \BibitemOpen
  \bibfield  {author} {\bibinfo {author} {\bibfnamefont {J.}~\bibnamefont {Shee}}, \bibinfo {author} {\bibfnamefont {M.}~\bibnamefont {Loipersberger}}, \bibinfo {author} {\bibfnamefont {A.}~\bibnamefont {Rettig}}, \bibinfo {author} {\bibfnamefont {J.}~\bibnamefont {Lee}},\ and\ \bibinfo {author} {\bibfnamefont {M.}~\bibnamefont {{Head-Gordon}}},\ }\bibfield  {title} {\enquote {\bibinfo {title} {Regularized {{Second-Order M{\o}ller}}--{{Plesset Theory}}: {{A More Accurate Alternative}} to {{Conventional MP2}} for {{Noncovalent Interactions}} and {{Transition Metal Thermochemistry}} for the {{Same Computational Cost}}},}\ }\href {https://doi.org/10.1021/acs.jpclett.1c03468} {\bibfield  {journal} {\bibinfo  {journal} {J. Phys. Chem. Lett.}\ }\textbf {\bibinfo {volume} {12}},\ \bibinfo {pages} {12084--12097} (\bibinfo {year} {2021}{\natexlab{a}})}\BibitemShut {NoStop}%
\bibitem [{\citenamefont {Kehrein}(2006)}]{kehrein2006}%
  \BibitemOpen
  \bibfield  {author} {\bibinfo {author} {\bibfnamefont {S.}~\bibnamefont {Kehrein}},\ }\href@noop {} {\emph {\bibinfo {title} {The {{Flow Equation Approach}} to {{Many-Particle Systems}}}}},\ \bibinfo {series} {Springer {{Tracts}} in {{Modern Physics Ser}}}\ No.\ \bibinfo {number} {v.217}\ (\bibinfo  {publisher} {Springer Berlin / Heidelberg},\ \bibinfo {address} {Berlin, Heidelberg},\ \bibinfo {year} {2006})\BibitemShut {NoStop}%
\bibitem [{\citenamefont {Cruz}\ \emph {et~al.}(2025{\natexlab{b}})\citenamefont {Cruz}, \citenamefont {Opoku}, \citenamefont {{\'S}miga}, \citenamefont {Grabowski}, \citenamefont {Ortiz},\ and\ \citenamefont {Hirata}}]{hirata_scs_gf2_2025}%
  \BibitemOpen
  \bibfield  {author} {\bibinfo {author} {\bibfnamefont {J.~C.}\ \bibnamefont {Cruz}}, \bibinfo {author} {\bibfnamefont {E.}~\bibnamefont {Opoku}}, \bibinfo {author} {\bibfnamefont {S.}~\bibnamefont {{\'S}miga}}, \bibinfo {author} {\bibfnamefont {I.}~\bibnamefont {Grabowski}}, \bibinfo {author} {\bibfnamefont {J.~V.}\ \bibnamefont {Ortiz}},\ and\ \bibinfo {author} {\bibfnamefont {S.}~\bibnamefont {Hirata}},\ }\bibfield  {title} {\enquote {\bibinfo {title} {Performance of the spin-component-scaled methods for energy bands},}\ }\href {https://doi.org/10.1080/00268976.2025.2504545} {\bibfield  {journal} {\bibinfo  {journal} {Mol. Phys.}\ ,\ \bibinfo {pages} {e2504545}} (\bibinfo {year} {2025}{\natexlab{b}})}\BibitemShut {NoStop}%
\bibitem [{\citenamefont {Knowles}\ and\ \citenamefont {Handy}(1988)}]{knowlesConvergence1988}%
  \BibitemOpen
  \bibfield  {author} {\bibinfo {author} {\bibfnamefont {P.~J.}\ \bibnamefont {Knowles}}\ and\ \bibinfo {author} {\bibfnamefont {N.~C.}\ \bibnamefont {Handy}},\ }\bibfield  {title} {\enquote {\bibinfo {title} {Convergence of projected unrestricted {{Hartee-Fock Moeller-Plesset}} series.}}\ }\href {https://doi.org/10.1021/j100322a018} {\bibfield  {journal} {\bibinfo  {journal} {J. Phys. Chem.}\ }\textbf {\bibinfo {volume} {92}},\ \bibinfo {pages} {3097--3100} (\bibinfo {year} {1988})}\BibitemShut {NoStop}%
\bibitem [{\citenamefont {Knowles}(2022)}]{knowles2022perturbation}%
  \BibitemOpen
  \bibfield  {author} {\bibinfo {author} {\bibfnamefont {P.~J.}\ \bibnamefont {Knowles}},\ }\bibfield  {title} {\enquote {\bibinfo {title} {Perturbation-adapted perturbation theory},}\ }\href {https://doi.org/10.1063/5.0079853} {\bibfield  {journal} {\bibinfo  {journal} {J. Chem. Phys.}\ }\textbf {\bibinfo {volume} {156}} (\bibinfo {year} {2022}),\ 10.1063/5.0079853}\BibitemShut {NoStop}%
\bibitem [{\citenamefont {Gombas}, \citenamefont {Surjan},\ and\ \citenamefont {Szabados}(2024)}]{gombas2024analysis}%
  \BibitemOpen
  \bibfield  {author} {\bibinfo {author} {\bibfnamefont {A.}~\bibnamefont {Gombas}}, \bibinfo {author} {\bibfnamefont {P.~R.}\ \bibnamefont {Surjan}},\ and\ \bibinfo {author} {\bibfnamefont {{\'A}.}~\bibnamefont {Szabados}},\ }\bibfield  {title} {\enquote {\bibinfo {title} {Analysis and assessment of knowles' partitioning in many-body perturbation theory},}\ }\href {https://doi.org/10.1021/acs.jctc.4c00166} {\bibfield  {journal} {\bibinfo  {journal} {J. Chem. Theory Comput.}\ }\textbf {\bibinfo {volume} {20}},\ \bibinfo {pages} {5094--5104} (\bibinfo {year} {2024})}\BibitemShut {NoStop}%
\bibitem [{\citenamefont {Szabados}, \citenamefont {Gomb{\'a}s},\ and\ \citenamefont {Surj{\'a}n}(2024)}]{szabados2024knowles}%
  \BibitemOpen
  \bibfield  {author} {\bibinfo {author} {\bibfnamefont {{\'A}.}~\bibnamefont {Szabados}}, \bibinfo {author} {\bibfnamefont {A.}~\bibnamefont {Gomb{\'a}s}},\ and\ \bibinfo {author} {\bibfnamefont {P.~R.}\ \bibnamefont {Surj{\'a}n}},\ }\bibfield  {title} {\enquote {\bibinfo {title} {Knowles partitioning at the multireference level},}\ }\href {https://doi.org/10.1021/acs.jpca.4c04279} {\bibfield  {journal} {\bibinfo  {journal} {J. Phys. Chem. A}\ }\textbf {\bibinfo {volume} {128}},\ \bibinfo {pages} {9311--9322} (\bibinfo {year} {2024})}\BibitemShut {NoStop}%
\bibitem [{\citenamefont {Dittmer}\ and\ \citenamefont {Head-Gordon}(2025)}]{dittmer2025repartitioning}%
  \BibitemOpen
  \bibfield  {author} {\bibinfo {author} {\bibfnamefont {L.~B.}\ \bibnamefont {Dittmer}}\ and\ \bibinfo {author} {\bibfnamefont {M.}~\bibnamefont {Head-Gordon}},\ }\bibfield  {title} {\enquote {\bibinfo {title} {Repartitioning the hamiltonian in many-body second-order brillouin--wigner perturbation theory: Uncovering new size-consistent models},}\ }\href {https://doi.org/https://doi.org/10.1063/5.0242211} {\bibfield  {journal} {\bibinfo  {journal} {J. Chem. Phys.}\ }\textbf {\bibinfo {volume} {162}} (\bibinfo {year} {2025}),\ https://doi.org/10.1063/5.0242211}\BibitemShut {NoStop}%
\bibitem [{\citenamefont {Bernholdt}\ and\ \citenamefont {Harrison}(1996)}]{densfit_1}%
  \BibitemOpen
  \bibfield  {author} {\bibinfo {author} {\bibfnamefont {D.~E.}\ \bibnamefont {Bernholdt}}\ and\ \bibinfo {author} {\bibfnamefont {R.~J.}\ \bibnamefont {Harrison}},\ }\bibfield  {title} {\enquote {\bibinfo {title} {Large-scale correlated electronic structure calculations: The {{RI-MP2}} method on parallel computers},}\ }\href {https://doi.org/10.1016/0009-2614(96)00054-1} {\bibfield  {journal} {\bibinfo  {journal} {Chem. Phys. Lett.}\ }\textbf {\bibinfo {volume} {250}},\ \bibinfo {pages} {477--484} (\bibinfo {year} {1996})}\BibitemShut {NoStop}%
\bibitem [{\citenamefont {Bauernschmitt}\ \emph {et~al.}(1997)\citenamefont {Bauernschmitt}, \citenamefont {H{\"a}ser}, \citenamefont {Treutler},\ and\ \citenamefont {Ahlrichs}}]{densfit_2}%
  \BibitemOpen
  \bibfield  {author} {\bibinfo {author} {\bibfnamefont {R.}~\bibnamefont {Bauernschmitt}}, \bibinfo {author} {\bibfnamefont {M.}~\bibnamefont {H{\"a}ser}}, \bibinfo {author} {\bibfnamefont {O.}~\bibnamefont {Treutler}},\ and\ \bibinfo {author} {\bibfnamefont {R.}~\bibnamefont {Ahlrichs}},\ }\bibfield  {title} {\enquote {\bibinfo {title} {Calculation of excitation energies within time-dependent density functional theory using auxiliary basis set expansions},}\ }\href {https://doi.org/10.1016/S0009-2614(96)01343-7} {\bibfield  {journal} {\bibinfo  {journal} {Chem. Phys. Lett.}\ }\textbf {\bibinfo {volume} {264}},\ \bibinfo {pages} {573--578} (\bibinfo {year} {1997})}\BibitemShut {NoStop}%
\bibitem [{\citenamefont {Unsleber}\ \emph {et~al.}(2018)\citenamefont {Unsleber}, \citenamefont {Dresselhaus}, \citenamefont {Klahr}, \citenamefont {Schnieders}, \citenamefont {B{\"o}ckers}, \citenamefont {Barton},\ and\ \citenamefont {Neugebauer}}]{serenity_1}%
  \BibitemOpen
  \bibfield  {author} {\bibinfo {author} {\bibfnamefont {J.~P.}\ \bibnamefont {Unsleber}}, \bibinfo {author} {\bibfnamefont {T.}~\bibnamefont {Dresselhaus}}, \bibinfo {author} {\bibfnamefont {K.}~\bibnamefont {Klahr}}, \bibinfo {author} {\bibfnamefont {D.}~\bibnamefont {Schnieders}}, \bibinfo {author} {\bibfnamefont {M.}~\bibnamefont {B{\"o}ckers}}, \bibinfo {author} {\bibfnamefont {D.}~\bibnamefont {Barton}},\ and\ \bibinfo {author} {\bibfnamefont {J.}~\bibnamefont {Neugebauer}},\ }\bibfield  {title} {\enquote {\bibinfo {title} {Serenity : {{A}} subsystem quantum chemistry program},}\ }\href {https://doi.org/10.1002/jcc.25162} {\bibfield  {journal} {\bibinfo  {journal} {J. Comput. Chem.}\ }\textbf {\bibinfo {volume} {39}},\ \bibinfo {pages} {788--798} (\bibinfo {year} {2018})}\BibitemShut {NoStop}%
\bibitem [{\citenamefont {Niemeyer}\ \emph {et~al.}(2023)\citenamefont {Niemeyer}, \citenamefont {Eschenbach}, \citenamefont {Bensberg}, \citenamefont {T{\"o}lle}, \citenamefont {Hellmann}, \citenamefont {Lampe}, \citenamefont {Massolle}, \citenamefont {Rikus}, \citenamefont {Schnieders}, \citenamefont {Unsleber},\ and\ \citenamefont {Neugebauer}}]{serenity_2}%
  \BibitemOpen
  \bibfield  {author} {\bibinfo {author} {\bibfnamefont {N.}~\bibnamefont {Niemeyer}}, \bibinfo {author} {\bibfnamefont {P.}~\bibnamefont {Eschenbach}}, \bibinfo {author} {\bibfnamefont {M.}~\bibnamefont {Bensberg}}, \bibinfo {author} {\bibfnamefont {J.}~\bibnamefont {T{\"o}lle}}, \bibinfo {author} {\bibfnamefont {L.}~\bibnamefont {Hellmann}}, \bibinfo {author} {\bibfnamefont {L.}~\bibnamefont {Lampe}}, \bibinfo {author} {\bibfnamefont {A.}~\bibnamefont {Massolle}}, \bibinfo {author} {\bibfnamefont {A.}~\bibnamefont {Rikus}}, \bibinfo {author} {\bibfnamefont {D.}~\bibnamefont {Schnieders}}, \bibinfo {author} {\bibfnamefont {J.~P.}\ \bibnamefont {Unsleber}},\ and\ \bibinfo {author} {\bibfnamefont {J.}~\bibnamefont {Neugebauer}},\ }\bibfield  {title} {\enquote {\bibinfo {title} {The subsystem quantum chemistry program {{{\textsc{Serenity}}}}},}\ }\href {https://doi.org/10.1002/wcms.1647} {\bibfield  {journal} {\bibinfo  {journal} {WIREs Comput. Mole. Sci.}\ }\textbf {\bibinfo {volume} {13}},\ \bibinfo {pages} {e1647} (\bibinfo {year} {2023})}\BibitemShut {NoStop}%
\bibitem [{\citenamefont {Barton}\ \emph {et~al.}(2024)\citenamefont {Barton}, \citenamefont {Bensberg}, \citenamefont {B{\"o}ckers}, \citenamefont {Dresselhaus}, \citenamefont {Eschenbach}, \citenamefont {G{\"o}llmann}, \citenamefont {Hellmann}, \citenamefont {Lampe}, \citenamefont {Massolle}, \citenamefont {Niemeyer}, \citenamefont {Ramez}, \citenamefont {Rikus}, \citenamefont {Schnieders}, \citenamefont {T{\"o}lle}, \citenamefont {Unsleber}, \citenamefont {Wegner},\ and\ \citenamefont {Neugebauer}}]{serenity_1.6.1}%
  \BibitemOpen
  \bibfield  {author} {\bibinfo {author} {\bibfnamefont {D.}~\bibnamefont {Barton}}, \bibinfo {author} {\bibfnamefont {M.}~\bibnamefont {Bensberg}}, \bibinfo {author} {\bibfnamefont {M.}~\bibnamefont {B{\"o}ckers}}, \bibinfo {author} {\bibfnamefont {T.}~\bibnamefont {Dresselhaus}}, \bibinfo {author} {\bibfnamefont {P.}~\bibnamefont {Eschenbach}}, \bibinfo {author} {\bibfnamefont {N.}~\bibnamefont {G{\"o}llmann}}, \bibinfo {author} {\bibfnamefont {L.}~\bibnamefont {Hellmann}}, \bibinfo {author} {\bibfnamefont {L.}~\bibnamefont {Lampe}}, \bibinfo {author} {\bibfnamefont {A.}~\bibnamefont {Massolle}}, \bibinfo {author} {\bibfnamefont {N.}~\bibnamefont {Niemeyer}}, \bibinfo {author} {\bibfnamefont {N.}~\bibnamefont {Ramez}}, \bibinfo {author} {\bibfnamefont {A.}~\bibnamefont {Rikus}}, \bibinfo {author} {\bibfnamefont {D.}~\bibnamefont {Schnieders}}, \bibinfo {author} {\bibfnamefont {J.}~\bibnamefont {T{\"o}lle}}, \bibinfo {author} {\bibfnamefont {J.~P.}\ \bibnamefont {Unsleber}}, \bibinfo {author} {\bibfnamefont {K.}~\bibnamefont {Wegner}},\ and\ \bibinfo {author} {\bibfnamefont {J.}~\bibnamefont {Neugebauer}},\ }\href {https://doi.org/10.5281/ZENODO.10838411} {\enquote {\bibinfo {title} {Qcserenity/serenity: {{Release}} 1.6.1},}\ }\bibinfo {howpublished} {Zenodo} (\bibinfo {year} {2024})\BibitemShut {NoStop}%
\bibitem [{\citenamefont {K{\'a}llay}(2014)}]{nafs_1}%
  \BibitemOpen
  \bibfield  {author} {\bibinfo {author} {\bibfnamefont {M.}~\bibnamefont {K{\'a}llay}},\ }\bibfield  {title} {\enquote {\bibinfo {title} {A systematic way for the cost reduction of density fitting methods},}\ }\href {https://doi.org/10.1063/1.4905005} {\bibfield  {journal} {\bibinfo  {journal} {J. Chem. Phys.}\ }\textbf {\bibinfo {volume} {141}},\ \bibinfo {pages} {244113} (\bibinfo {year} {2014})}\BibitemShut {NoStop}%
\bibitem [{\citenamefont {Mester}, \citenamefont {Nagy},\ and\ \citenamefont {K{\'a}llay}(2017)}]{nafs_2}%
  \BibitemOpen
  \bibfield  {author} {\bibinfo {author} {\bibfnamefont {D.}~\bibnamefont {Mester}}, \bibinfo {author} {\bibfnamefont {P.~R.}\ \bibnamefont {Nagy}},\ and\ \bibinfo {author} {\bibfnamefont {M.}~\bibnamefont {K{\'a}llay}},\ }\bibfield  {title} {\enquote {\bibinfo {title} {Reduced-cost linear-response {{CC2}} method based on natural orbitals and natural auxiliary functions},}\ }\href {https://doi.org/10.1063/1.4983277} {\bibfield  {journal} {\bibinfo  {journal} {J. Chem. Phys.}\ }\textbf {\bibinfo {volume} {146}},\ \bibinfo {pages} {194102} (\bibinfo {year} {2017})}\BibitemShut {NoStop}%
\bibitem [{\citenamefont {T\"olle}, \citenamefont {Niemeyer},\ and\ \citenamefont {Neugebauer}(2024)}]{tolle2024accelerating}%
  \BibitemOpen
  \bibfield  {author} {\bibinfo {author} {\bibfnamefont {J.}~\bibnamefont {T\"olle}}, \bibinfo {author} {\bibfnamefont {N.}~\bibnamefont {Niemeyer}},\ and\ \bibinfo {author} {\bibfnamefont {J.}~\bibnamefont {Neugebauer}},\ }\bibfield  {title} {\enquote {\bibinfo {title} {Accelerating analytic-continuation gw calculations with a laplace transform and natural auxiliary functions},}\ }\href {https://doi.org/10.1021/acs.jctc.3c01264} {\bibfield  {journal} {\bibinfo  {journal} {J. Chem. Theory Comput.}\ }\textbf {\bibinfo {volume} {20}},\ \bibinfo {pages} {2022--2032} (\bibinfo {year} {2024})}\BibitemShut {NoStop}%
\bibitem [{\citenamefont {Pulay}(1980)}]{diis1}%
  \BibitemOpen
  \bibfield  {author} {\bibinfo {author} {\bibfnamefont {P.}~\bibnamefont {Pulay}},\ }\bibfield  {title} {\enquote {\bibinfo {title} {Convergence acceleration of iterative sequences. the case of scf iteration},}\ }\href {https://doi.org/10.1016/0009-2614(80)80396-4} {\bibfield  {journal} {\bibinfo  {journal} {Chem. Phys. Lett.}\ }\textbf {\bibinfo {volume} {73}},\ \bibinfo {pages} {393--398} (\bibinfo {year} {1980})}\BibitemShut {NoStop}%
\bibitem [{\citenamefont {Pulay}(1982)}]{diis2}%
  \BibitemOpen
  \bibfield  {author} {\bibinfo {author} {\bibfnamefont {P.}~\bibnamefont {Pulay}},\ }\bibfield  {title} {\enquote {\bibinfo {title} {Improved {{{\textsc{SCF}}}} convergence acceleration},}\ }\href {https://doi.org/10.1002/jcc.540030413} {\bibfield  {journal} {\bibinfo  {journal} {J. Comput. Chem.}\ }\textbf {\bibinfo {volume} {3}},\ \bibinfo {pages} {556--560} (\bibinfo {year} {1982})}\BibitemShut {NoStop}%
\bibitem [{\citenamefont {Hamilton}\ and\ \citenamefont {Pulay}(1986)}]{diis3}%
  \BibitemOpen
  \bibfield  {author} {\bibinfo {author} {\bibfnamefont {T.~P.}\ \bibnamefont {Hamilton}}\ and\ \bibinfo {author} {\bibfnamefont {P.}~\bibnamefont {Pulay}},\ }\bibfield  {title} {\enquote {\bibinfo {title} {Direct inversion in the iterative subspace ({{DIIS}}) optimization of open-shell, excited-state, and small multiconfiguration {{SCF}} wave functions},}\ }\href {https://doi.org/10.1063/1.449880} {\bibfield  {journal} {\bibinfo  {journal} {J. Chem. Phys.}\ }\textbf {\bibinfo {volume} {84}},\ \bibinfo {pages} {5728--5734} (\bibinfo {year} {1986})}\BibitemShut {NoStop}%
\bibitem [{\citenamefont {Dunning}(1989)}]{augccpvtz_1}%
  \BibitemOpen
  \bibfield  {author} {\bibinfo {author} {\bibfnamefont {T.~H.}\ \bibnamefont {Dunning}},\ }\bibfield  {title} {\enquote {\bibinfo {title} {Gaussian basis sets for use in correlated molecular calculations. {{I}}. {{The}} atoms boron through neon and hydrogen},}\ }\href {https://doi.org/10.1063/1.456153} {\bibfield  {journal} {\bibinfo  {journal} {J. Chem. Phys.}\ }\textbf {\bibinfo {volume} {90}},\ \bibinfo {pages} {1007--1023} (\bibinfo {year} {1989})}\BibitemShut {NoStop}%
\bibitem [{\citenamefont {Kendall}, \citenamefont {Dunning},\ and\ \citenamefont {Harrison}(1992)}]{augccpvtz_2}%
  \BibitemOpen
  \bibfield  {author} {\bibinfo {author} {\bibfnamefont {R.~A.}\ \bibnamefont {Kendall}}, \bibinfo {author} {\bibfnamefont {T.~H.}\ \bibnamefont {Dunning}},\ and\ \bibinfo {author} {\bibfnamefont {R.~J.}\ \bibnamefont {Harrison}},\ }\bibfield  {title} {\enquote {\bibinfo {title} {Electron affinities of the first-row atoms revisited. {{Systematic}} basis sets and wave functions},}\ }\href {https://doi.org/10.1063/1.462569} {\bibfield  {journal} {\bibinfo  {journal} {J. Chem. Phys.}\ }\textbf {\bibinfo {volume} {96}},\ \bibinfo {pages} {6796--6806} (\bibinfo {year} {1992})}\BibitemShut {NoStop}%
\bibitem [{\citenamefont {Prascher}\ \emph {et~al.}(2011)\citenamefont {Prascher}, \citenamefont {Woon}, \citenamefont {Peterson}, \citenamefont {Dunning},\ and\ \citenamefont {Wilson}}]{augccpvtz_3}%
  \BibitemOpen
  \bibfield  {author} {\bibinfo {author} {\bibfnamefont {B.~P.}\ \bibnamefont {Prascher}}, \bibinfo {author} {\bibfnamefont {D.~E.}\ \bibnamefont {Woon}}, \bibinfo {author} {\bibfnamefont {K.~A.}\ \bibnamefont {Peterson}}, \bibinfo {author} {\bibfnamefont {T.~H.}\ \bibnamefont {Dunning}},\ and\ \bibinfo {author} {\bibfnamefont {A.~K.}\ \bibnamefont {Wilson}},\ }\bibfield  {title} {\enquote {\bibinfo {title} {Gaussian basis sets for use in correlated molecular calculations. {{VII}}. {{Valence}}, core-valence, and scalar relativistic basis sets for {{Li}}, {{Be}}, {{Na}}, and {{Mg}}},}\ }\href {https://doi.org/10.1007/s00214-010-0764-0} {\bibfield  {journal} {\bibinfo  {journal} {Theor. Chem. Acc.}\ }\textbf {\bibinfo {volume} {128}},\ \bibinfo {pages} {69--82} (\bibinfo {year} {2011})}\BibitemShut {NoStop}%
\bibitem [{\citenamefont {Woon}\ and\ \citenamefont {Dunning}(1994)}]{augccpvtz_4}%
  \BibitemOpen
  \bibfield  {author} {\bibinfo {author} {\bibfnamefont {D.~E.}\ \bibnamefont {Woon}}\ and\ \bibinfo {author} {\bibfnamefont {T.~H.}\ \bibnamefont {Dunning}},\ }\bibfield  {title} {\enquote {\bibinfo {title} {Gaussian basis sets for use in correlated molecular calculations. {{IV}}. {{Calculation}} of static electrical response properties},}\ }\href {https://doi.org/10.1063/1.466439} {\bibfield  {journal} {\bibinfo  {journal} {J. Chem. Phys.}\ }\textbf {\bibinfo {volume} {100}},\ \bibinfo {pages} {2975--2988} (\bibinfo {year} {1994})}\BibitemShut {NoStop}%
\bibitem [{\citenamefont {Weigend}, \citenamefont {Köhn},\ and\ \citenamefont {H{\"a}ttig}(2002)}]{pVTZ_RI_1}%
  \BibitemOpen
  \bibfield  {author} {\bibinfo {author} {\bibfnamefont {F.}~\bibnamefont {Weigend}}, \bibinfo {author} {\bibfnamefont {A.}~\bibnamefont {Köhn}},\ and\ \bibinfo {author} {\bibfnamefont {C.}~\bibnamefont {H{\"a}ttig}},\ }\bibfield  {title} {\enquote {\bibinfo {title} {Efficient use of the correlation consistent basis sets in resolution of the identity mp2 calculations},}\ }\href {https://doi.org/10.1063/1.1445115} {\bibfield  {journal} {\bibinfo  {journal} {J. Chem. Phys.}\ }\textbf {\bibinfo {volume} {116}},\ \bibinfo {pages} {3175--3183} (\bibinfo {year} {2002})}\BibitemShut {NoStop}%
\bibitem [{\citenamefont {H{\"a}ttig}(2005)}]{pvtz_RI_2}%
  \BibitemOpen
  \bibfield  {author} {\bibinfo {author} {\bibfnamefont {C.}~\bibnamefont {H{\"a}ttig}},\ }\bibfield  {title} {\enquote {\bibinfo {title} {Optimization of auxiliary basis sets for ri-mp2 and ri-cc2 calculations: Core-valence and quintuple-{$\zeta$} basis sets for h to ar and qzvpp basis sets for li to kr},}\ }\href {https://doi.org/10.1039/b415208e} {\bibfield  {journal} {\bibinfo  {journal} {Phys. Chem. Chem. Phys.}\ }\textbf {\bibinfo {volume} {7}},\ \bibinfo {pages} {59--66} (\bibinfo {year} {2005})}\BibitemShut {NoStop}%
\bibitem [{\citenamefont {Weigend}\ \emph {et~al.}(1998)\citenamefont {Weigend}, \citenamefont {H{\"a}ser}, \citenamefont {Patzelt},\ and\ \citenamefont {Ahlrichs}}]{def2_RI}%
  \BibitemOpen
  \bibfield  {author} {\bibinfo {author} {\bibfnamefont {F.}~\bibnamefont {Weigend}}, \bibinfo {author} {\bibfnamefont {M.}~\bibnamefont {H{\"a}ser}}, \bibinfo {author} {\bibfnamefont {H.}~\bibnamefont {Patzelt}},\ and\ \bibinfo {author} {\bibfnamefont {R.}~\bibnamefont {Ahlrichs}},\ }\bibfield  {title} {\enquote {\bibinfo {title} {{{RI-MP2}}: Optimized auxiliary basis sets and demonstration of efficiency},}\ }\href {https://doi.org/10.1016/S0009-2614(98)00862-8} {\bibfield  {journal} {\bibinfo  {journal} {Chem. Phys. Lett.}\ }\textbf {\bibinfo {volume} {294}},\ \bibinfo {pages} {143--152} (\bibinfo {year} {1998})}\BibitemShut {NoStop}%
\bibitem [{\citenamefont {Hellweg}\ and\ \citenamefont {Rappoport}(2015)}]{def2svpd_tzvppd_qzvppd_RI}%
  \BibitemOpen
  \bibfield  {author} {\bibinfo {author} {\bibfnamefont {A.}~\bibnamefont {Hellweg}}\ and\ \bibinfo {author} {\bibfnamefont {D.}~\bibnamefont {Rappoport}},\ }\bibfield  {title} {\enquote {\bibinfo {title} {Development of new auxiliary basis functions of the {{Karlsruhe}} segmented contracted basis sets including diffuse basis functions (def2-{{SVPD}}, def2-{{TZVPPD}}, and def2-{{QVPPD}}) for {{RI-MP2}} and {{RI-CC}} calculations},}\ }\href {https://doi.org/10.1039/C4CP04286G} {\bibfield  {journal} {\bibinfo  {journal} {Phys. Chem. Chem. Phys.}\ }\textbf {\bibinfo {volume} {17}},\ \bibinfo {pages} {1010--1017} (\bibinfo {year} {2015})}\BibitemShut {NoStop}%
\bibitem [{\citenamefont {Jung}\ \emph {et~al.}(2004)\citenamefont {Jung}, \citenamefont {Lochan}, \citenamefont {Dutoi},\ and\ \citenamefont {{Head-Gordon}}}]{sos}%
  \BibitemOpen
  \bibfield  {author} {\bibinfo {author} {\bibfnamefont {Y.}~\bibnamefont {Jung}}, \bibinfo {author} {\bibfnamefont {R.~C.}\ \bibnamefont {Lochan}}, \bibinfo {author} {\bibfnamefont {A.~D.}\ \bibnamefont {Dutoi}},\ and\ \bibinfo {author} {\bibfnamefont {M.}~\bibnamefont {{Head-Gordon}}},\ }\bibfield  {title} {\enquote {\bibinfo {title} {Scaled opposite-spin second order {{M\o ller}}--{{Plesset}} correlation energy: {{An}} economical electronic structure method},}\ }\href {https://doi.org/10.1063/1.1809602} {\bibfield  {journal} {\bibinfo  {journal} {J. Chem. Phys.}\ }\textbf {\bibinfo {volume} {121}},\ \bibinfo {pages} {9793--9802} (\bibinfo {year} {2004})}\BibitemShut {NoStop}%
\bibitem [{\citenamefont {{\'S}miga}\ and\ \citenamefont {Grabowski}(2018)}]{deltamp2_1}%
  \BibitemOpen
  \bibfield  {author} {\bibinfo {author} {\bibfnamefont {S.}~\bibnamefont {{\'S}miga}}\ and\ \bibinfo {author} {\bibfnamefont {I.}~\bibnamefont {Grabowski}},\ }\bibfield  {title} {\enquote {\bibinfo {title} {Spin-{{Component-Scaled $\Delta$MP2 Parametrization}}: {{Toward}} a {{Simple}} and {{Reliable Method}} for {{Ionization Energies}}},}\ }\href {https://doi.org/10.1021/acs.jctc.8b00638} {\bibfield  {journal} {\bibinfo  {journal} {J. Chem. Theory Comput.}\ }\textbf {\bibinfo {volume} {14}},\ \bibinfo {pages} {4780--4790} (\bibinfo {year} {2018})}\BibitemShut {NoStop}%
\bibitem [{\citenamefont {{\'S}miga}, \citenamefont {Sieci{\'n}ska},\ and\ \citenamefont {Grabowski}(2020)}]{deltamp2_2}%
  \BibitemOpen
  \bibfield  {author} {\bibinfo {author} {\bibfnamefont {S.}~\bibnamefont {{\'S}miga}}, \bibinfo {author} {\bibfnamefont {S.}~\bibnamefont {Sieci{\'n}ska}},\ and\ \bibinfo {author} {\bibfnamefont {I.}~\bibnamefont {Grabowski}},\ }\bibfield  {title} {\enquote {\bibinfo {title} {From simple molecules to nanotubes. {{Reliable}} predictions of ionization potentials from the {{$\Delta$MP2-SCS}} methods},}\ }\href {https://doi.org/10.1088/1367-2630/abaa00} {\bibfield  {journal} {\bibinfo  {journal} {New J. Phys.}\ }\textbf {\bibinfo {volume} {22}},\ \bibinfo {pages} {083084} (\bibinfo {year} {2020})}\BibitemShut {NoStop}%
\bibitem [{\citenamefont {Opoku}, \citenamefont {Paw{\l}owski},\ and\ \citenamefont {Ortiz}(2021)}]{deltamp2_3}%
  \BibitemOpen
  \bibfield  {author} {\bibinfo {author} {\bibfnamefont {E.}~\bibnamefont {Opoku}}, \bibinfo {author} {\bibfnamefont {F.}~\bibnamefont {Paw{\l}owski}},\ and\ \bibinfo {author} {\bibfnamefont {J.~V.}\ \bibnamefont {Ortiz}},\ }\bibfield  {title} {\enquote {\bibinfo {title} {A new generation of diagonal self-energies for the calculation of electron removal energies},}\ }\href {https://doi.org/10.1063/5.0070849} {\bibfield  {journal} {\bibinfo  {journal} {J. Chem. Phys.}\ }\textbf {\bibinfo {volume} {155}},\ \bibinfo {pages} {204107} (\bibinfo {year} {2021})}\BibitemShut {NoStop}%
\bibitem [{\citenamefont {Opoku}, \citenamefont {Paw{\l}owski},\ and\ \citenamefont {Ortiz}(2022)}]{deltamp2_4}%
  \BibitemOpen
  \bibfield  {author} {\bibinfo {author} {\bibfnamefont {E.}~\bibnamefont {Opoku}}, \bibinfo {author} {\bibfnamefont {F.}~\bibnamefont {Paw{\l}owski}},\ and\ \bibinfo {author} {\bibfnamefont {J.~V.}\ \bibnamefont {Ortiz}},\ }\bibfield  {title} {\enquote {\bibinfo {title} {Electron {{Propagator Self-Energies}} versus {{Improved GW100 Vertical Ionization Energies}}},}\ }\href {https://doi.org/10.1021/acs.jctc.2c00502} {\bibfield  {journal} {\bibinfo  {journal} {J. Chem. Theory Comput.}\ }\textbf {\bibinfo {volume} {18}},\ \bibinfo {pages} {4927--4944} (\bibinfo {year} {2022})}\BibitemShut {NoStop}%
\bibitem [{\citenamefont {Weigend}\ and\ \citenamefont {Ahlrichs}(2005)}]{def2svp_tzvp_tzvppd}%
  \BibitemOpen
  \bibfield  {author} {\bibinfo {author} {\bibfnamefont {F.}~\bibnamefont {Weigend}}\ and\ \bibinfo {author} {\bibfnamefont {R.}~\bibnamefont {Ahlrichs}},\ }\bibfield  {title} {\enquote {\bibinfo {title} {Balanced basis sets of split valence, triple zeta valence and quadruple zeta valence quality for {{H}} to {{Rn}}: {{Design}} and assessment of accuracy},}\ }\href {https://doi.org/10.1039/b508541a} {\bibfield  {journal} {\bibinfo  {journal} {Phys. Chem. Chem. Phys.}\ }\textbf {\bibinfo {volume} {7}},\ \bibinfo {pages} {3297} (\bibinfo {year} {2005})}\BibitemShut {NoStop}%
\bibitem [{\citenamefont {Van~Setten}\ \emph {et~al.}(2015{\natexlab{b}})\citenamefont {Van~Setten}, \citenamefont {Caruso}, \citenamefont {Sharifzadeh}, \citenamefont {Ren}, \citenamefont {Scheffler}, \citenamefont {Liu}, \citenamefont {Lischner}, \citenamefont {Lin}, \citenamefont {Deslippe}, \citenamefont {Louie}, \citenamefont {Yang}, \citenamefont {Weigend}, \citenamefont {Neaton}, \citenamefont {Evers},\ and\ \citenamefont {Rinke}}]{GW100}%
  \BibitemOpen
  \bibfield  {author} {\bibinfo {author} {\bibfnamefont {M.~J.}\ \bibnamefont {Van~Setten}}, \bibinfo {author} {\bibfnamefont {F.}~\bibnamefont {Caruso}}, \bibinfo {author} {\bibfnamefont {S.}~\bibnamefont {Sharifzadeh}}, \bibinfo {author} {\bibfnamefont {X.}~\bibnamefont {Ren}}, \bibinfo {author} {\bibfnamefont {M.}~\bibnamefont {Scheffler}}, \bibinfo {author} {\bibfnamefont {F.}~\bibnamefont {Liu}}, \bibinfo {author} {\bibfnamefont {J.}~\bibnamefont {Lischner}}, \bibinfo {author} {\bibfnamefont {L.}~\bibnamefont {Lin}}, \bibinfo {author} {\bibfnamefont {J.~R.}\ \bibnamefont {Deslippe}}, \bibinfo {author} {\bibfnamefont {S.~G.}\ \bibnamefont {Louie}}, \bibinfo {author} {\bibfnamefont {C.}~\bibnamefont {Yang}}, \bibinfo {author} {\bibfnamefont {F.}~\bibnamefont {Weigend}}, \bibinfo {author} {\bibfnamefont {J.~B.}\ \bibnamefont {Neaton}}, \bibinfo {author} {\bibfnamefont {F.}~\bibnamefont {Evers}},\ and\ \bibinfo {author} {\bibfnamefont {P.}~\bibnamefont {Rinke}},\ }\bibfield  {title} {\enquote {\bibinfo {title} {{{{\emph{GW}}}} 100: {{Benchmarking}} {{{\emph{G}}}} {\textsubscript{0}} {{{\emph{W}}}} {\textsubscript{0}} for {{Molecular Systems}}},}\ }\href {https://doi.org/10.1021/acs.jctc.5b00453} {\bibfield  {journal} {\bibinfo  {journal} {J. Chem. Theory Comput.}\ }\textbf {\bibinfo {volume} {11}},\ \bibinfo {pages} {5665--5687} (\bibinfo {year} {2015}{\natexlab{b}})}\BibitemShut {NoStop}%
\bibitem [{\citenamefont {Sekino}\ and\ \citenamefont {Bartlett}(1993)}]{sekinoMolecularHyperpolarizabilities1993}%
  \BibitemOpen
  \bibfield  {author} {\bibinfo {author} {\bibfnamefont {H.}~\bibnamefont {Sekino}}\ and\ \bibinfo {author} {\bibfnamefont {R.~J.}\ \bibnamefont {Bartlett}},\ }\bibfield  {title} {\enquote {\bibinfo {title} {Molecular hyperpolarizabilities},}\ }\href {https://doi.org/10.1063/1.464129} {\bibfield  {journal} {\bibinfo  {journal} {J. Chem. Phys.}\ }\textbf {\bibinfo {volume} {98}},\ \bibinfo {pages} {3022--3037} (\bibinfo {year} {1993})}\BibitemShut {NoStop}%
\bibitem [{\citenamefont {Scuseria}\ \emph {et~al.}(1991)\citenamefont {Scuseria}, \citenamefont {Miller}, \citenamefont {Jensen},\ and\ \citenamefont {Geertsen}}]{dipoleCO}%
  \BibitemOpen
  \bibfield  {author} {\bibinfo {author} {\bibfnamefont {G.~E.}\ \bibnamefont {Scuseria}}, \bibinfo {author} {\bibfnamefont {M.~D.}\ \bibnamefont {Miller}}, \bibinfo {author} {\bibfnamefont {F.}~\bibnamefont {Jensen}},\ and\ \bibinfo {author} {\bibfnamefont {J.}~\bibnamefont {Geertsen}},\ }\bibfield  {title} {\enquote {\bibinfo {title} {The dipole moment of carbon monoxide},}\ }\href {https://doi.org/10.1063/1.460293} {\bibfield  {journal} {\bibinfo  {journal} {J. Chem. Phys.}\ }\textbf {\bibinfo {volume} {94}},\ \bibinfo {pages} {6660--6663} (\bibinfo {year} {1991})}\BibitemShut {NoStop}%
\bibitem [{\citenamefont {Adamo}\ \emph {et~al.}(1999)\citenamefont {Adamo}, \citenamefont {Cossi}, \citenamefont {Scalmani},\ and\ \citenamefont {Barone}}]{adamoAccurateStaticPolarizabilities1999}%
  \BibitemOpen
  \bibfield  {author} {\bibinfo {author} {\bibfnamefont {C.}~\bibnamefont {Adamo}}, \bibinfo {author} {\bibfnamefont {M.}~\bibnamefont {Cossi}}, \bibinfo {author} {\bibfnamefont {G.}~\bibnamefont {Scalmani}},\ and\ \bibinfo {author} {\bibfnamefont {V.}~\bibnamefont {Barone}},\ }\bibfield  {title} {\enquote {\bibinfo {title} {Accurate static polarizabilities by density functional theory: Assessment of the {{PBE0}} model},}\ }\href {https://doi.org/10.1016/S0009-2614(99)00515-1} {\bibfield  {journal} {\bibinfo  {journal} {Chem. Phys. Lett.}\ }\textbf {\bibinfo {volume} {307}},\ \bibinfo {pages} {265--271} (\bibinfo {year} {1999})}\BibitemShut {NoStop}%
\bibitem [{\citenamefont {Hickey}\ and\ \citenamefont {Rowley}(2014)}]{dipole_benchmark}%
  \BibitemOpen
  \bibfield  {author} {\bibinfo {author} {\bibfnamefont {A.~L.}\ \bibnamefont {Hickey}}\ and\ \bibinfo {author} {\bibfnamefont {C.~N.}\ \bibnamefont {Rowley}},\ }\bibfield  {title} {\enquote {\bibinfo {title} {Benchmarking {{Quantum Chemical Methods}} for the {{Calculation}} of {{Molecular Dipole Moments}} and {{Polarizabilities}}},}\ }\href {https://doi.org/10.1021/jp502475e} {\bibfield  {journal} {\bibinfo  {journal} {J. Phys. Chem. A}\ }\textbf {\bibinfo {volume} {118}},\ \bibinfo {pages} {3678--3687} (\bibinfo {year} {2014})}\BibitemShut {NoStop}%
\bibitem [{\citenamefont {Neese}(2022{\natexlab{a}})}]{orca_1}%
  \BibitemOpen
  \bibfield  {author} {\bibinfo {author} {\bibfnamefont {F.}~\bibnamefont {Neese}},\ }\bibfield  {title} {\enquote {\bibinfo {title} {Software update: the orca program system, version 5.0},}\ }\href {https://doi.org/10.1002/wcms.1606} {\bibfield  {journal} {\bibinfo  {journal} {WIRES Comput. Molec. Sci.}\ }\textbf {\bibinfo {volume} {12}},\ \bibinfo {pages} {e1606} (\bibinfo {year} {2022}{\natexlab{a}})}\BibitemShut {NoStop}%
\bibitem [{\citenamefont {Neese}(2022{\natexlab{b}})}]{orca_2}%
  \BibitemOpen
  \bibfield  {author} {\bibinfo {author} {\bibfnamefont {F.}~\bibnamefont {Neese}},\ }\bibfield  {title} {\enquote {\bibinfo {title} {The shark integral generation and digestion system},}\ }\href {https://doi.org/10.1002/jcc.26942} {\bibfield  {journal} {\bibinfo  {journal} {J. Comp. Chem.}\ ,\ \bibinfo {pages} {1--16}} (\bibinfo {year} {2022}{\natexlab{b}})}\BibitemShut {NoStop}%
\bibitem [{\citenamefont {Neese}(2012)}]{orca_3}%
  \BibitemOpen
  \bibfield  {author} {\bibinfo {author} {\bibfnamefont {F.}~\bibnamefont {Neese}},\ }\bibfield  {title} {\enquote {\bibinfo {title} {The orca program system},}\ }\href {https://doi.org/10.1002/wcms.81} {\bibfield  {journal} {\bibinfo  {journal} {WIRES Comput. Molec. Sci.}\ }\textbf {\bibinfo {volume} {2}},\ \bibinfo {pages} {73--78} (\bibinfo {year} {2012})}\BibitemShut {NoStop}%
\bibitem [{\citenamefont {Neese}(2018)}]{orca_4}%
  \BibitemOpen
  \bibfield  {author} {\bibinfo {author} {\bibfnamefont {F.}~\bibnamefont {Neese}},\ }\bibfield  {title} {\enquote {\bibinfo {title} {Software update: the orca program system, version 4.0},}\ }\href {https://doi.org/10.1002/wcms.1327} {\bibfield  {journal} {\bibinfo  {journal} {WIRES Comput. Molec. Sci.}\ }\textbf {\bibinfo {volume} {8}},\ \bibinfo {pages} {1--6} (\bibinfo {year} {2018})}\BibitemShut {NoStop}%
\bibitem [{\citenamefont {Neese}\ \emph {et~al.}(2020)\citenamefont {Neese}, \citenamefont {Wennmohs}, \citenamefont {Becker},\ and\ \citenamefont {Riplinger}}]{orca_5}%
  \BibitemOpen
  \bibfield  {author} {\bibinfo {author} {\bibfnamefont {F.}~\bibnamefont {Neese}}, \bibinfo {author} {\bibfnamefont {F.}~\bibnamefont {Wennmohs}}, \bibinfo {author} {\bibfnamefont {U.}~\bibnamefont {Becker}},\ and\ \bibinfo {author} {\bibfnamefont {C.}~\bibnamefont {Riplinger}},\ }\bibfield  {title} {\enquote {\bibinfo {title} {The orca quantum chemistry program package},}\ }\href {https://doi.org/10.1063/5.0004608} {\bibfield  {journal} {\bibinfo  {journal} {J. Chem. Phys.}\ }\textbf {\bibinfo {volume} {152}},\ \bibinfo {pages} {Art. No. L224108} (\bibinfo {year} {2020})}\BibitemShut {NoStop}%
\bibitem [{\citenamefont {Pracht}, \citenamefont {Bohle},\ and\ \citenamefont {Grimme}(2020)}]{crest_1}%
  \BibitemOpen
  \bibfield  {author} {\bibinfo {author} {\bibfnamefont {P.}~\bibnamefont {Pracht}}, \bibinfo {author} {\bibfnamefont {F.}~\bibnamefont {Bohle}},\ and\ \bibinfo {author} {\bibfnamefont {S.}~\bibnamefont {Grimme}},\ }\bibfield  {title} {\enquote {\bibinfo {title} {Automated exploration of the low-energy chemical space with fast quantum chemical methods},}\ }\href {https://doi.org/10.1039/C9CP06869D} {\bibfield  {journal} {\bibinfo  {journal} {Phys. Chem. Chem. Phys.}\ }\textbf {\bibinfo {volume} {22}},\ \bibinfo {pages} {7169--7192} (\bibinfo {year} {2020})}\BibitemShut {NoStop}%
\bibitem [{\citenamefont {Grimme}(2019)}]{crest_2}%
  \BibitemOpen
  \bibfield  {author} {\bibinfo {author} {\bibfnamefont {S.}~\bibnamefont {Grimme}},\ }\bibfield  {title} {\enquote {\bibinfo {title} {Exploration of {{Chemical Compound}}, {{Conformer}}, and {{Reaction Space}} with {{Meta-Dynamics Simulations Based}} on {{Tight-Binding Quantum Chemical Calculations}}},}\ }\href {https://doi.org/10.1021/acs.jctc.9b00143} {\bibfield  {journal} {\bibinfo  {journal} {J. Chem. Theory Comput.}\ }\textbf {\bibinfo {volume} {15}},\ \bibinfo {pages} {2847--2862} (\bibinfo {year} {2019})}\BibitemShut {NoStop}%
\bibitem [{\citenamefont {Pracht}\ \emph {et~al.}(2024)\citenamefont {Pracht}, \citenamefont {Grimme}, \citenamefont {Bannwarth}, \citenamefont {Bohle}, \citenamefont {Ehlert}, \citenamefont {Feldmann}, \citenamefont {Gorges}, \citenamefont {M{\"u}ller}, \citenamefont {Neudecker}, \citenamefont {Plett}, \citenamefont {Spicher}, \citenamefont {Steinbach}, \citenamefont {Weso{\l}owski},\ and\ \citenamefont {Zeller}}]{crest_3}%
  \BibitemOpen
  \bibfield  {author} {\bibinfo {author} {\bibfnamefont {P.}~\bibnamefont {Pracht}}, \bibinfo {author} {\bibfnamefont {S.}~\bibnamefont {Grimme}}, \bibinfo {author} {\bibfnamefont {C.}~\bibnamefont {Bannwarth}}, \bibinfo {author} {\bibfnamefont {F.}~\bibnamefont {Bohle}}, \bibinfo {author} {\bibfnamefont {S.}~\bibnamefont {Ehlert}}, \bibinfo {author} {\bibfnamefont {G.}~\bibnamefont {Feldmann}}, \bibinfo {author} {\bibfnamefont {J.}~\bibnamefont {Gorges}}, \bibinfo {author} {\bibfnamefont {M.}~\bibnamefont {M{\"u}ller}}, \bibinfo {author} {\bibfnamefont {T.}~\bibnamefont {Neudecker}}, \bibinfo {author} {\bibfnamefont {C.}~\bibnamefont {Plett}}, \bibinfo {author} {\bibfnamefont {S.}~\bibnamefont {Spicher}}, \bibinfo {author} {\bibfnamefont {P.}~\bibnamefont {Steinbach}}, \bibinfo {author} {\bibfnamefont {P.~A.}\ \bibnamefont {Weso{\l}owski}},\ and\ \bibinfo {author} {\bibfnamefont {F.}~\bibnamefont {Zeller}},\ }\bibfield  {title} {\enquote {\bibinfo {title} {{{CREST}}---{{A}} program for the exploration of low-energy molecular chemical space},}\ }\href {https://doi.org/10.1063/5.0197592} {\bibfield  {journal} {\bibinfo  {journal} {J. Chem. Phys.}\ }\textbf {\bibinfo {volume} {160}},\ \bibinfo {pages} {114110} (\bibinfo {year} {2024})}\BibitemShut {NoStop}%
\bibitem [{\citenamefont {Pulay}\ and\ \citenamefont {Saebo}(1986)}]{OOMP2_1}%
  \BibitemOpen
  \bibfield  {author} {\bibinfo {author} {\bibfnamefont {P.}~\bibnamefont {Pulay}}\ and\ \bibinfo {author} {\bibfnamefont {S.}~\bibnamefont {Saebo}},\ }\bibfield  {title} {\enquote {\bibinfo {title} {Orbital-invariant formulation and second-order gradient evaluation in {{M{\o}ller}}--{{Plesset}} perturbation theory},}\ }\href {https://doi.org/10.1007/BF00526697} {\bibfield  {journal} {\bibinfo  {journal} {Theor. Chem. Acc.}\ }\textbf {\bibinfo {volume} {69}},\ \bibinfo {pages} {357--368} (\bibinfo {year} {1986})}\BibitemShut {NoStop}%
\bibitem [{\citenamefont {Neese}\ \emph {et~al.}(2009)\citenamefont {Neese}, \citenamefont {Schwabe}, \citenamefont {Kossmann}, \citenamefont {Schirmer},\ and\ \citenamefont {Grimme}}]{OO-MP2_1}%
  \BibitemOpen
  \bibfield  {author} {\bibinfo {author} {\bibfnamefont {F.}~\bibnamefont {Neese}}, \bibinfo {author} {\bibfnamefont {T.}~\bibnamefont {Schwabe}}, \bibinfo {author} {\bibfnamefont {S.}~\bibnamefont {Kossmann}}, \bibinfo {author} {\bibfnamefont {B.}~\bibnamefont {Schirmer}},\ and\ \bibinfo {author} {\bibfnamefont {S.}~\bibnamefont {Grimme}},\ }\bibfield  {title} {\enquote {\bibinfo {title} {Assessment of {{Orbital-Optimized}}, {{Spin-Component Scaled Second-Order Many-Body Perturbation Theory}} for {{Thermochemistry}} and {{Kinetics}}},}\ }\href {https://doi.org/10.1021/ct9003299} {\bibfield  {journal} {\bibinfo  {journal} {J. Chem. Theory Comput.}\ }\textbf {\bibinfo {volume} {5}},\ \bibinfo {pages} {3060--3073} (\bibinfo {year} {2009})}\BibitemShut {NoStop}%
\bibitem [{\citenamefont {Shee}\ \emph {et~al.}(2021{\natexlab{b}})\citenamefont {Shee}, \citenamefont {Loipersberger}, \citenamefont {Rettig}, \citenamefont {Lee},\ and\ \citenamefont {{Head-Gordon}}}]{kappaSigmaOOMP2}%
  \BibitemOpen
  \bibfield  {author} {\bibinfo {author} {\bibfnamefont {J.}~\bibnamefont {Shee}}, \bibinfo {author} {\bibfnamefont {M.}~\bibnamefont {Loipersberger}}, \bibinfo {author} {\bibfnamefont {A.}~\bibnamefont {Rettig}}, \bibinfo {author} {\bibfnamefont {J.}~\bibnamefont {Lee}},\ and\ \bibinfo {author} {\bibfnamefont {M.}~\bibnamefont {{Head-Gordon}}},\ }\bibfield  {title} {\enquote {\bibinfo {title} {Regularized {{Second-Order M{\o}ller}}--{{Plesset Theory}}: {{A More Accurate Alternative}} to {{Conventional MP2}} for {{Noncovalent Interactions}} and {{Transition Metal Thermochemistry}} for the {{Same Computational Cost}}},}\ }\href {https://doi.org/10.1021/acs.jpclett.1c03468} {\bibfield  {journal} {\bibinfo  {journal} {J. Phys. Chem. Lett.}\ }\textbf {\bibinfo {volume} {12}},\ \bibinfo {pages} {12084--12097} (\bibinfo {year} {2021}{\natexlab{b}})}\BibitemShut {NoStop}%
\bibitem [{\citenamefont {Stewart}(1970)}]{sto-Ng}%
  \BibitemOpen
  \bibfield  {author} {\bibinfo {author} {\bibfnamefont {R.~F.}\ \bibnamefont {Stewart}},\ }\bibfield  {title} {\enquote {\bibinfo {title} {Small {{Gaussian Expansions}} of {{Slater-Type Orbitals}}},}\ }\href {https://doi.org/10.1063/1.1672702} {\bibfield  {journal} {\bibinfo  {journal} {J. Chem. Phys.}\ }\textbf {\bibinfo {volume} {52}},\ \bibinfo {pages} {431--438} (\bibinfo {year} {1970})}\BibitemShut {NoStop}%
\bibitem [{\citenamefont {Hehre}, \citenamefont {Stewart},\ and\ \citenamefont {Pople}(1969)}]{sto-ng_1}%
  \BibitemOpen
  \bibfield  {author} {\bibinfo {author} {\bibfnamefont {W.~J.}\ \bibnamefont {Hehre}}, \bibinfo {author} {\bibfnamefont {R.~F.}\ \bibnamefont {Stewart}},\ and\ \bibinfo {author} {\bibfnamefont {J.~A.}\ \bibnamefont {Pople}},\ }\bibfield  {title} {\enquote {\bibinfo {title} {Self-consistent molecular-orbital methods. i. use of gaussian expansions of slater-type atomic orbitals},}\ }\href {https://doi.org/10.1063/1.1672392} {\bibfield  {journal} {\bibinfo  {journal} {J. Chem. Phys.}\ }\textbf {\bibinfo {volume} {51}},\ \bibinfo {pages} {2657--2664} (\bibinfo {year} {1969})}\BibitemShut {NoStop}%
\bibitem [{\citenamefont {Hehre}\ \emph {et~al.}(1970)\citenamefont {Hehre}, \citenamefont {Ditchfield}, \citenamefont {Stewart},\ and\ \citenamefont {Pople}}]{sto-ng_2}%
  \BibitemOpen
  \bibfield  {author} {\bibinfo {author} {\bibfnamefont {W.~J.}\ \bibnamefont {Hehre}}, \bibinfo {author} {\bibfnamefont {R.}~\bibnamefont {Ditchfield}}, \bibinfo {author} {\bibfnamefont {R.~F.}\ \bibnamefont {Stewart}},\ and\ \bibinfo {author} {\bibfnamefont {J.~A.}\ \bibnamefont {Pople}},\ }\bibfield  {title} {\enquote {\bibinfo {title} {Self-consistent molecular orbital methods. iv. use of gaussian expansions of slater-type orbitals. extension to second-row molecules},}\ }\href {https://doi.org/10.1063/1.1673374} {\bibfield  {journal} {\bibinfo  {journal} {J. Chem. Phys.}\ }\textbf {\bibinfo {volume} {52}},\ \bibinfo {pages} {2769--2773} (\bibinfo {year} {1970})}\BibitemShut {NoStop}%
\bibitem [{\citenamefont {Epifanovsky}\ \emph {et~al.}(2021)\citenamefont {Epifanovsky}, \citenamefont {Gilbert}, \citenamefont {Feng}, \citenamefont {Lee}, \citenamefont {Mao}, \citenamefont {Mardirossian}, \citenamefont {Pokhilko}, \citenamefont {White}, \citenamefont {Coons}, \citenamefont {Dempwolff}, \citenamefont {Gan}, \citenamefont {Hait}, \citenamefont {Horn}, \citenamefont {Jacobson}, \citenamefont {Kaliman}, \citenamefont {Kussmann}, \citenamefont {Lange}, \citenamefont {Lao}, \citenamefont {Levine}, \citenamefont {Liu}, \citenamefont {McKenzie}, \citenamefont {Morrison}, \citenamefont {Nanda}, \citenamefont {Plasser}, \citenamefont {Rehn}, \citenamefont {Vidal}, \citenamefont {You}, \citenamefont {Zhu}, \citenamefont {Alam}, \citenamefont {Albrecht}, \citenamefont {Aldossary}, \citenamefont {Alguire}, \citenamefont {Andersen}, \citenamefont {Athavale}, \citenamefont {Barton}, \citenamefont {Begam}, \citenamefont {Behn}, \citenamefont {Bellonzi}, \citenamefont {Bernard}, \citenamefont {Berquist}, \citenamefont {Burton}, \citenamefont {Carreras}, \citenamefont {{Carter-Fenk}}, \citenamefont {Chakraborty}, \citenamefont {Chien}, \citenamefont {Closser}, \citenamefont {{Cofer-Shabica}}, \citenamefont {Dasgupta}, \citenamefont {De~Wergifosse}, \citenamefont {Deng}, \citenamefont {Diedenhofen}, \citenamefont {Do}, \citenamefont {Ehlert}, \citenamefont {Fang}, \citenamefont {Fatehi}, \citenamefont {Feng}, \citenamefont {Friedhoff}, \citenamefont {Gayvert}, \citenamefont {Ge}, \citenamefont {Gidofalvi}, \citenamefont {Goldey}, \citenamefont {Gomes}, \citenamefont {{Gonz{\'a}lez-Espinoza}}, \citenamefont {Gulania}, \citenamefont {Gunina}, \citenamefont {{Hanson-Heine}}, \citenamefont {Harbach}, \citenamefont {Hauser}, \citenamefont {Herbst}, \citenamefont {Hern{\'a}ndez~Vera}, \citenamefont {Hodecker}, \citenamefont {Holden}, \citenamefont {Houck}, \citenamefont {Huang}, \citenamefont {Hui}, \citenamefont {Huynh}, \citenamefont {Ivanov}, \citenamefont {J{\'a}sz}, \citenamefont {Ji}, \citenamefont {Jiang}, \citenamefont {Kaduk}, \citenamefont {K{\"a}hler}, \citenamefont {Khistyaev}, \citenamefont {Kim}, \citenamefont {Kis}, \citenamefont {Klunzinger}, \citenamefont {{Koczor-Benda}}, \citenamefont {Koh}, \citenamefont {Kosenkov}, \citenamefont {Koulias}, \citenamefont {Kowalczyk}, \citenamefont {Krauter}, \citenamefont {Kue}, \citenamefont {Kunitsa}, \citenamefont {Kus}, \citenamefont {Ladj{\'a}nszki}, \citenamefont {Landau}, \citenamefont {Lawler}, \citenamefont {Lefrancois}, \citenamefont {Lehtola}, \citenamefont {Li}, \citenamefont {Li}, \citenamefont {Liang}, \citenamefont {Liebenthal}, \citenamefont {Lin}, \citenamefont {Lin}, \citenamefont {Liu}, \citenamefont {Liu}, \citenamefont {Loipersberger}, \citenamefont {Luenser}, \citenamefont {Manjanath}, \citenamefont {Manohar}, \citenamefont {Mansoor}, \citenamefont {Manzer}, \citenamefont {Mao}, \citenamefont {Marenich}, \citenamefont {Markovich}, \citenamefont {Mason}, \citenamefont {Maurer}, \citenamefont {McLaughlin}, \citenamefont {Menger}, \citenamefont {Mewes}, \citenamefont {Mewes}, \citenamefont {Morgante}, \citenamefont {Mullinax}, \citenamefont {Oosterbaan}, \citenamefont {Paran}, \citenamefont {Paul}, \citenamefont {Paul}, \citenamefont {Pavo{\v s}evi{\'c}}, \citenamefont {Pei}, \citenamefont {Prager}, \citenamefont {Proynov}, \citenamefont {R{\'a}k}, \citenamefont {{Ramos-Cordoba}}, \citenamefont {Rana}, \citenamefont {Rask}, \citenamefont {Rettig}, \citenamefont {Richard}, \citenamefont {Rob}, \citenamefont {Rossomme}, \citenamefont {Scheele}, \citenamefont {Scheurer}, \citenamefont {Schneider}, \citenamefont {Sergueev}, \citenamefont {Sharada}, \citenamefont {Skomorowski}, \citenamefont {Small}, \citenamefont {Stein}, \citenamefont {Su}, \citenamefont {Sundstrom}, \citenamefont {Tao}, \citenamefont {Thirman}, \citenamefont {Tornai}, \citenamefont {Tsuchimochi}, \citenamefont {Tubman}, \citenamefont {Veccham}, \citenamefont {Vydrov}, \citenamefont {Wenzel}, \citenamefont {Witte}, \citenamefont {Yamada}, \citenamefont {Yao}, \citenamefont {Yeganeh}, \citenamefont {Yost}, \citenamefont {Zech}, \citenamefont {Zhang}, \citenamefont {Zhang}, \citenamefont {Zhang}, \citenamefont {Zuev}, \citenamefont {{Aspuru-Guzik}}, \citenamefont {Bell}, \citenamefont {Besley}, \citenamefont {Bravaya}, \citenamefont {Brooks}, \citenamefont {Casanova}, \citenamefont {Chai}, \citenamefont {Coriani}, \citenamefont {Cramer}, \citenamefont {Cserey}, \citenamefont {DePrince}, \citenamefont {DiStasio}, \citenamefont {Dreuw}, \citenamefont {Dunietz}, \citenamefont {Furlani}, \citenamefont {Goddard}, \citenamefont {{Hammes-Schiffer}}, \citenamefont {{Head-Gordon}}, \citenamefont {Hehre}, \citenamefont {Hsu}, \citenamefont {Jagau}, \citenamefont {Jung}, \citenamefont {Klamt}, \citenamefont {Kong}, \citenamefont {Lambrecht}, \citenamefont {Liang}, \citenamefont {Mayhall}, \citenamefont {McCurdy}, \citenamefont {Neaton}, \citenamefont {Ochsenfeld}, \citenamefont {Parkhill}, \citenamefont {Peverati}, \citenamefont {Rassolov}, \citenamefont {Shao}, \citenamefont {Slipchenko}, \citenamefont {Stauch}, \citenamefont {Steele}, \citenamefont {Subotnik}, \citenamefont {Thom}, \citenamefont {Tkatchenko}, \citenamefont {Truhlar}, \citenamefont {Van~Voorhis}, \citenamefont {Wesolowski}, \citenamefont {Whaley}, \citenamefont {Woodcock}, \citenamefont {Zimmerman}, \citenamefont {Faraji}, \citenamefont {Gill}, \citenamefont {{Head-Gordon}}, \citenamefont {Herbert},\ and\ \citenamefont {Krylov}}]{qchem}%
  \BibitemOpen
  \bibfield  {author} {\bibinfo {author} {\bibfnamefont {E.}~\bibnamefont {Epifanovsky}}, \bibinfo {author} {\bibfnamefont {A.~T.~B.}\ \bibnamefont {Gilbert}}, \bibinfo {author} {\bibfnamefont {X.}~\bibnamefont {Feng}}, \bibinfo {author} {\bibfnamefont {J.}~\bibnamefont {Lee}}, \bibinfo {author} {\bibfnamefont {Y.}~\bibnamefont {Mao}}, \bibinfo {author} {\bibfnamefont {N.}~\bibnamefont {Mardirossian}}, \bibinfo {author} {\bibfnamefont {P.}~\bibnamefont {Pokhilko}}, \bibinfo {author} {\bibfnamefont {A.~F.}\ \bibnamefont {White}}, \bibinfo {author} {\bibfnamefont {M.~P.}\ \bibnamefont {Coons}}, \bibinfo {author} {\bibfnamefont {A.~L.}\ \bibnamefont {Dempwolff}}, \bibinfo {author} {\bibfnamefont {Z.}~\bibnamefont {Gan}}, \bibinfo {author} {\bibfnamefont {D.}~\bibnamefont {Hait}}, \bibinfo {author} {\bibfnamefont {P.~R.}\ \bibnamefont {Horn}}, \bibinfo {author} {\bibfnamefont {L.~D.}\ \bibnamefont {Jacobson}}, \bibinfo {author} {\bibfnamefont {I.}~\bibnamefont {Kaliman}}, \bibinfo {author} {\bibfnamefont {J.}~\bibnamefont {Kussmann}}, \bibinfo {author} {\bibfnamefont {A.~W.}\ \bibnamefont {Lange}}, \bibinfo {author} {\bibfnamefont {K.~U.}\ \bibnamefont {Lao}}, \bibinfo {author} {\bibfnamefont {D.~S.}\ \bibnamefont {Levine}}, \bibinfo {author} {\bibfnamefont {J.}~\bibnamefont {Liu}}, \bibinfo {author} {\bibfnamefont {S.~C.}\ \bibnamefont {McKenzie}}, \bibinfo {author} {\bibfnamefont {A.~F.}\ \bibnamefont {Morrison}}, \bibinfo {author} {\bibfnamefont {K.~D.}\ \bibnamefont {Nanda}}, \bibinfo {author} {\bibfnamefont {F.}~\bibnamefont {Plasser}}, \bibinfo {author} {\bibfnamefont {D.~R.}\ \bibnamefont {Rehn}}, \bibinfo {author} {\bibfnamefont {M.~L.}\ \bibnamefont {Vidal}}, \bibinfo {author} {\bibfnamefont {Z.-Q.}\ \bibnamefont {You}}, \bibinfo {author} {\bibfnamefont {Y.}~\bibnamefont {Zhu}}, \bibinfo {author} {\bibfnamefont {B.}~\bibnamefont {Alam}}, \bibinfo {author} {\bibfnamefont {B.~J.}\ \bibnamefont {Albrecht}}, \bibinfo {author} {\bibfnamefont {A.}~\bibnamefont {Aldossary}}, \bibinfo {author} {\bibfnamefont {E.}~\bibnamefont {Alguire}}, \bibinfo {author} {\bibfnamefont {J.~H.}\ \bibnamefont {Andersen}}, \bibinfo {author} {\bibfnamefont {V.}~\bibnamefont {Athavale}}, \bibinfo {author} {\bibfnamefont {D.}~\bibnamefont {Barton}}, \bibinfo {author} {\bibfnamefont {K.}~\bibnamefont {Begam}}, \bibinfo {author} {\bibfnamefont {A.}~\bibnamefont {Behn}}, \bibinfo {author} {\bibfnamefont {N.}~\bibnamefont {Bellonzi}}, \bibinfo {author} {\bibfnamefont {Y.~A.}\ \bibnamefont {Bernard}}, \bibinfo {author} {\bibfnamefont {E.~J.}\ \bibnamefont {Berquist}}, \bibinfo {author} {\bibfnamefont {H.~G.~A.}\ \bibnamefont {Burton}}, \bibinfo {author} {\bibfnamefont {A.}~\bibnamefont {Carreras}}, \bibinfo {author} {\bibfnamefont {K.}~\bibnamefont {{Carter-Fenk}}}, \bibinfo {author} {\bibfnamefont {R.}~\bibnamefont {Chakraborty}}, \bibinfo {author} {\bibfnamefont {A.~D.}\ \bibnamefont {Chien}}, \bibinfo {author} {\bibfnamefont {K.~D.}\ \bibnamefont {Closser}}, \bibinfo {author} {\bibfnamefont {V.}~\bibnamefont {{Cofer-Shabica}}}, \bibinfo {author} {\bibfnamefont {S.}~\bibnamefont {Dasgupta}}, \bibinfo {author} {\bibfnamefont {M.}~\bibnamefont {De~Wergifosse}}, \bibinfo {author} {\bibfnamefont {J.}~\bibnamefont {Deng}}, \bibinfo {author} {\bibfnamefont {M.}~\bibnamefont {Diedenhofen}}, \bibinfo {author} {\bibfnamefont {H.}~\bibnamefont {Do}}, \bibinfo {author} {\bibfnamefont {S.}~\bibnamefont {Ehlert}}, \bibinfo {author} {\bibfnamefont {P.-T.}\ \bibnamefont {Fang}}, \bibinfo {author} {\bibfnamefont {S.}~\bibnamefont {Fatehi}}, \bibinfo {author} {\bibfnamefont {Q.}~\bibnamefont {Feng}}, \bibinfo {author} {\bibfnamefont {T.}~\bibnamefont {Friedhoff}}, \bibinfo {author} {\bibfnamefont {J.}~\bibnamefont {Gayvert}}, \bibinfo {author} {\bibfnamefont {Q.}~\bibnamefont {Ge}}, \bibinfo {author} {\bibfnamefont {G.}~\bibnamefont {Gidofalvi}}, \bibinfo {author} {\bibfnamefont {M.}~\bibnamefont {Goldey}}, \bibinfo {author} {\bibfnamefont {J.}~\bibnamefont {Gomes}}, \bibinfo {author} {\bibfnamefont {C.~E.}\ \bibnamefont {{Gonz{\'a}lez-Espinoza}}}, \bibinfo {author} {\bibfnamefont {S.}~\bibnamefont {Gulania}}, \bibinfo {author} {\bibfnamefont {A.~O.}\ \bibnamefont {Gunina}}, \bibinfo {author} {\bibfnamefont {M.~W.~D.}\ \bibnamefont {{Hanson-Heine}}}, \bibinfo {author} {\bibfnamefont {P.~H.~P.}\ \bibnamefont {Harbach}}, \bibinfo {author} {\bibfnamefont {A.}~\bibnamefont {Hauser}}, \bibinfo {author} {\bibfnamefont {M.~F.}\ \bibnamefont {Herbst}}, \bibinfo {author} {\bibfnamefont {M.}~\bibnamefont {Hern{\'a}ndez~Vera}}, \bibinfo {author} {\bibfnamefont {M.}~\bibnamefont {Hodecker}}, \bibinfo {author} {\bibfnamefont {Z.~C.}\ \bibnamefont {Holden}}, \bibinfo {author} {\bibfnamefont {S.}~\bibnamefont {Houck}}, \bibinfo {author} {\bibfnamefont {X.}~\bibnamefont {Huang}}, \bibinfo {author} {\bibfnamefont {K.}~\bibnamefont {Hui}}, \bibinfo {author} {\bibfnamefont {B.~C.}\ \bibnamefont {Huynh}}, \bibinfo {author} {\bibfnamefont {M.}~\bibnamefont {Ivanov}}, \bibinfo {author} {\bibfnamefont {{\'A}.}~\bibnamefont {J{\'a}sz}}, \bibinfo {author} {\bibfnamefont {H.}~\bibnamefont {Ji}}, \bibinfo {author} {\bibfnamefont {H.}~\bibnamefont {Jiang}}, \bibinfo {author} {\bibfnamefont {B.}~\bibnamefont {Kaduk}}, \bibinfo {author} {\bibfnamefont {S.}~\bibnamefont {K{\"a}hler}}, \bibinfo {author} {\bibfnamefont {K.}~\bibnamefont {Khistyaev}}, \bibinfo {author} {\bibfnamefont {J.}~\bibnamefont {Kim}}, \bibinfo {author} {\bibfnamefont {G.}~\bibnamefont {Kis}}, \bibinfo {author} {\bibfnamefont {P.}~\bibnamefont {Klunzinger}}, \bibinfo {author} {\bibfnamefont {Z.}~\bibnamefont {{Koczor-Benda}}}, \bibinfo {author} {\bibfnamefont {J.~H.}\ \bibnamefont {Koh}}, \bibinfo {author} {\bibfnamefont {D.}~\bibnamefont {Kosenkov}}, \bibinfo {author} {\bibfnamefont {L.}~\bibnamefont {Koulias}}, \bibinfo {author} {\bibfnamefont {T.}~\bibnamefont {Kowalczyk}}, \bibinfo {author} {\bibfnamefont {C.~M.}\ \bibnamefont {Krauter}}, \bibinfo {author} {\bibfnamefont {K.}~\bibnamefont {Kue}}, \bibinfo {author} {\bibfnamefont {A.}~\bibnamefont {Kunitsa}}, \bibinfo {author} {\bibfnamefont {T.}~\bibnamefont {Kus}}, \bibinfo {author} {\bibfnamefont {I.}~\bibnamefont {Ladj{\'a}nszki}}, \bibinfo {author} {\bibfnamefont {A.}~\bibnamefont {Landau}}, \bibinfo {author} {\bibfnamefont {K.~V.}\ \bibnamefont {Lawler}}, \bibinfo {author} {\bibfnamefont {D.}~\bibnamefont {Lefrancois}}, \bibinfo {author} {\bibfnamefont {S.}~\bibnamefont {Lehtola}}, \bibinfo {author} {\bibfnamefont {R.~R.}\ \bibnamefont {Li}}, \bibinfo {author} {\bibfnamefont {Y.-P.}\ \bibnamefont {Li}}, \bibinfo {author} {\bibfnamefont {J.}~\bibnamefont {Liang}}, \bibinfo {author} {\bibfnamefont {M.}~\bibnamefont {Liebenthal}}, \bibinfo {author} {\bibfnamefont {H.-H.}\ \bibnamefont {Lin}}, \bibinfo {author} {\bibfnamefont {Y.-S.}\ \bibnamefont {Lin}}, \bibinfo {author} {\bibfnamefont {F.}~\bibnamefont {Liu}}, \bibinfo {author} {\bibfnamefont {K.-Y.}\ \bibnamefont {Liu}}, \bibinfo {author} {\bibfnamefont {M.}~\bibnamefont {Loipersberger}}, \bibinfo {author} {\bibfnamefont {A.}~\bibnamefont {Luenser}}, \bibinfo {author} {\bibfnamefont {A.}~\bibnamefont {Manjanath}}, \bibinfo {author} {\bibfnamefont {P.}~\bibnamefont {Manohar}}, \bibinfo {author} {\bibfnamefont {E.}~\bibnamefont {Mansoor}}, \bibinfo {author} {\bibfnamefont {S.~F.}\ \bibnamefont {Manzer}}, \bibinfo {author} {\bibfnamefont {S.-P.}\ \bibnamefont {Mao}}, \bibinfo {author} {\bibfnamefont {A.~V.}\ \bibnamefont {Marenich}}, \bibinfo {author} {\bibfnamefont {T.}~\bibnamefont {Markovich}}, \bibinfo {author} {\bibfnamefont {S.}~\bibnamefont {Mason}}, \bibinfo {author} {\bibfnamefont {S.~A.}\ \bibnamefont {Maurer}}, \bibinfo {author} {\bibfnamefont {P.~F.}\ \bibnamefont {McLaughlin}}, \bibinfo {author} {\bibfnamefont {M.~F. S.~J.}\ \bibnamefont {Menger}}, \bibinfo {author} {\bibfnamefont {J.-M.}\ \bibnamefont {Mewes}}, \bibinfo {author} {\bibfnamefont {S.~A.}\ \bibnamefont {Mewes}}, \bibinfo {author} {\bibfnamefont {P.}~\bibnamefont {Morgante}}, \bibinfo {author} {\bibfnamefont {J.~W.}\ \bibnamefont {Mullinax}}, \bibinfo {author} {\bibfnamefont {K.~J.}\ \bibnamefont {Oosterbaan}}, \bibinfo {author} {\bibfnamefont {G.}~\bibnamefont {Paran}}, \bibinfo {author} {\bibfnamefont {A.~C.}\ \bibnamefont {Paul}}, \bibinfo {author} {\bibfnamefont {S.~K.}\ \bibnamefont {Paul}}, \bibinfo {author} {\bibfnamefont {F.}~\bibnamefont {Pavo{\v s}evi{\'c}}}, \bibinfo {author} {\bibfnamefont {Z.}~\bibnamefont {Pei}}, \bibinfo {author} {\bibfnamefont {S.}~\bibnamefont {Prager}}, \bibinfo {author} {\bibfnamefont {E.~I.}\ \bibnamefont {Proynov}}, \bibinfo {author} {\bibfnamefont {{\'A}.}~\bibnamefont {R{\'a}k}}, \bibinfo {author} {\bibfnamefont {E.}~\bibnamefont {{Ramos-Cordoba}}}, \bibinfo {author} {\bibfnamefont {B.}~\bibnamefont {Rana}}, \bibinfo {author} {\bibfnamefont {A.~E.}\ \bibnamefont {Rask}}, \bibinfo {author} {\bibfnamefont {A.}~\bibnamefont {Rettig}}, \bibinfo {author} {\bibfnamefont {R.~M.}\ \bibnamefont {Richard}}, \bibinfo {author} {\bibfnamefont {F.}~\bibnamefont {Rob}}, \bibinfo {author} {\bibfnamefont {E.}~\bibnamefont {Rossomme}}, \bibinfo {author} {\bibfnamefont {T.}~\bibnamefont {Scheele}}, \bibinfo {author} {\bibfnamefont {M.}~\bibnamefont {Scheurer}}, \bibinfo {author} {\bibfnamefont {M.}~\bibnamefont {Schneider}}, \bibinfo {author} {\bibfnamefont {N.}~\bibnamefont {Sergueev}}, \bibinfo {author} {\bibfnamefont {S.~M.}\ \bibnamefont {Sharada}}, \bibinfo {author} {\bibfnamefont {W.}~\bibnamefont {Skomorowski}}, \bibinfo {author} {\bibfnamefont {D.~W.}\ \bibnamefont {Small}}, \bibinfo {author} {\bibfnamefont {C.~J.}\ \bibnamefont {Stein}}, \bibinfo {author} {\bibfnamefont {Y.-C.}\ \bibnamefont {Su}}, \bibinfo {author} {\bibfnamefont {E.~J.}\ \bibnamefont {Sundstrom}}, \bibinfo {author} {\bibfnamefont {Z.}~\bibnamefont {Tao}}, \bibinfo {author} {\bibfnamefont {J.}~\bibnamefont {Thirman}}, \bibinfo {author} {\bibfnamefont {G.~J.}\ \bibnamefont {Tornai}}, \bibinfo {author} {\bibfnamefont {T.}~\bibnamefont
  {Tsuchimochi}}, \bibinfo {author} {\bibfnamefont {N.~M.}\ \bibnamefont {Tubman}}, \bibinfo {author} {\bibfnamefont {S.~P.}\ \bibnamefont {Veccham}}, \bibinfo {author} {\bibfnamefont {O.}~\bibnamefont {Vydrov}}, \bibinfo {author} {\bibfnamefont {J.}~\bibnamefont {Wenzel}}, \bibinfo {author} {\bibfnamefont {J.}~\bibnamefont {Witte}}, \bibinfo {author} {\bibfnamefont {A.}~\bibnamefont {Yamada}}, \bibinfo {author} {\bibfnamefont {K.}~\bibnamefont {Yao}}, \bibinfo {author} {\bibfnamefont {S.}~\bibnamefont {Yeganeh}}, \bibinfo {author} {\bibfnamefont {S.~R.}\ \bibnamefont {Yost}}, \bibinfo {author} {\bibfnamefont {A.}~\bibnamefont {Zech}}, \bibinfo {author} {\bibfnamefont {I.~Y.}\ \bibnamefont {Zhang}}, \bibinfo {author} {\bibfnamefont {X.}~\bibnamefont {Zhang}}, \bibinfo {author} {\bibfnamefont {Y.}~\bibnamefont {Zhang}}, \bibinfo {author} {\bibfnamefont {D.}~\bibnamefont {Zuev}}, \bibinfo {author} {\bibfnamefont {A.}~\bibnamefont {{Aspuru-Guzik}}}, \bibinfo {author} {\bibfnamefont {A.~T.}\ \bibnamefont {Bell}}, \bibinfo {author} {\bibfnamefont {N.~A.}\ \bibnamefont {Besley}}, \bibinfo {author} {\bibfnamefont {K.~B.}\ \bibnamefont {Bravaya}}, \bibinfo {author} {\bibfnamefont {B.~R.}\ \bibnamefont {Brooks}}, \bibinfo {author} {\bibfnamefont {D.}~\bibnamefont {Casanova}}, \bibinfo {author} {\bibfnamefont {J.-D.}\ \bibnamefont {Chai}}, \bibinfo {author} {\bibfnamefont {S.}~\bibnamefont {Coriani}}, \bibinfo {author} {\bibfnamefont {C.~J.}\ \bibnamefont {Cramer}}, \bibinfo {author} {\bibfnamefont {G.}~\bibnamefont {Cserey}}, \bibinfo {author} {\bibfnamefont {A.~E.}\ \bibnamefont {DePrince}}, \bibinfo {author} {\bibfnamefont {R.~A.}\ \bibnamefont {DiStasio}}, \bibinfo {author} {\bibfnamefont {A.}~\bibnamefont {Dreuw}}, \bibinfo {author} {\bibfnamefont {B.~D.}\ \bibnamefont {Dunietz}}, \bibinfo {author} {\bibfnamefont {T.~R.}\ \bibnamefont {Furlani}}, \bibinfo {author} {\bibfnamefont {W.~A.}\ \bibnamefont {Goddard}}, \bibinfo {author} {\bibfnamefont {S.}~\bibnamefont {{Hammes-Schiffer}}}, \bibinfo {author} {\bibfnamefont {T.}~\bibnamefont {{Head-Gordon}}}, \bibinfo {author} {\bibfnamefont {W.~J.}\ \bibnamefont {Hehre}}, \bibinfo {author} {\bibfnamefont {C.-P.}\ \bibnamefont {Hsu}}, \bibinfo {author} {\bibfnamefont {T.-C.}\ \bibnamefont {Jagau}}, \bibinfo {author} {\bibfnamefont {Y.}~\bibnamefont {Jung}}, \bibinfo {author} {\bibfnamefont {A.}~\bibnamefont {Klamt}}, \bibinfo {author} {\bibfnamefont {J.}~\bibnamefont {Kong}}, \bibinfo {author} {\bibfnamefont {D.~S.}\ \bibnamefont {Lambrecht}}, \bibinfo {author} {\bibfnamefont {W.}~\bibnamefont {Liang}}, \bibinfo {author} {\bibfnamefont {N.~J.}\ \bibnamefont {Mayhall}}, \bibinfo {author} {\bibfnamefont {C.~W.}\ \bibnamefont {McCurdy}}, \bibinfo {author} {\bibfnamefont {J.~B.}\ \bibnamefont {Neaton}}, \bibinfo {author} {\bibfnamefont {C.}~\bibnamefont {Ochsenfeld}}, \bibinfo {author} {\bibfnamefont {J.~A.}\ \bibnamefont {Parkhill}}, \bibinfo {author} {\bibfnamefont {R.}~\bibnamefont {Peverati}}, \bibinfo {author} {\bibfnamefont {V.~A.}\ \bibnamefont {Rassolov}}, \bibinfo {author} {\bibfnamefont {Y.}~\bibnamefont {Shao}}, \bibinfo {author} {\bibfnamefont {L.~V.}\ \bibnamefont {Slipchenko}}, \bibinfo {author} {\bibfnamefont {T.}~\bibnamefont {Stauch}}, \bibinfo {author} {\bibfnamefont {R.~P.}\ \bibnamefont {Steele}}, \bibinfo {author} {\bibfnamefont {J.~E.}\ \bibnamefont {Subotnik}}, \bibinfo {author} {\bibfnamefont {A.~J.~W.}\ \bibnamefont {Thom}}, \bibinfo {author} {\bibfnamefont {A.}~\bibnamefont {Tkatchenko}}, \bibinfo {author} {\bibfnamefont {D.~G.}\ \bibnamefont {Truhlar}}, \bibinfo {author} {\bibfnamefont {T.}~\bibnamefont {Van~Voorhis}}, \bibinfo {author} {\bibfnamefont {T.~A.}\ \bibnamefont {Wesolowski}}, \bibinfo {author} {\bibfnamefont {K.~B.}\ \bibnamefont {Whaley}}, \bibinfo {author} {\bibfnamefont {H.~L.}\ \bibnamefont {Woodcock}}, \bibinfo {author} {\bibfnamefont {P.~M.}\ \bibnamefont {Zimmerman}}, \bibinfo {author} {\bibfnamefont {S.}~\bibnamefont {Faraji}}, \bibinfo {author} {\bibfnamefont {P.~M.~W.}\ \bibnamefont {Gill}}, \bibinfo {author} {\bibfnamefont {M.}~\bibnamefont {{Head-Gordon}}}, \bibinfo {author} {\bibfnamefont {J.~M.}\ \bibnamefont {Herbert}},\ and\ \bibinfo {author} {\bibfnamefont {A.~I.}\ \bibnamefont {Krylov}},\ }\bibfield  {title} {\enquote {\bibinfo {title} {Software for the frontiers of quantum chemistry: {{An}} overview of developments in the {{Q-Chem}} 5 package},}\ }\href {https://doi.org/10.1063/5.0055522} {\bibfield  {journal} {\bibinfo  {journal} {The . Chem. Phys.}\ }\textbf {\bibinfo {volume} {155}},\ \bibinfo {pages} {084801} (\bibinfo {year} {2021})}\BibitemShut {NoStop}%
\bibitem [{\citenamefont {Sun}\ \emph {et~al.}(2020)\citenamefont {Sun}, \citenamefont {Zhang}, \citenamefont {Banerjee}, \citenamefont {Bao}, \citenamefont {Barbry}, \citenamefont {Blunt}, \citenamefont {Bogdanov}, \citenamefont {Booth}, \citenamefont {Chen}, \citenamefont {Cui}, \citenamefont {Eriksen}, \citenamefont {Gao}, \citenamefont {Guo}, \citenamefont {Hermann}, \citenamefont {Hermes}, \citenamefont {Koh}, \citenamefont {Koval}, \citenamefont {Lehtola}, \citenamefont {Li}, \citenamefont {Liu}, \citenamefont {Mardirossian}, \citenamefont {McClain}, \citenamefont {Motta}, \citenamefont {Mussard}, \citenamefont {Pham}, \citenamefont {Pulkin}, \citenamefont {Purwanto}, \citenamefont {Robinson}, \citenamefont {Ronca}, \citenamefont {Sayfutyarova}, \citenamefont {Scheurer}, \citenamefont {Schurkus}, \citenamefont {Smith}, \citenamefont {Sun}, \citenamefont {Sun}, \citenamefont {Upadhyay}, \citenamefont {Wagner}, \citenamefont {Wang}, \citenamefont {White}, \citenamefont {Whitfield}, \citenamefont {Williamson}, \citenamefont {Wouters}, \citenamefont {Yang}, \citenamefont {Yu}, \citenamefont {Zhu}, \citenamefont {Berkelbach}, \citenamefont {Sharma}, \citenamefont {Sokolov},\ and\ \citenamefont {Chan}}]{pyscf_1}%
  \BibitemOpen
  \bibfield  {author} {\bibinfo {author} {\bibfnamefont {Q.}~\bibnamefont {Sun}}, \bibinfo {author} {\bibfnamefont {X.}~\bibnamefont {Zhang}}, \bibinfo {author} {\bibfnamefont {S.}~\bibnamefont {Banerjee}}, \bibinfo {author} {\bibfnamefont {P.}~\bibnamefont {Bao}}, \bibinfo {author} {\bibfnamefont {M.}~\bibnamefont {Barbry}}, \bibinfo {author} {\bibfnamefont {N.~S.}\ \bibnamefont {Blunt}}, \bibinfo {author} {\bibfnamefont {N.~A.}\ \bibnamefont {Bogdanov}}, \bibinfo {author} {\bibfnamefont {G.~H.}\ \bibnamefont {Booth}}, \bibinfo {author} {\bibfnamefont {J.}~\bibnamefont {Chen}}, \bibinfo {author} {\bibfnamefont {Z.-H.}\ \bibnamefont {Cui}}, \bibinfo {author} {\bibfnamefont {J.~J.}\ \bibnamefont {Eriksen}}, \bibinfo {author} {\bibfnamefont {Y.}~\bibnamefont {Gao}}, \bibinfo {author} {\bibfnamefont {S.}~\bibnamefont {Guo}}, \bibinfo {author} {\bibfnamefont {J.}~\bibnamefont {Hermann}}, \bibinfo {author} {\bibfnamefont {M.~R.}\ \bibnamefont {Hermes}}, \bibinfo {author} {\bibfnamefont {K.}~\bibnamefont {Koh}}, \bibinfo {author} {\bibfnamefont {P.}~\bibnamefont {Koval}}, \bibinfo {author} {\bibfnamefont {S.}~\bibnamefont {Lehtola}}, \bibinfo {author} {\bibfnamefont {Z.}~\bibnamefont {Li}}, \bibinfo {author} {\bibfnamefont {J.}~\bibnamefont {Liu}}, \bibinfo {author} {\bibfnamefont {N.}~\bibnamefont {Mardirossian}}, \bibinfo {author} {\bibfnamefont {J.~D.}\ \bibnamefont {McClain}}, \bibinfo {author} {\bibfnamefont {M.}~\bibnamefont {Motta}}, \bibinfo {author} {\bibfnamefont {B.}~\bibnamefont {Mussard}}, \bibinfo {author} {\bibfnamefont {H.~Q.}\ \bibnamefont {Pham}}, \bibinfo {author} {\bibfnamefont {A.}~\bibnamefont {Pulkin}}, \bibinfo {author} {\bibfnamefont {W.}~\bibnamefont {Purwanto}}, \bibinfo {author} {\bibfnamefont {P.~J.}\ \bibnamefont {Robinson}}, \bibinfo {author} {\bibfnamefont {E.}~\bibnamefont {Ronca}}, \bibinfo {author} {\bibfnamefont {E.~R.}\ \bibnamefont {Sayfutyarova}}, \bibinfo {author} {\bibfnamefont {M.}~\bibnamefont {Scheurer}}, \bibinfo {author} {\bibfnamefont {H.~F.}\ \bibnamefont {Schurkus}}, \bibinfo {author} {\bibfnamefont {J.~E.~T.}\ \bibnamefont {Smith}}, \bibinfo {author} {\bibfnamefont {C.}~\bibnamefont {Sun}}, \bibinfo {author} {\bibfnamefont {S.-N.}\ \bibnamefont {Sun}}, \bibinfo {author} {\bibfnamefont {S.}~\bibnamefont {Upadhyay}}, \bibinfo {author} {\bibfnamefont {L.~K.}\ \bibnamefont {Wagner}}, \bibinfo {author} {\bibfnamefont {X.}~\bibnamefont {Wang}}, \bibinfo {author} {\bibfnamefont {A.}~\bibnamefont {White}}, \bibinfo {author} {\bibfnamefont {J.~D.}\ \bibnamefont {Whitfield}}, \bibinfo {author} {\bibfnamefont {M.~J.}\ \bibnamefont {Williamson}}, \bibinfo {author} {\bibfnamefont {S.}~\bibnamefont {Wouters}}, \bibinfo {author} {\bibfnamefont {J.}~\bibnamefont {Yang}}, \bibinfo {author} {\bibfnamefont {J.~M.}\ \bibnamefont {Yu}}, \bibinfo {author} {\bibfnamefont {T.}~\bibnamefont {Zhu}}, \bibinfo {author} {\bibfnamefont {T.~C.}\ \bibnamefont {Berkelbach}}, \bibinfo {author} {\bibfnamefont {S.}~\bibnamefont {Sharma}}, \bibinfo {author} {\bibfnamefont {A.~Y.}\ \bibnamefont {Sokolov}},\ and\ \bibinfo {author} {\bibfnamefont {G.~K.-L.}\ \bibnamefont {Chan}},\ }\bibfield  {title} {\enquote {\bibinfo {title} {Recent developments in the {{P}} {\textsc{y}} {{SCF}} program package},}\ }\href {https://doi.org/10.1063/5.0006074} {\bibfield  {journal} {\bibinfo  {journal} {J. Chem. Phys.}\ }\textbf {\bibinfo {volume} {153}},\ \bibinfo {pages} {024109} (\bibinfo {year} {2020})}\BibitemShut {NoStop}%
\bibitem [{\citenamefont {Sun}\ \emph {et~al.}(2018)\citenamefont {Sun}, \citenamefont {Berkelbach}, \citenamefont {Blunt}, \citenamefont {Booth}, \citenamefont {Guo}, \citenamefont {Li}, \citenamefont {Liu}, \citenamefont {McClain}, \citenamefont {Sayfutyarova}, \citenamefont {Sharma}, \citenamefont {Wouters},\ and\ \citenamefont {Chan}}]{pyscf_2}%
  \BibitemOpen
  \bibfield  {author} {\bibinfo {author} {\bibfnamefont {Q.}~\bibnamefont {Sun}}, \bibinfo {author} {\bibfnamefont {T.~C.}\ \bibnamefont {Berkelbach}}, \bibinfo {author} {\bibfnamefont {N.~S.}\ \bibnamefont {Blunt}}, \bibinfo {author} {\bibfnamefont {G.~H.}\ \bibnamefont {Booth}}, \bibinfo {author} {\bibfnamefont {S.}~\bibnamefont {Guo}}, \bibinfo {author} {\bibfnamefont {Z.}~\bibnamefont {Li}}, \bibinfo {author} {\bibfnamefont {J.}~\bibnamefont {Liu}}, \bibinfo {author} {\bibfnamefont {J.~D.}\ \bibnamefont {McClain}}, \bibinfo {author} {\bibfnamefont {E.~R.}\ \bibnamefont {Sayfutyarova}}, \bibinfo {author} {\bibfnamefont {S.}~\bibnamefont {Sharma}}, \bibinfo {author} {\bibfnamefont {S.}~\bibnamefont {Wouters}},\ and\ \bibinfo {author} {\bibfnamefont {G.~K.-L.}\ \bibnamefont {Chan}},\ }\bibfield  {title} {\enquote {\bibinfo {title} {P {\textsc{y}} {{SCF}}: The {{Python}}-based simulations of chemistry framework},}\ }\href {https://doi.org/10.1002/wcms.1340} {\bibfield  {journal} {\bibinfo  {journal} {WIREs Comput. Molec. Sci.}\ }\textbf {\bibinfo {volume} {8}},\ \bibinfo {pages} {e1340} (\bibinfo {year} {2018})}\BibitemShut {NoStop}%
\bibitem [{\citenamefont {Sun}(2015)}]{libcint_pyscf_3}%
  \BibitemOpen
  \bibfield  {author} {\bibinfo {author} {\bibfnamefont {Q.}~\bibnamefont {Sun}},\ }\bibfield  {title} {\enquote {\bibinfo {title} {Libcint: {{An}} efficient general integral library for {{Gaussian}} basis functions},}\ }\href {https://doi.org/10.1002/jcc.23981} {\bibfield  {journal} {\bibinfo  {journal} {J. Comput. Chem.}\ }\textbf {\bibinfo {volume} {36}},\ \bibinfo {pages} {1664--1671} (\bibinfo {year} {2015})}\BibitemShut {NoStop}%
\bibitem [{\citenamefont {Motta}\ \emph {et~al.}(2017)\citenamefont {Motta}, \citenamefont {Ceperley}, \citenamefont {Chan}, \citenamefont {Gomez}, \citenamefont {Gull}, \citenamefont {Guo}, \citenamefont {{Jim{\'e}nez-Hoyos}}, \citenamefont {Lan}, \citenamefont {Li}, \citenamefont {Ma}, \citenamefont {Millis}, \citenamefont {Prokof'ev}, \citenamefont {Ray}, \citenamefont {Scuseria}, \citenamefont {Sorella}, \citenamefont {Stoudenmire}, \citenamefont {Sun}, \citenamefont {Tupitsyn}, \citenamefont {White}, \citenamefont {Zgid}, \citenamefont {Zhang},\ and\ \citenamefont {{Simons Collaboration on the Many-Electron Problem}}}]{mrci}%
  \BibitemOpen
  \bibfield  {author} {\bibinfo {author} {\bibfnamefont {M.}~\bibnamefont {Motta}}, \bibinfo {author} {\bibfnamefont {D.~M.}\ \bibnamefont {Ceperley}}, \bibinfo {author} {\bibfnamefont {G.~K.-L.}\ \bibnamefont {Chan}}, \bibinfo {author} {\bibfnamefont {J.~A.}\ \bibnamefont {Gomez}}, \bibinfo {author} {\bibfnamefont {E.}~\bibnamefont {Gull}}, \bibinfo {author} {\bibfnamefont {S.}~\bibnamefont {Guo}}, \bibinfo {author} {\bibfnamefont {C.~A.}\ \bibnamefont {{Jim{\'e}nez-Hoyos}}}, \bibinfo {author} {\bibfnamefont {T.~N.}\ \bibnamefont {Lan}}, \bibinfo {author} {\bibfnamefont {J.}~\bibnamefont {Li}}, \bibinfo {author} {\bibfnamefont {F.}~\bibnamefont {Ma}}, \bibinfo {author} {\bibfnamefont {A.~J.}\ \bibnamefont {Millis}}, \bibinfo {author} {\bibfnamefont {N.~V.}\ \bibnamefont {Prokof'ev}}, \bibinfo {author} {\bibfnamefont {U.}~\bibnamefont {Ray}}, \bibinfo {author} {\bibfnamefont {G.~E.}\ \bibnamefont {Scuseria}}, \bibinfo {author} {\bibfnamefont {S.}~\bibnamefont {Sorella}}, \bibinfo {author} {\bibfnamefont {E.~M.}\ \bibnamefont {Stoudenmire}}, \bibinfo {author} {\bibfnamefont {Q.}~\bibnamefont {Sun}}, \bibinfo {author} {\bibfnamefont {I.~S.}\ \bibnamefont {Tupitsyn}}, \bibinfo {author} {\bibfnamefont {S.~R.}\ \bibnamefont {White}}, \bibinfo {author} {\bibfnamefont {D.}~\bibnamefont {Zgid}}, \bibinfo {author} {\bibfnamefont {S.}~\bibnamefont {Zhang}},\ and\ \bibinfo {author} {\bibnamefont {{Simons Collaboration on the Many-Electron Problem}}},\ }\bibfield  {title} {\enquote {\bibinfo {title} {Towards the {{Solution}} of the {{Many-Electron Problem}} in {{Real Materials}}: {{Equation}} of {{State}} of the {{Hydrogen Chain}} with {{State-of-the-Art Many-Body Methods}}},}\ }\href {https://doi.org/10.1103/PhysRevX.7.031059} {\bibfield  {journal} {\bibinfo  {journal} {Phys. Rev. X}\ }\textbf {\bibinfo {volume} {7}},\ \bibinfo {pages} {031059} (\bibinfo {year} {2017})}\BibitemShut {NoStop}%
\bibitem [{\citenamefont {Gilbert}, \citenamefont {Besley},\ and\ \citenamefont {Gill}(2008)}]{gilbert2008self}%
  \BibitemOpen
  \bibfield  {author} {\bibinfo {author} {\bibfnamefont {A.~T.}\ \bibnamefont {Gilbert}}, \bibinfo {author} {\bibfnamefont {N.~A.}\ \bibnamefont {Besley}},\ and\ \bibinfo {author} {\bibfnamefont {P.~M.}\ \bibnamefont {Gill}},\ }\bibfield  {title} {\enquote {\bibinfo {title} {Self-consistent field calculations of excited states using the maximum overlap method (mom)},}\ }\href {https://doi.org/10.1021/jp801738f} {\bibfield  {journal} {\bibinfo  {journal} {J. Phys. Chem. A}\ }\textbf {\bibinfo {volume} {112}},\ \bibinfo {pages} {13164--13171} (\bibinfo {year} {2008})}\BibitemShut {NoStop}%
\bibitem [{\citenamefont {Paetow}\ and\ \citenamefont {Neugebauer}(2025)}]{paetow2025excited}%
  \BibitemOpen
  \bibfield  {author} {\bibinfo {author} {\bibfnamefont {L.}~\bibnamefont {Paetow}}\ and\ \bibinfo {author} {\bibfnamefont {J.}~\bibnamefont {Neugebauer}},\ }\bibfield  {title} {\enquote {\bibinfo {title} {Excited state dipole moments from $\delta$scf: a benchmark},}\ }\href {https://doi.org/10.1039/D5CP01695A} {\bibfield  {journal} {\bibinfo  {journal} {Phys. Chem. Chem. Phys.}\ }\textbf {\bibinfo {volume} {27}},\ \bibinfo {pages} {16354--16370} (\bibinfo {year} {2025})}\BibitemShut {NoStop}%
\bibitem [{\citenamefont {Christiansen}, \citenamefont {Koch},\ and\ \citenamefont {J{\o}rgensen}(1995)}]{christiansen1995second}%
  \BibitemOpen
  \bibfield  {author} {\bibinfo {author} {\bibfnamefont {O.}~\bibnamefont {Christiansen}}, \bibinfo {author} {\bibfnamefont {H.}~\bibnamefont {Koch}},\ and\ \bibinfo {author} {\bibfnamefont {P.}~\bibnamefont {J{\o}rgensen}},\ }\bibfield  {title} {\enquote {\bibinfo {title} {The second-order approximate coupled cluster singles and doubles model cc2},}\ }\href {https://doi.org/10.1016/0009-2614(95)00841-Q} {\bibfield  {journal} {\bibinfo  {journal} {Chem. Phys. Lett.}\ }\textbf {\bibinfo {volume} {243}},\ \bibinfo {pages} {409--418} (\bibinfo {year} {1995})}\BibitemShut {NoStop}%
\bibitem [{\citenamefont {Trofimov}\ and\ \citenamefont {Schirmer}(1995)}]{trofimov1995efficient}%
  \BibitemOpen
  \bibfield  {author} {\bibinfo {author} {\bibfnamefont {A.~B.}\ \bibnamefont {Trofimov}}\ and\ \bibinfo {author} {\bibfnamefont {J.}~\bibnamefont {Schirmer}},\ }\bibfield  {title} {\enquote {\bibinfo {title} {An efficient polarization propagator approach to valence electron excitation spectra},}\ }\href {https://doi.org/10.1088/0953-4075/28/12/003} {\bibfield  {journal} {\bibinfo  {journal} {J. Phys. B: At. Mol. Opt. Phys.}\ }\textbf {\bibinfo {volume} {28}},\ \bibinfo {pages} {2299} (\bibinfo {year} {1995})}\BibitemShut {NoStop}%
\bibitem [{\citenamefont {Hohenstein}, \citenamefont {Parrish},\ and\ \citenamefont {Mart{\'\i}nez}(2012)}]{hohenstein2012tensor}%
  \BibitemOpen
  \bibfield  {author} {\bibinfo {author} {\bibfnamefont {E.~G.}\ \bibnamefont {Hohenstein}}, \bibinfo {author} {\bibfnamefont {R.~M.}\ \bibnamefont {Parrish}},\ and\ \bibinfo {author} {\bibfnamefont {T.~J.}\ \bibnamefont {Mart{\'\i}nez}},\ }\bibfield  {title} {\enquote {\bibinfo {title} {Tensor hypercontraction density fitting. i. quartic scaling second-and third-order m{\o}ller-plesset perturbation theory},}\ }\href {https://doi.org/10.1063/1.4732310} {\bibfield  {journal} {\bibinfo  {journal} {J. Chem. Phys.}\ }\textbf {\bibinfo {volume} {137}} (\bibinfo {year} {2012}),\ 10.1063/1.4732310}\BibitemShut {NoStop}%
\bibitem [{\citenamefont {Parrish}\ \emph {et~al.}(2012)\citenamefont {Parrish}, \citenamefont {Hohenstein}, \citenamefont {Mart{\'\i}nez},\ and\ \citenamefont {Sherrill}}]{parrish2012tensor}%
  \BibitemOpen
  \bibfield  {author} {\bibinfo {author} {\bibfnamefont {R.~M.}\ \bibnamefont {Parrish}}, \bibinfo {author} {\bibfnamefont {E.~G.}\ \bibnamefont {Hohenstein}}, \bibinfo {author} {\bibfnamefont {T.~J.}\ \bibnamefont {Mart{\'\i}nez}},\ and\ \bibinfo {author} {\bibfnamefont {C.~D.}\ \bibnamefont {Sherrill}},\ }\bibfield  {title} {\enquote {\bibinfo {title} {Tensor hypercontraction. ii. least-squares renormalization},}\ }\href {https://doi.org/10.1063/1.4768233} {\bibfield  {journal} {\bibinfo  {journal} {J. Chem. Phys.}\ }\textbf {\bibinfo {volume} {137}} (\bibinfo {year} {2012}),\ 10.1063/1.4768233}\BibitemShut {NoStop}%
\bibitem [{\citenamefont {Matthews}(2020)}]{matthews2020improved}%
  \BibitemOpen
  \bibfield  {author} {\bibinfo {author} {\bibfnamefont {D.~A.}\ \bibnamefont {Matthews}},\ }\bibfield  {title} {\enquote {\bibinfo {title} {Improved grid optimization and fitting in least squares tensor hypercontraction},}\ }\href {https://doi.org/10.1021/acs.jctc.9b01205} {\bibfield  {journal} {\bibinfo  {journal} {J. Chem. Theory Comput.}\ }\textbf {\bibinfo {volume} {16}},\ \bibinfo {pages} {1382--1385} (\bibinfo {year} {2020})}\BibitemShut {NoStop}%
\bibitem [{\citenamefont {Scott}, \citenamefont {Backhouse},\ and\ \citenamefont {Booth}(2023)}]{scott2023moment}%
  \BibitemOpen
  \bibfield  {author} {\bibinfo {author} {\bibfnamefont {C.~J.}\ \bibnamefont {Scott}}, \bibinfo {author} {\bibfnamefont {O.~J.}\ \bibnamefont {Backhouse}},\ and\ \bibinfo {author} {\bibfnamefont {G.~H.}\ \bibnamefont {Booth}},\ }\bibfield  {title} {\enquote {\bibinfo {title} {A “moment-conserving” reformulation of gw theory},}\ }\href {https://doi.org/10.1063/5.0143291} {\bibfield  {journal} {\bibinfo  {journal} {J. Chem. Phys.}\ }\textbf {\bibinfo {volume} {158}} (\bibinfo {year} {2023}),\ 10.1063/5.0143291}\BibitemShut {NoStop}%
\bibitem [{\citenamefont {Backhouse}\ \emph {et~al.}(2025)\citenamefont {Backhouse}, \citenamefont {Allen}, \citenamefont {Scott},\ and\ \citenamefont {Booth}}]{backhouse2025self}%
  \BibitemOpen
  \bibfield  {author} {\bibinfo {author} {\bibfnamefont {O.~J.}\ \bibnamefont {Backhouse}}, \bibinfo {author} {\bibfnamefont {M.~K.}\ \bibnamefont {Allen}}, \bibinfo {author} {\bibfnamefont {C.~J.}\ \bibnamefont {Scott}},\ and\ \bibinfo {author} {\bibfnamefont {G.~H.}\ \bibnamefont {Booth}},\ }\bibfield  {title} {\enquote {\bibinfo {title} {Self-consistent gw via conservation of spectral moments},}\ }\href {https://doi.org/10.1021/acs.jctc.5c00657} {\bibfield  {journal} {\bibinfo  {journal} {J. Chem. Theory Comput.}\ }\textbf {\bibinfo {volume} {21}},\ \bibinfo {pages} {8963--8981} (\bibinfo {year} {2025})}\BibitemShut {NoStop}%
\bibitem [{\citenamefont {T{\"o}lle}\ and\ \citenamefont {Kin-Lic~Chan}(2024)}]{tolle2024ab}%
  \BibitemOpen
  \bibfield  {author} {\bibinfo {author} {\bibfnamefont {J.}~\bibnamefont {T{\"o}lle}}\ and\ \bibinfo {author} {\bibfnamefont {G.}~\bibnamefont {Kin-Lic~Chan}},\ }\bibfield  {title} {\enquote {\bibinfo {title} {Ab-g0w0: A practical g0w0 method without frequency integration based on an auxiliary boson expansion},}\ }\href {https://doi.org/10.1021/jp801738f} {\bibfield  {journal} {\bibinfo  {journal} {J. Chem. Phys.}\ }\textbf {\bibinfo {volume} {160}} (\bibinfo {year} {2024}),\ 10.1021/jp801738f}\BibitemShut {NoStop}%
\end{thebibliography}
\end{document}